\documentclass[fleqn,usenatbib]{mnras}

\usepackage{newtxtext,newtxmath}

\usepackage[T1]{fontenc}

\DeclareRobustCommand{\VAN}[3]{#2}
\let\VANthebibliography\thebibliography
\def\thebibliography{\DeclareRobustCommand{\VAN}[3]{##3}\VANthebibliography}

\usepackage{graphicx}	
\usepackage{amsmath}	

\usepackage{acronym}    
\usepackage{xspace}

\newcommand{\gwcosmo}{\texttt{gwcosmo}\xspace} 
\newcommand{\kmsMpc}{km s$^{-1}$ Mpc$^{-1}$\xspace}
\newcommand{\sqdeg}{\ensuremath{\text{deg}^2}\xspace}
\newcommand{\eg}{\emph{e.g.~}} 

\newcommand{\HubbleMeasCombinedCounterpartFLATLOG}{\ensuremath{68.7^{+17.0}_{-7.8}}\xspace}
\newcommand{\HubbleMeasCombinedCounterpartFLATLOGpixel}{\ensuremath{68.8^{+15.9}_{-7.8}}}

\makeatletter
\g@addto@macro\bfseries{\boldmath}
\makeatother

\acrodef{H0}[$H_0$]{the Hubble constant}
\acrodef{GW}[GW]{gravitational wave}
\acrodef{EM}[EM]{electromagnetic}
\acrodef{BNS}[BNS]{binary neutron star}
\acrodef{BBH}[BBH]{binary black hole}
\acrodef{LCDM}[$\Lambda$CDM]{$\Lambda$-cold-dark-matter}
\acrodef{SNR}[SNR]{signal-to-noise ratio}
\acrodef{ra}[RA]{right ascension}
\acrodef{dec}[dec]{declination}
\acrodef{los}[LOS]{line-of-sight}
\acrodef{KDE}[KDE]{kernel density estimate}

\acrodef{DES}[DES]{Dark Energy Survey}
\acrodef{GLADE}[GLADE]{Galaxy List for the Advanced Detector Era}

\acrodef{mth}[$m_\text{th}$]{apparent magnitude threshold}

\title[A Pixelated Approach to Galaxy Catalogue Incompleteness]{A Pixelated Approach to Galaxy Catalogue Incompleteness: Improving the Dark Siren Measurement of the Hubble Constant}

\author[R. Gray et al.]{
R. Gray,$^{1,2}$\thanks{E-mail: rachel.gray@ligo.org}
C. Messenger,$^{2}$
J. Veitch$^{2}$
\\
$^{1}$Department of Physics and Astronomy, Queen Mary University of London, Mile End Road, London, E1 4NS, United Kingdom\\
$^{2}$SUPA, School of Physics and Astronomy, University of Glasgow, Glasgow G12 8QQ, United Kingdom
}

\date{Accepted 2022-02-08. Received 2022-02-04; in original form 2021-12-16}

\pubyear{2022}

\begin{document}
\label{firstpage}
\pagerange{\pageref{firstpage}--\pageref{lastpage}}
\maketitle

\begin{abstract}
The use of gravitational wave standard sirens for cosmological analyses is becoming well known, with particular interest in measuring the Hubble constant, $H_0$, 
and in shedding light on the current tension between early- and late-time measurements. The current tension is over $4\sigma$ and standard sirens will be able to provide a completely independent measurement. Dark sirens (binary black hole or neutron star mergers with no electromagnetic counterparts) can be informative if the missing redshift information is provided through the use of galaxy catalogues to identify potential host galaxies of the merger. However, galaxy catalogue incompleteness affects this analysis, and accurate modelling of it is essential for obtaining an unbiased measurement of $H_0$. Previously most methods have assumed uniform completeness within the sky area of a gravitational wave event. This paper presents an updated methodology in which the completeness of the galaxy catalogue is estimated in a directionally-dependent matter, by pixelating the sky and computing the completeness of the galaxy catalogue along each line of sight. The $H_0$ inference for a single event is carried out on a pixel-by-pixel basis, and the pixels are combined for the final result. A reanalysis of the events in the first gravitational wave transient catalogue (GWTC-1) leads to an improvement on the measured value of $H_0$ of approximately 5\% compared to the 68.3\% highest-density interval of the equivalent LIGO and Virgo result, with $H_0=$ {\HubbleMeasCombinedCounterpartFLATLOGpixel} \kmsMpc.
\end{abstract}



\acresetall
\section{Introduction}

The use of \ac{GW} signals from the mergers of compact binaries (black holes or neutron stars) to constrain cosmological parameters has gained traction in recent years. The tension between early- and late-time measurements of \ac{H0} remains above $4\sigma$ at the time of writing \citep{Aghanim:2018eyx,Riess:2019cxk}, making novel and independent measurements of \ac{H0} of particular interest at present. 
\acp{GW} have the luminosity distance of the source directly encoded within them, without requiring external calibration, making them independent distance measures \citep{Schutz:1986}. Preferring the early-time measurements would indicate that the source of the tension lies in systematic errors between the different measurement techniques, while preferring the local, late-time measurement would indicate that the source of the tension lies in fundamental physics, and that the current cosmological model, \ac{LCDM}, does not adequately fit our universe to the level of accuracy at which it is now being measured.

The first 3 observing runs of Advanced LIGO and Virgo have provided over 50 \ac{GW} detections \citep{O2:catalog,PhysRevX.11.021053,NSBH2021}. Of these, only one (the well-known \ac{BNS} GW170817) has been observed with a confirmed \ac{EM} counterpart \citep{GW170817:discovery,GW170817:MMA}. In order to use the remaining detections for a measurement of \ac{H0}, a different method of obtaining redshift information about the sources is required.
Of particular interest is the \emph{statistical} or \emph{galaxy catalogue} dark siren method, in which galaxy catalogues are used to identify the galaxies within the localisation volume of a \ac{GW} event, and they are all treated as potential hosts \citep{Schutz:1986,DelPozzo:2012}. The contribution from an individual event is less informative than in the counterpart case, but combining information from multiple events reduces the uncertainty and allows an additional constraint on \ac{H0} to be made (see, \eg \citet{MacLeod:2007jd,Chen:2017rfc,Fishbach:2018gjp,Soares-Santos:2019irc,Gray:2019ksv,GW1908142020,GW190814:DES,Vasylyev2020,Finke:2021aom}. An alternative, cross-correlating \ac{GW} events with galaxies of known redshift, is presented in \citet{2021PhRvD.103d3520M}). One important aspect of this method is acknowledging that galaxy catalogues are incomplete, and therefore may not contain the real host galaxy of the \ac{GW} event. In order to account for this, an incompleteness correction must be applied. The galaxy catalogue methodology, as implemented in the \gwcosmo codebase \citep{Gray:2019ksv} was applied to the \ac{GW} detections from the first \ac{GW} transient catalogue (GWTC-1) in \citet{O2H0paper}, leading to a measurement of $H_0=$ \HubbleMeasCombinedCounterpartFLATLOG \kmsMpc.\footnote{\gwcosmo is available at \url{https://git.ligo.org/lscsoft/gwcosmo}.}

The result in \citet{O2H0paper} showed proof-of-principle and highlighted many of the upcoming challenges within the field. The analysis  also had several limitations. One of note was the assumption that galaxy catalogue completeness (due to the limited sensitivity of the telescope(s) carrying out the survey) is uniform with the sky area of each \ac{GW} event. However, the GLADE catalogue \citep{Dalya:2018cnd} (used to inform the redshift prior for the majority of the GWTC-1 events in \citet{O2H0paper}) is a composite catalogue, made up of multiple different surveys, and as a result has highly variable completeness across the sky. Even for a galaxy catalogue that is made from a single survey, limited by a single telescope's sensitivity, the Milky Way band means that completeness cannot ever be truly uniform across the whole sky. Another approximation of note was that the sky and distance information from a \ac{GW} event were treated as uncorrelated, rather than being treated as fully 3D. This was done in order to aid computational efficiency, at the expense of losing some constraining power from the \ac{GW} events.

The issue of non-uniform catalogue incompleteness was addressed in \citet{Finke:2021aom}, in which areas of the sky with similar levels of completeness were grouped together, and a completeness correction was applied. The methods of completeness correction used in \citet{Finke:2021aom} 
differ to the one used here (and in \citet{Gray:2019ksv,O2H0paper}), where the limiting magnitude of the telescope is used to assess the probability of any galaxy being `seen'. Under the same set of assumptions these two methods should give consistent results, however the methodology outlined here allows for a more complete treatment of the \ac{GW} mass priors (see section \ref{sec:los distance} for details), including consistency between the prior applied to individual events and the prior used to compute \ac{GW} selection effects, which is necessary for an unbiased estimate of \ac{H0}.

The assumption of uniform catalogue completeness is problematic in two ways. When a uniform completeness correction is applied across a non-uniform patch of sky, there there will be areas for which the completeness is over-estimated, and areas for which it is under-estimated. The contribution from galaxies in areas where it is over-estimated will be artificially inflated, which risks biasing the result if the host galaxy is not inside the galaxy catalogue. Where it is under-estimated, useful information about the redshift distribution of galaxies at higher redshifts will be diluted unnecessarily by the completeness correction. 

Fortunately there is an extension to the method which addresses both of these limitations at once. This extension entails pixelating the sky into equally sized pieces, and analysing each independently, using a line-of-sight distance distribution for the \ac{GW} event within each pixel, and an estimate of the completeness within  that section of the sky. Both of the approximations described above are removed, making the analysis more robust and, theoretically, more informative.

This paper builds upon the work in \citet{Gray:2019ksv} to allow variations in galaxy catalogue completeness to be robustly estimated. 
Section \ref{sec:pixelmethod} outlines the Bayesian implementation of the pixelated method. Section \ref{sec:pixelpracticalities} discusses the practicalities of implementing it in the \gwcosmo codebase and applying it to real data. Section \ref{sec:pixelO2H0} reanalyses the GWTC-1 detections using the pixelated method, while keeping all other assumptions the same as in \citet{O2H0paper}, and quantifies the improvement to the results. Section \ref{sec:pixelconclusion} discusses potential extensions for the pixelated method in the future, and concludes the paper.

\section{Methodology}
\label{sec:pixelmethod}
A measurement of \ac{H0} can be made using a set of $N_{\text{det}}$ detected \ac{GW} events. Expressing this in a Bayesian form, the posterior probability on \ac{H0} can be written as follows:
\begin{equation}\label{Eq:posteriormain}
\begin{aligned}
p(H_0|\{x_{\text{GW}}\},&\{D_{\text{GW}}\},I)\propto \\ &p(H_0|I)p(N_{\text{det}}|H_0,I)\prod_i^{N_{\text{det}}} p({x_{\text{GW}}}_i|{D_{\text{GW}}}_i,H_0,I).
\end{aligned}
\end{equation}
Here $\{x_{\text{GW}}\}$ is a set of \ac{GW} data, and each $x_{\text{GW}}$ corresponds to a \ac{GW} detection (denoted by $D_{\text{GW}}$). In general an event is deemed detected if some detection statistic associated with $x_{\text{GW}}$ passes a threshold, such as the \ac{SNR} of the event passing some \ac{SNR} threshold for the \ac{GW} detector network. The term $p(H_0|I)$ is the prior on \ac{H0}, and $p(N_{\text{det}}|H_0,I)$ is the probability of detecting $N_{\text{det}}$ events for a given value of \ac{H0} (intrinsically linked to the astrophysical rate of events, $R$, but independent of \ac{H0} when a prior of $1/R$ is used \citep{Fishbach:2018edt}). The final term is a product over $N_{\text{det}}$ individual event likelihoods. The parameter $I$ is a placeholder which contains any additional information which is not explicitly stated (such as the underlying cosmological and population models).

In \citet{Gray:2019ksv}, when considering the likelihood for a single \ac{GW} event, the first step was to marginalise over the probability that the host galaxy of the event is, or is not, inside the catalogue. Here it is first necessary to consider the likelihood's dependence on sky direction, $\Omega$. This can be written as follows:
\begin{equation}
\begin{aligned} \label{Eq:likelihood_continuous}
p(x_{\text{GW}}|D_{\text{GW}},H_0,I) = \int p(x_{\text{GW}},\Omega|D_{\text{GW}},H_0,I) \;d\Omega.
\end{aligned} 
\end{equation}
Instead of considering the continuous variable $\Omega$, it is necessary to switch to a discrete approximation: that of splitting the sky into $N_\text{pix}$ equally-sized pieces, which are later summed. Doing so, Eq. \ref{Eq:likelihood_continuous} becomes
\begin{equation}
\begin{aligned}
p&(x_{\text{GW}}|D_{\text{GW}},H_0,I) \\&= \sum^{N_{\text{pix}}}_i p(x_{\text{GW}}|\Omega_i,D_{\text{GW}},H_0,I) p(\Omega_i|D_{\text{GW}},H_0,I) ,
\\ & = \sum^{N_{\text{pix}}}_i p(x_{\text{GW}}|\Omega_i,D_{\text{GW}},H_0,I) \dfrac{p(D_{\text{GW}}|\Omega_i,H_0,I)p(\Omega_i|H_0,I)}{p(D_{\text{GW}}|H_0,I)} .
\end{aligned} 
\end{equation}
It is possible to further simplify this expression if the probability of detection is assumed to be uniform over the sky --- a reasonable assumption, as it is averaged over the full length of an observing run, and so the rotation of the Earth blurs out much of the sky-dependence \citep{2017ApJ...835...31C}. The term $p(D_{\text{GW}}|\Omega_i,H_0,I)$ loses its dependence on $\Omega_i$ and cancels with the denominator. Making this approximation ignores the mild declination-dependence that the probability of detection retains, which is expected to have only a minor impact on the result (a full investigation of which is left for the future).
Acknowledging that $p(\Omega_i|H_0,I)$ is independent of \ac{H0} in an isotropic universe gives the following:
\begin{equation}
\begin{aligned}
p(x_{\text{GW}}|D_{\text{GW}},H_0,I) &= \sum^{N_{\text{pix}}}_i p(x_{\text{GW}}|\Omega_i,D_{\text{GW}},H_0,I) p(\Omega_i|I), \\
&= \dfrac{1}{N_{\text{pix}}} \sum^{N_{\text{pix}}}_i p(x_{\text{GW}}|\Omega_i,D_{\text{GW}},H_0,I),
\end{aligned} 
\end{equation}
where the equally-sized pixels mean that $p(\Omega_i|I)$ becomes a constant corresponding to the fraction of the surface area of a sphere which one pixel covers.
It is worth noting that any pixels with zero \ac{GW} support will evaluate to zero, and so the sum over $N_{\text{pix}}$ can be reduced to a sum over $N_{\text{GWpix}}$ with no impact to the result (for clarity, the pre-factor would remain $1/N_\text{pix}$ in this case).

Now that the likelihood has been pixelated, it can be marginalised over the possibility of the host being inside ($G$) or outside ($\bar{G}$) the galaxy catalogue. This leads to the likelihood within a single pixel taking the following form:
\begin{equation}
\label{Eq:pixelG}
\begin{aligned}
p(x_{\text{GW}}|\Omega_i,&D_{\text{GW}},H_0,I) \\=  &p(x_{\text{GW}}|\Omega_i,G,D_{\text{GW}},H_0,I) p(G|\Omega_i,D_{\text{GW}},H_0,I) \\
&+ p(x_{\text{GW}}|\Omega_i,\bar{G},D_{\text{GW}},H_0,I) p(\bar{G}|\Omega_i,D_{\text{GW}},H_0,I),
\end{aligned}
\end{equation}
Here, $p(x_{\text{GW}}|\Omega_i,G,D_{\text{GW}},H_0,I)$ is the likelihood when the host galaxy is inside the catalogue, for which the redshift prior will consist of the galaxies in the galaxy catalogue which lie within pixel $i$. The redshift uncertainty of each galaxy is assumed to be Gaussian, the standard deviation of which is provided by the galaxy catalogue, and is marginalised over. The term $p(x_{\text{GW}}|\Omega_i,\bar{G},D_{\text{GW}},H_0,I)$ is the likelihood when the host galaxy is \emph{not} inside the catalogue, for which the redshift prior will be, in the simplest case, an uninformative uniform in co-moving volume distribution. Alternatively a more complex \ac{GW} rate evolution model, which allows the merger rate of binaries to be redshift-dependent, can be assumed. These likelihoods are weighted by the probability that the host galaxy is inside the catalogue or outside it (assuming it lies in pixel $i$), which is determined using the \ac{mth} of that pixel, as well as an assumption about the luminosity distribution of galaxies in the universe -- that they follow, \eg a Schechter function \citep{Schechter:1976}. See the appendix of \citet{Gray:2019ksv} for more details.

The value of \ac{mth} can now be chosen on a by-pixel basis, allowing varying catalogue incompleteness to be accurately taken into account, modulo the choice of pixel size.  There is also the possibility that some pixels will be empty (contain no galaxies due to \eg obscuration by the Milky Way band), in which case Eq. \ref{Eq:pixelG} simplifies to the ``empty catalogue'' case  in which $m_\text{th} \rightarrow -\infty$, leading to $p(G|\Omega_i,D_{\text{GW}},H_0,I)=0$ and $p(\bar{G}|\Omega_i,D_{\text{GW}},H_0,I)=1$. The contribution of that pixel then comes fully from the $p(x_{\text{GW}}|\Omega_i,\bar{G},D_{\text{GW}},H_0,I)$ term.

Additionally, for each pixel $p(x_{\text{GW}}|\Omega_i,H_0,I)$ can be approximated as the \ac{GW} information corresponding to the patch of sky covered by pixel $i$, meaning there is no longer a requirement to separate \ac{GW} sky and distance information.  The distance posterior along the line-of-sight of each pixel can be estimated, making this method inherently ``3D''.

\section{Pixelating \texttt{gwcosmo}: practicalities}
\label{sec:pixelpracticalities}

In order to implement the pixel-based method described in Section \ref{sec:pixelmethod}, \texttt{healpy}, a Python implementation of \texttt{HEALPix} which handles pixelated data on a sphere, is used \citep{Zonca2019,2005ApJ...622..759G}.\footnote{\url{http://healpix.sf.net}} It allows the user to split the sky into equally sized pixels, for a choice of resolutions, where the resolution is set by a parameter called nside. The total number of pixels the sky is divided into is set by $12 \times \text{nside}^2$, where nside is must be a power of 2. At its lowest resolution, the sky is divided into 12 pixels of equal area.  A one-step increase in resolution (which corresponds to doubling nside) divides each pixel into 4 further pixels.

In \gwcosmo's case, \texttt{healpy} allows data points with known \ac{ra} and \ac{dec} (\eg \ac{GW} posterior samples, or galaxies) to be uniquely associated to a specific pixel. As increasing or decreasing the resolution by one step involves either dividing one existing pixel into 4, or combining 4 pixels into one, this opens up the possibility of combining different resolution \texttt{healpy} maps within the same analysis, without the danger of double-counting data. This is useful because variations in the \acl{los} distance estimate for a \ac{GW} across its sky area, and variations in the \ac{mth} of a galaxy catalogue across the same area, are not (necessarily) on the same scale.

\begin{figure*}
\includegraphics[width=0.8\linewidth]{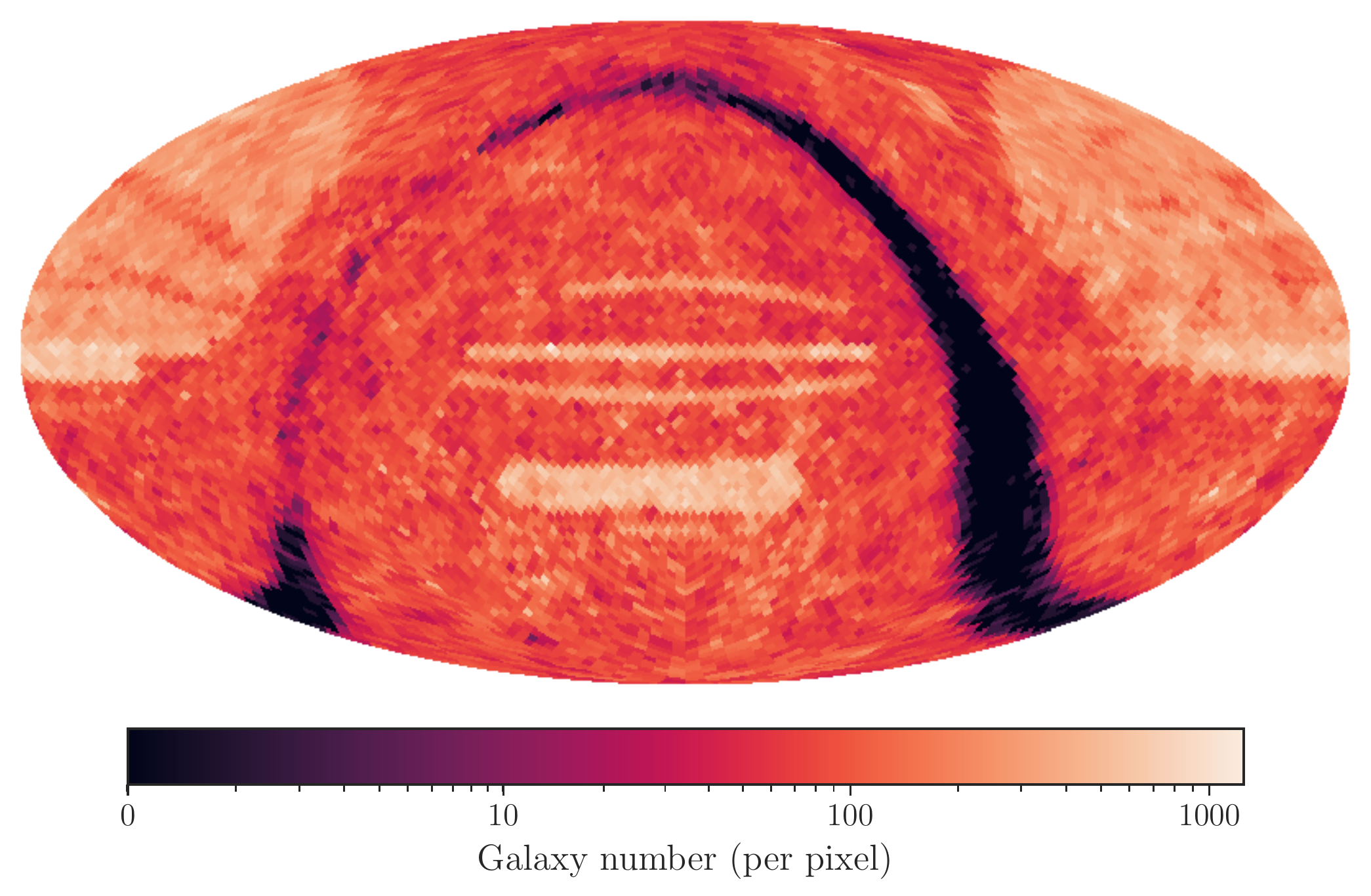}
\\
\includegraphics[width=0.49\linewidth]{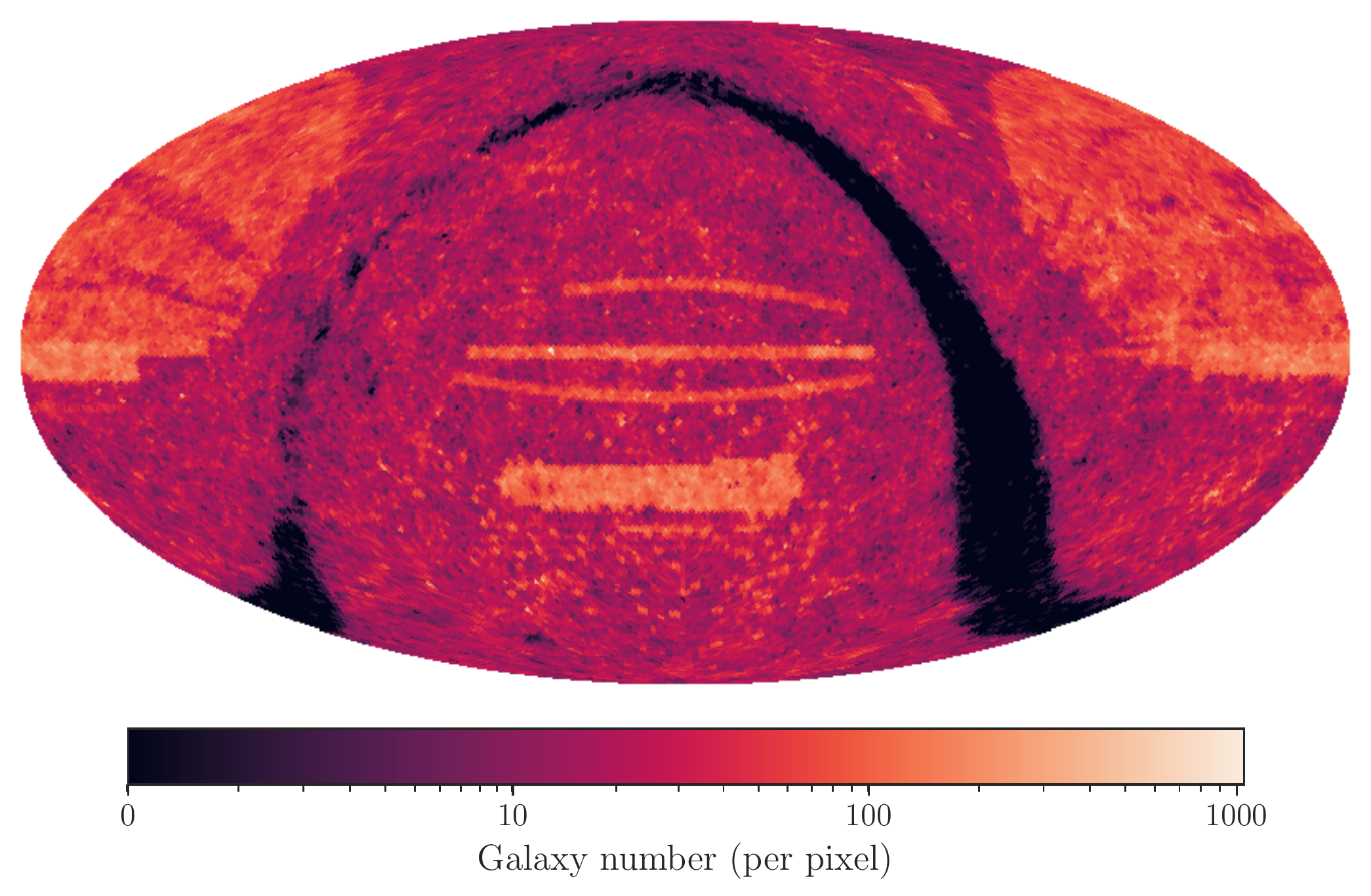}
\includegraphics[width=0.49\linewidth]{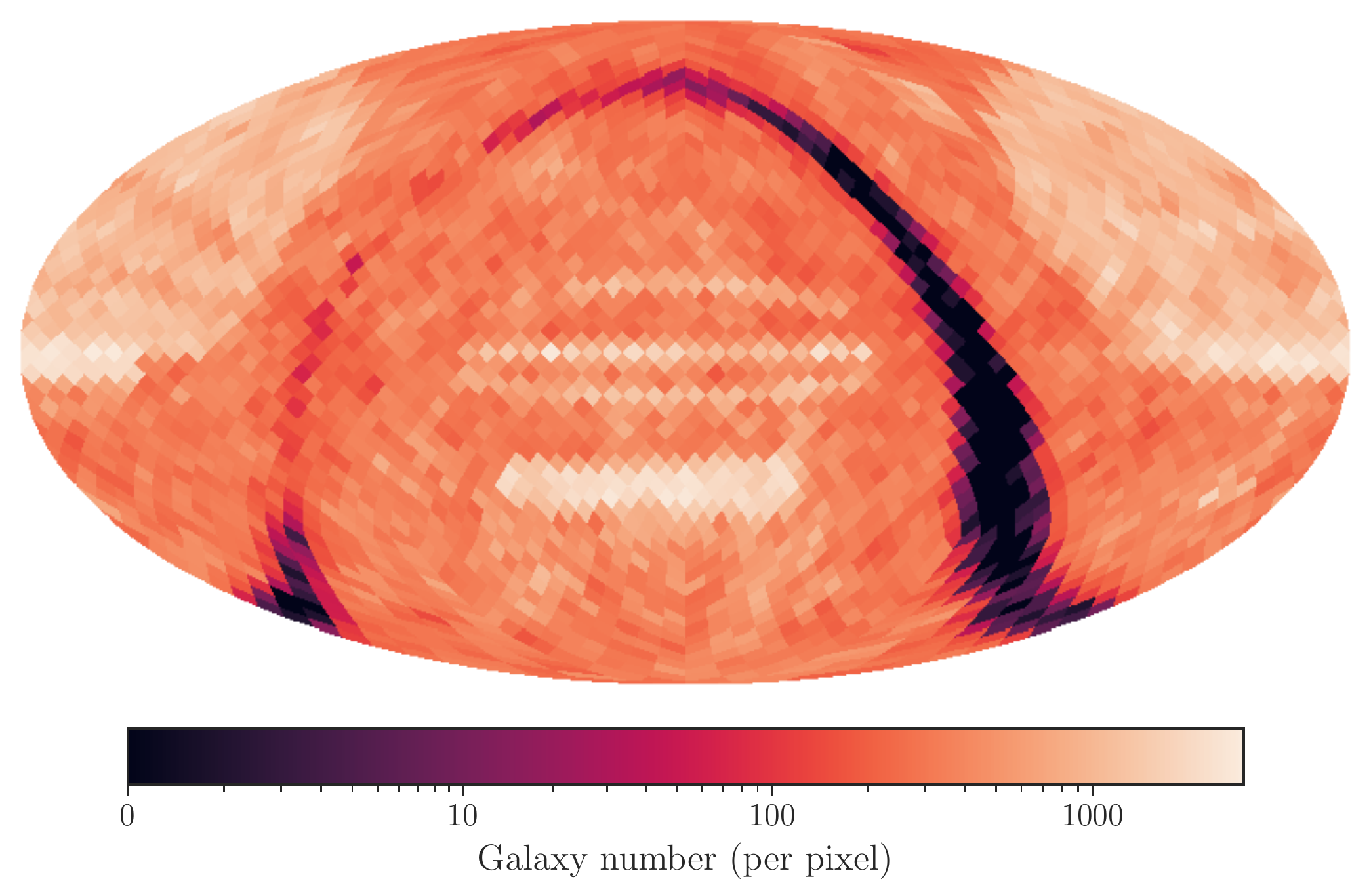}
\caption[Different resolutions of the galaxy number density map for GLADE 2.4.]{Different resolutions of the galaxy number density map for GLADE 2.4 \citep{Dalya:2018cnd}. Brighter colours correspond to a higher number density, darker colours to a lower one. Black pixels are empty, covered by the Milky Way band. \textit{Top panel:} nside=32. The sky is divided into 12,288 pixels, each covering an area of $3.36 \text{ deg}^2$. \textit{Bottom left panel:} nside=64. The sky is divided into 49,152 pixels, each covering an area of $0.84 \text{ deg}^2$. \textit{Bottom right panel:} nside=16. The sky is divided into 3072 pixels, each covering an area of $13.4 \text{ deg}^2$. }
\label{Fig:GLADE_galres}
\end{figure*}

Take, for example, the GLADE 2.4 catalogue \citep{Dalya:2018cnd}.  It is a composite catalogue, made up of many different surveys. Those surveys cover different patches of the sky, and have overlapped in some areas. Taking a \texttt{healpy} nside of 32 divides the catalogue into 12,288 pixels, each covering approximately $3.36 \text{ deg}^2$. The top panel of Fig. \ref{Fig:GLADE_galres} visualises this, and shows how the number density of galaxies varies on that scale. Many sharp features can be seen, and are adequately represented by this resolution, though they could be better represented with a higher one (\eg nside=64, the lower left panel), while a lower resolution starts to blur out useful information (\eg nside=16, the lower right panel).
 However, looking at Table \ref{table:gwcatsum} which summarises the GWTC-1 \ac{BBH} detections, it can be seen that GW151226, a relatively nearby event with reasonable catalogue support, has a 90\% sky area of $1033\text{ deg}^2$. So even a conservative analysis which only considered the 90\% most probable sky area would still require 308 pixels to represent it with an nside of 32. To cover the 99.9\% sky area would push this to $\mathcal{O}(1000)$. The computational time for this analysis increases approximately linearly with the number of pixels, meaning that it could take 1000 times longer to analyse the same event as it did when using original \citet{Gray:2019ksv} method, or would require 1000 computer cores to analyse it on a similar timescale, if the analysis was parallelised. However, dividing the \ac{GW} sky area into 1000 pixels is an unnecessarily high resolution for adequately representing the changes in the event's distance distribution. 
 
 \begin{table}
\centering
{\renewcommand{\arraystretch}{1.5}
\begin{tabular}{cccccccccc}
\hline
\hline
Event & $\Delta\Omega/$\sqdeg & $d_\mathrm{L}/$Mpc & $z_\text{event}$ & $V/$Mpc$^3$ \\
\hline
GW150914 & 182 & $440^{+150}_{-170}$ & $0.09^{+0.03}_{-0.03}$ & $3.5\times 10^6$ \\
GW151012 & 1523 & $1080^{+550}_{-490}$ & $0.21^{+0.09}_{-0.09}$ & $5.8\times 10^8$ \\
GW151226 & 1033 & $450^{+180}_{-190}$ & $0.09^{+0.04}_{-0.04}$ & $2.4\times 10^7$ \\
GW170104 & 921 & $990^{+440}_{-430}$ & $0.20^{+0.08}_{-0.08}$ & $2.4\times 10^8$ \\
GW170608 & 392 & $320^{+120}_{-110}$ & $0.07^{+0.02}_{-0.02}$ & $3.4\times 10^6$  \\
GW170729 & 1041 & $2840^{+1400}_{-1360}$ & $0.49^{+0.19}_{-0.21}$ & $8.7\times 10^9$  \\
GW170809 & 308 & $1030^{+320}_{-390}$ & $0.20^{+0.05}_{-0.07}$ & $9.1\times 10^7$  \\
GW170814 & 87 & $600^{+150}_{-220}$ & $0.12^{+0.03}_{-0.04}$ & $4.0\times 10^6$ \\
GW170818 & 39 & $1060^{+420}_{-380}$ & $0.21^{+0.07}_{-0.07}$ & $1.5\times 10^7$ \\ 
GW170823 & 1666 & $1940^{+970}_{-900}$ & $0.35^{+0.15}_{-0.15}$ & $3.5\times 10^9$ \\
\hline
\hline
\end{tabular}}
\caption[Relevant parameters of the \acp{BBH} from GWTC-1.]{Relevant parameters of the \acp{BBH} from GWTC-1: $90\%$ sky localization region $\Delta\Omega$ (\sqdeg), luminosity distance $d_{\text{L}}$ (Mpc, median with $90\%$ credible intervals), and estimated redshift $z_\text{event}$ (median with $90\%$ range assuming Planck 2015 cosmology) from \cite{O2:catalog}. The final column gives each event's $90\%$ 3D localisation comoving volume.}
\label{table:gwcatsum}
\end{table}

The method of choosing which resolution to treat the \ac{GW} data with, and computing the \acl{los} distance distributions for it, is discussed in Section \ref{sec:los distance}. The method of combining \ac{GW} data with a galaxy catalogue  represented by a higher resolution is presented in Section \ref{sec:varymth}.

\subsection{Line-of-sight luminosity distance estimates}
\label{sec:los distance}
The \ac{GW} data used for this cosmological analysis comes in the form of posterior samples, which are drawn from the posterior distributions of \ac{GW} parameters for each event, including the luminosity distance, sky location, and detector frame masses.
As was discussed in \citet{O2H0paper}, it is important to re-weight the \ac{GW} posterior samples in order to remove the detector-frame mass prior used for parameter estimation, and apply the desired source-frame mass prior in order to match the priors used to compute the normalising probability of detection term. This importance arises due to the cosmological information that is inherently encoded with a \ac{GW} detection, in the relation between source-frame and detector-frame mass. The intrinsic mass of a merger cannot be measured directly, but requires converting detector-frame (redshifted) masses to source-frame, which is $H_0$-dependent. This cosmological information can be particularly informative (with enough \ac{BBH} detections) when the source frame mass distribution of \acp{BBH} contains sharp features such as peaks or cut-offs \citep{Farr:2019twy,PhysRevD.104.062009}. The re-weighting of an event's posterior samples not only affects the shape of the event's distance distribution, but also its normalisation as a function of \ac{H0}, as samples may become inconsistent with the source-frame mass prior for certain values of \ac{H0}.  As such, when implementing the pixelated method into \gwcosmo, the \ac{GW} mass and distance information needs to be preserved for each pixel. This means that, even though 3D skymaps exist which contain the \ac{los} \ac{GW} distance information for each pixel, these cannot be used here because they include the marginalised-over priors applied during parameter estimation.

In order to retain the necessary information, the posterior samples are used directly to compute the \ac{los} distance estimate for each pixel. A crude way to do this would be to choose some resolution for which to grid up the sky, then divide up the posterior samples between pixels, and create the \ac{los} distance estimate for each pixel from the samples within it. However, this attempt fails at the first hurdle, where the finite number of samples comes into play.  Most events have $\mathcal{O}(100,000)$ samples, of which $\mathcal{O}(100)$ are required to get a reasonable distance estimate (and more samples is preferable, as the estimate will be more reliable).  As the samples cover the full parameter space, including \ac{ra} and \ac{dec}, the edges of the event's sky area are particularly poorly sampled.  Even for a relatively low-resolution pixelation of the sky, this leaves pixels in the event's 99.9\% sky area with as few as $\mathcal{O}(1)$ samples.

In order to avoid this, while still remaining compatible with a \texttt{healpy} implementation, the \ac{los} distance estimate for each pixel is found by selecting all the samples within a certain angular radius of the centre of a pixel (determined by the resolution of the pixels). If the number of samples exceeds some threshold to be deemed ``enough samples'' (taken to be 100 in this case), then a \ac{KDE} on these samples is used to create the \ac{los} distance estimate.  If there are not enough samples within the selected area, the angular radius is incrementally increased until the number of samples passes the threshold. \acp{KDE} are normalised by default, so each pixel's \ac{los} distance distribution needs to be additionally weighted by the sky probability within that pixel of the \ac{GW} skymap.
By selecting samples in this way, the necessary information required to re-weight the samples by their source-frame mass priors is retained. Pixels in the most probable sky areas will have $\mathcal{O}(10,000)$ samples, while those in the lowest probability regions will only have 100, taken from a larger sky patch. The decrease in reliability of these low-probability distance estimates will be compensated for by the down-weighting of their contribution to the final result.
Practically, \gwcosmo uses \texttt{SciPy}'s \texttt{gaussian\_kde} \citep{2020SciPy-NMeth} for the \ac{los} distance estimates. This \ac{KDE} is effectively a weighted sum of Gaussians centered at the location of each posterior sample in the parameter space. The width of these Gaussians, and hence the overall smoothing of the \ac{KDE}, which determines how accurately underlying structure is represented, is computed using the Scott method \citep{Scott}.

\begin{figure*}
\centering
\includegraphics[width=0.3\linewidth]{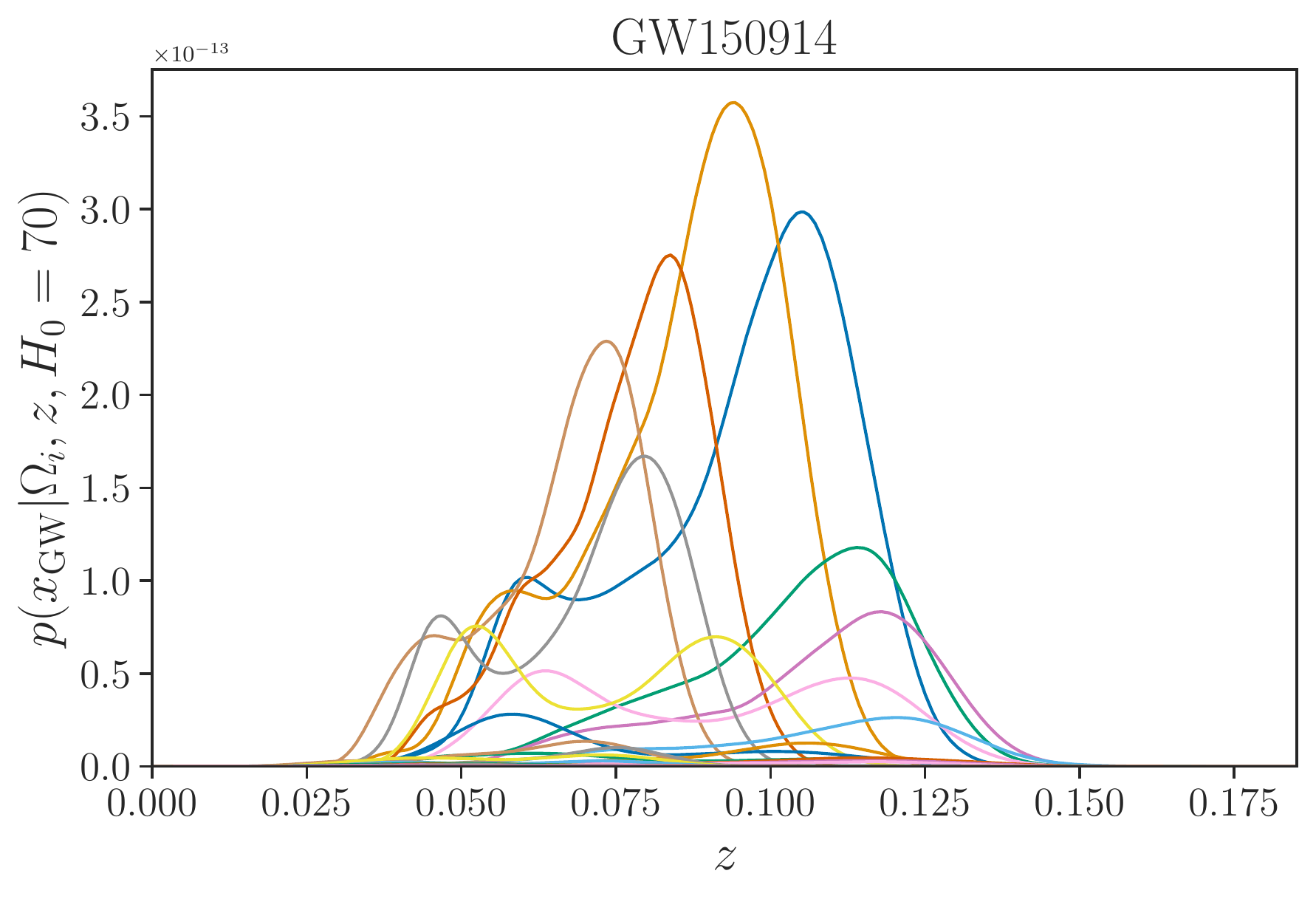}
\includegraphics[width=0.3\linewidth]{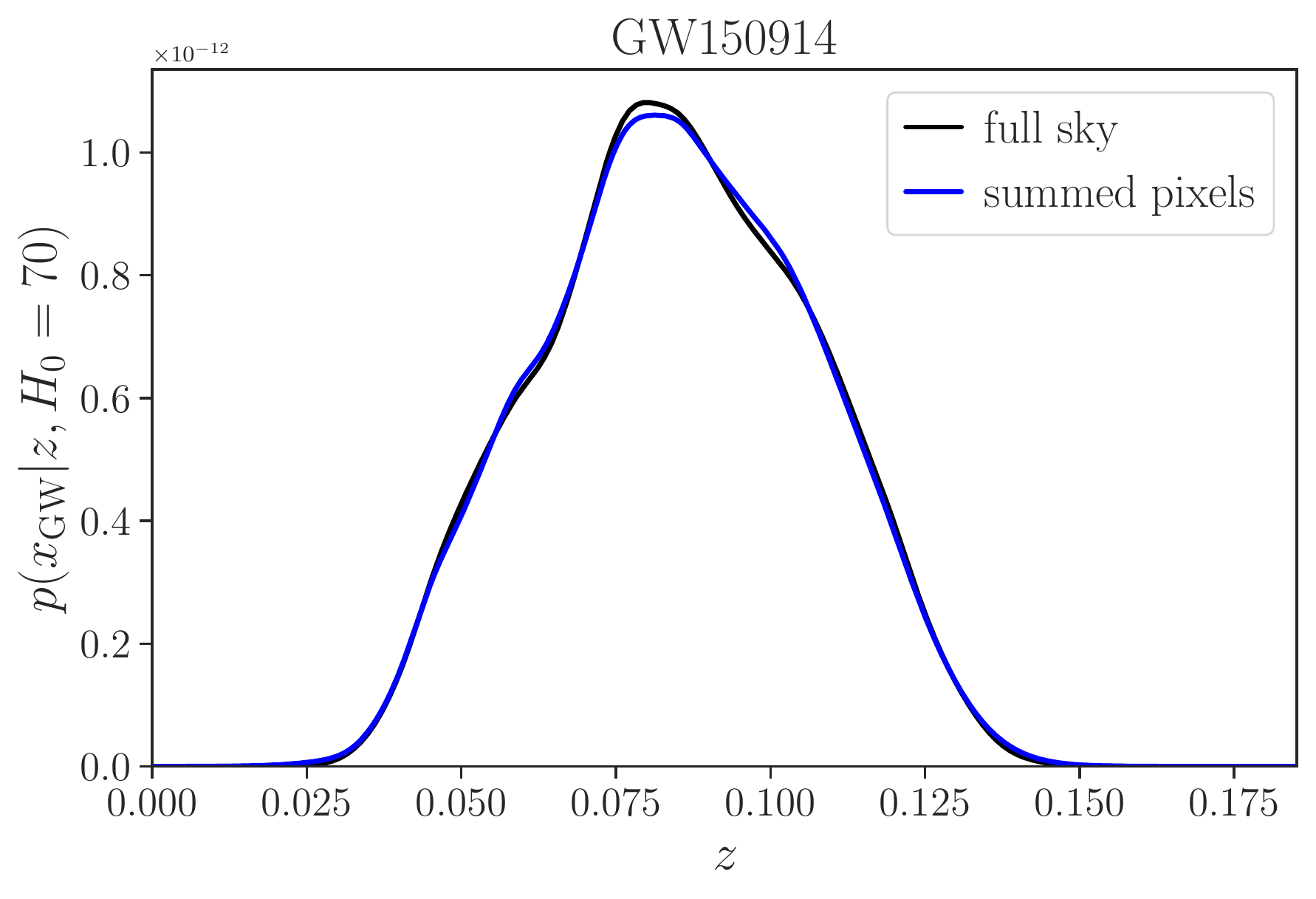}
\\
\includegraphics[width=0.3\linewidth]{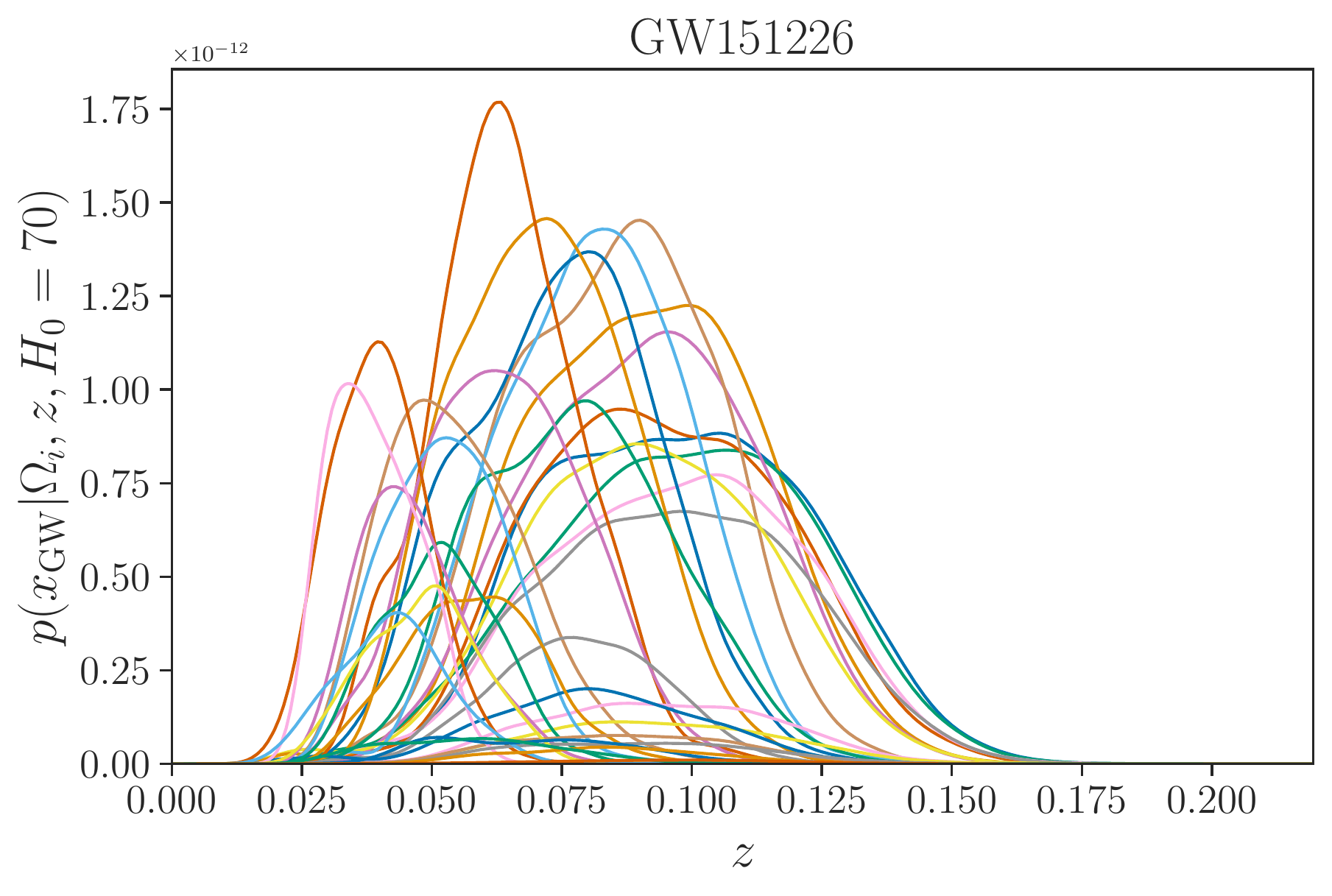}
\includegraphics[width=0.3\linewidth]{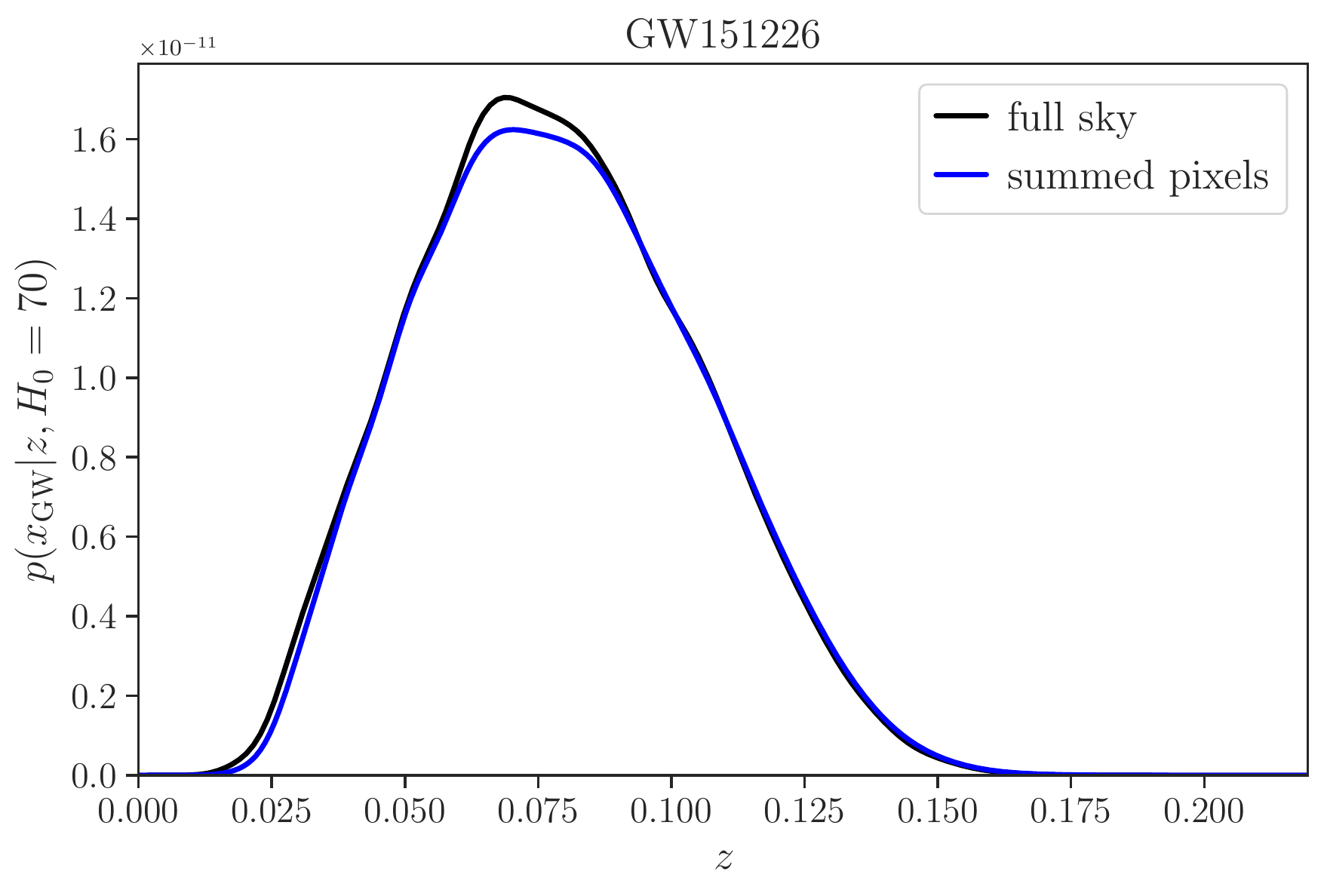}
\\
\includegraphics[width=0.3\linewidth]{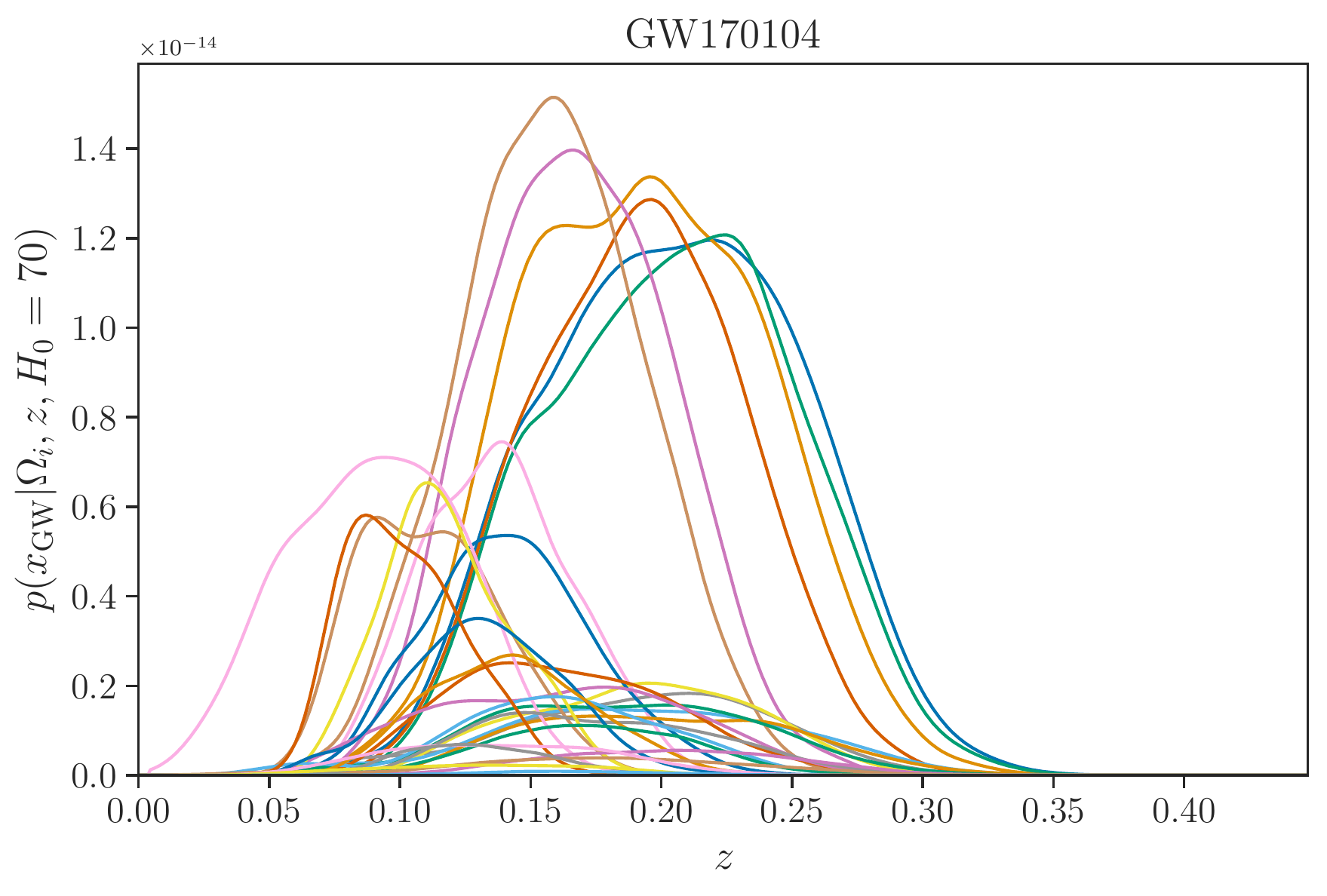}
\includegraphics[width=0.3\linewidth]{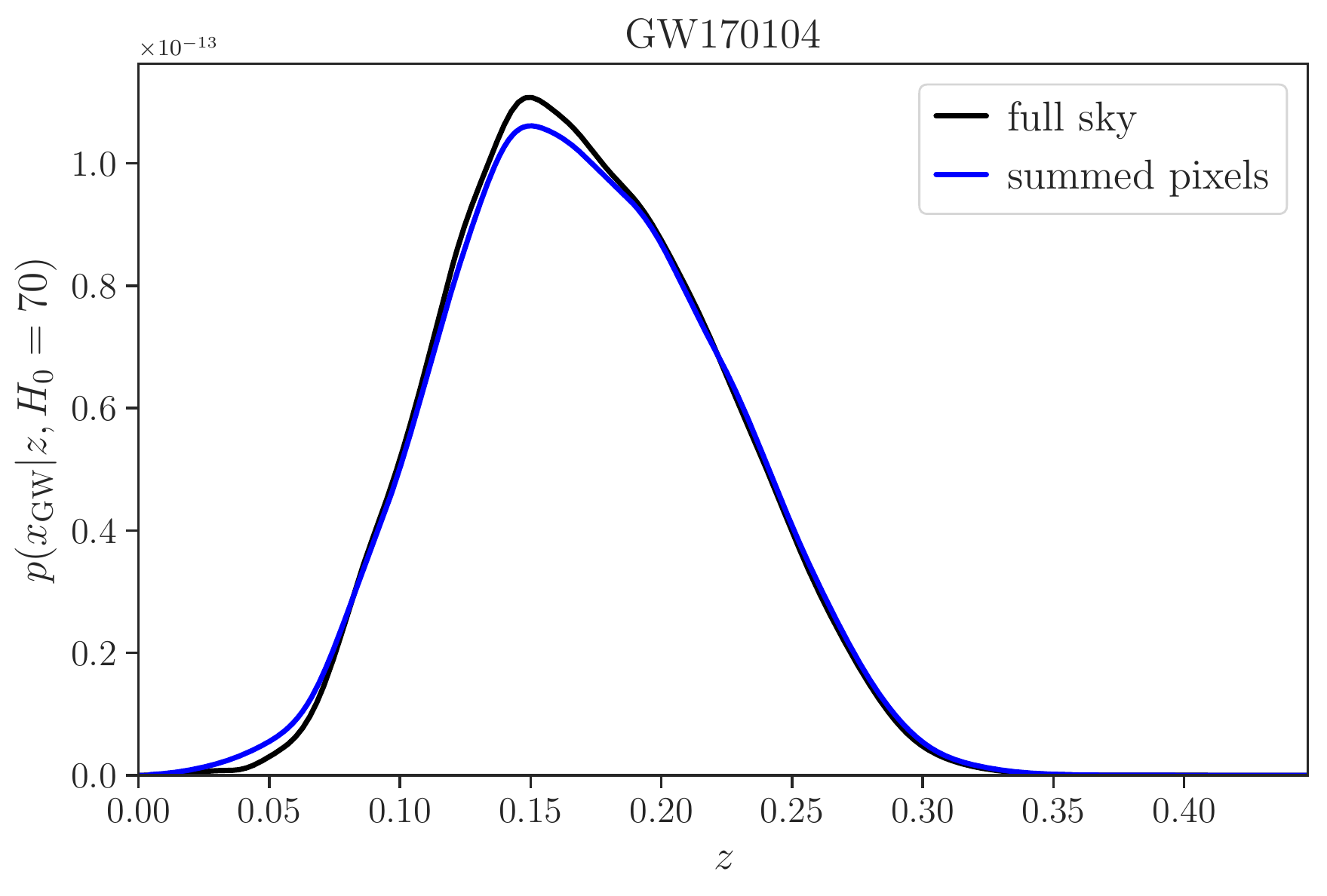}
\\
\includegraphics[width=0.3\linewidth]{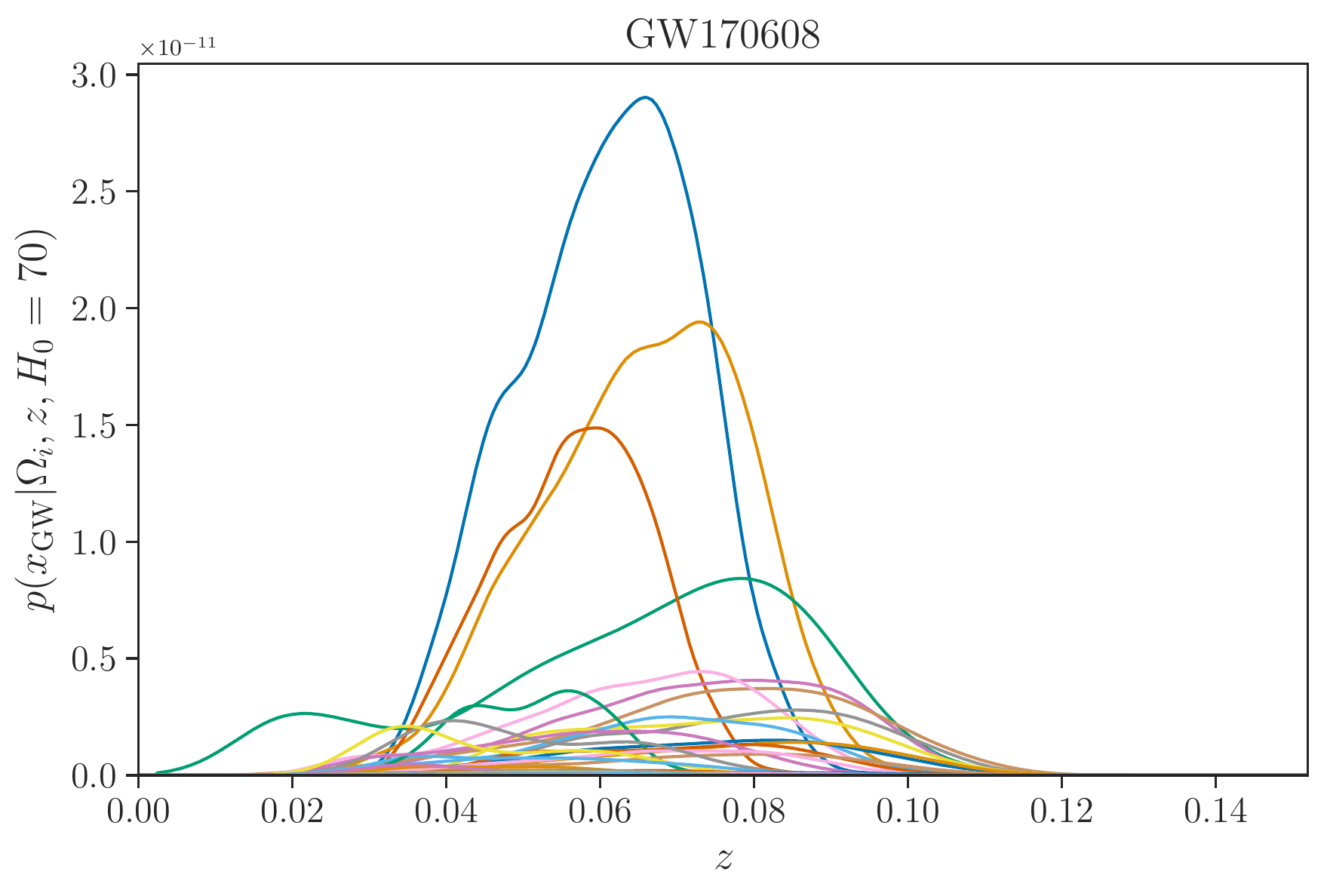}
\includegraphics[width=0.3\linewidth]{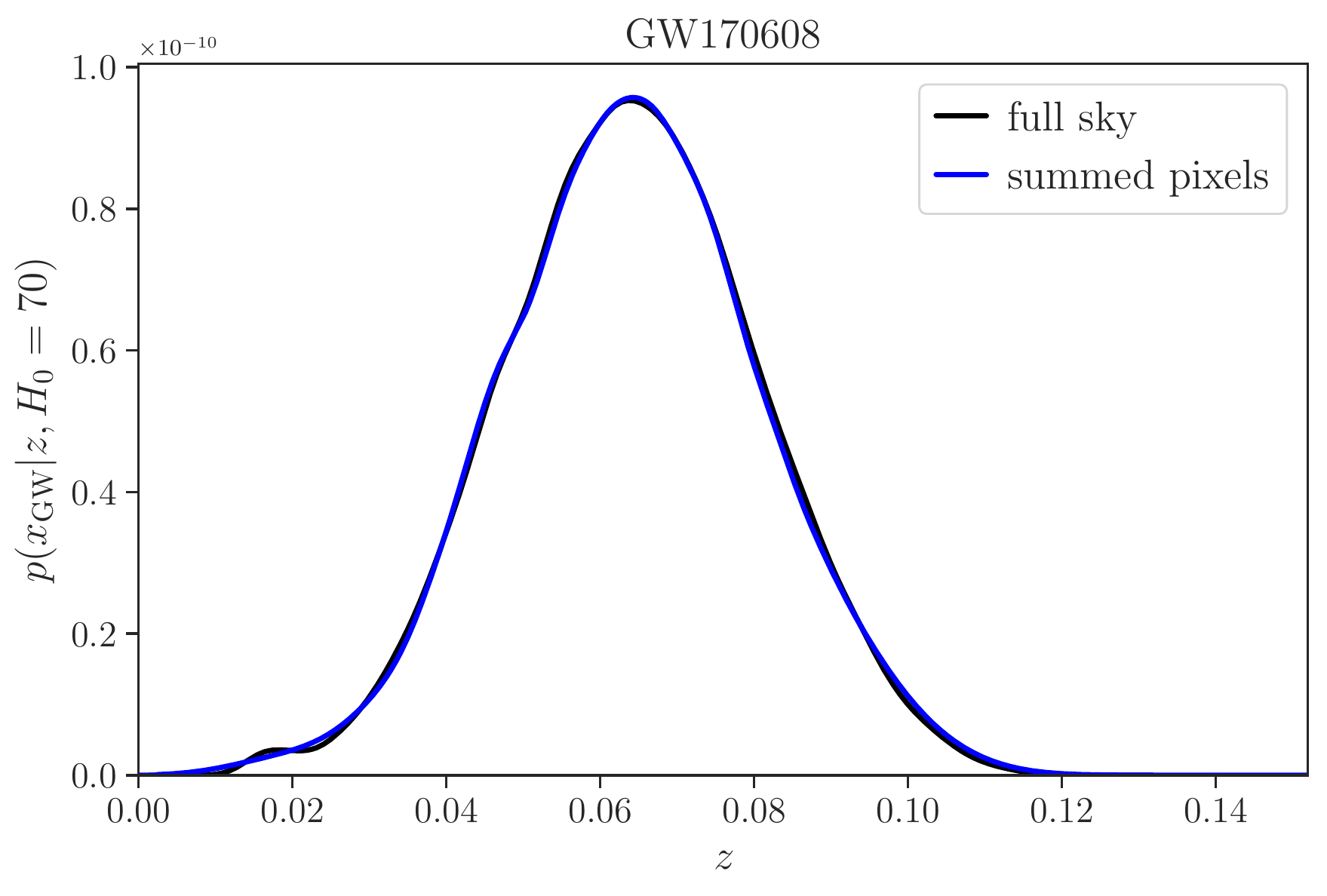}
\\
\includegraphics[width=0.3\linewidth]{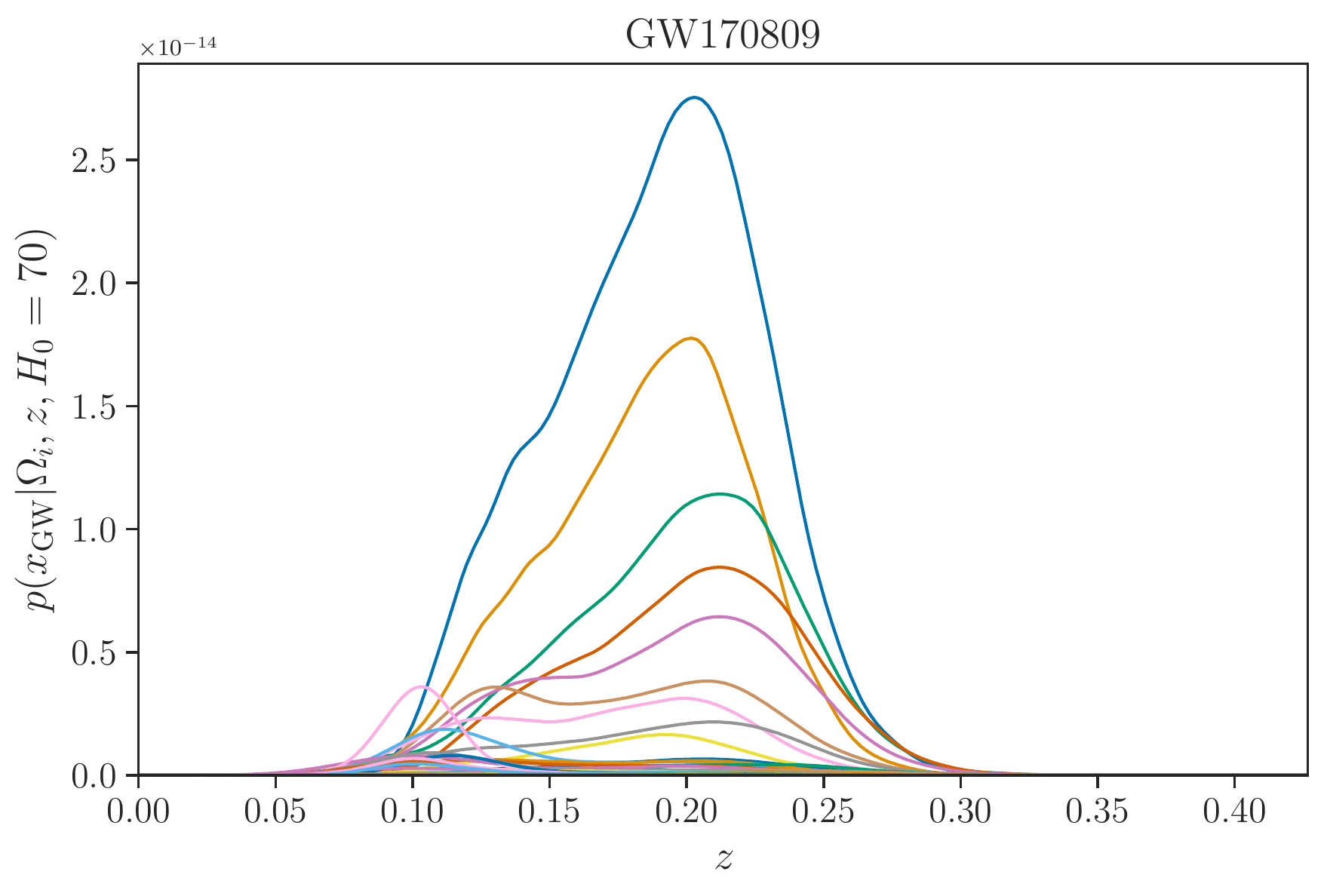}
\includegraphics[width=0.3\linewidth]{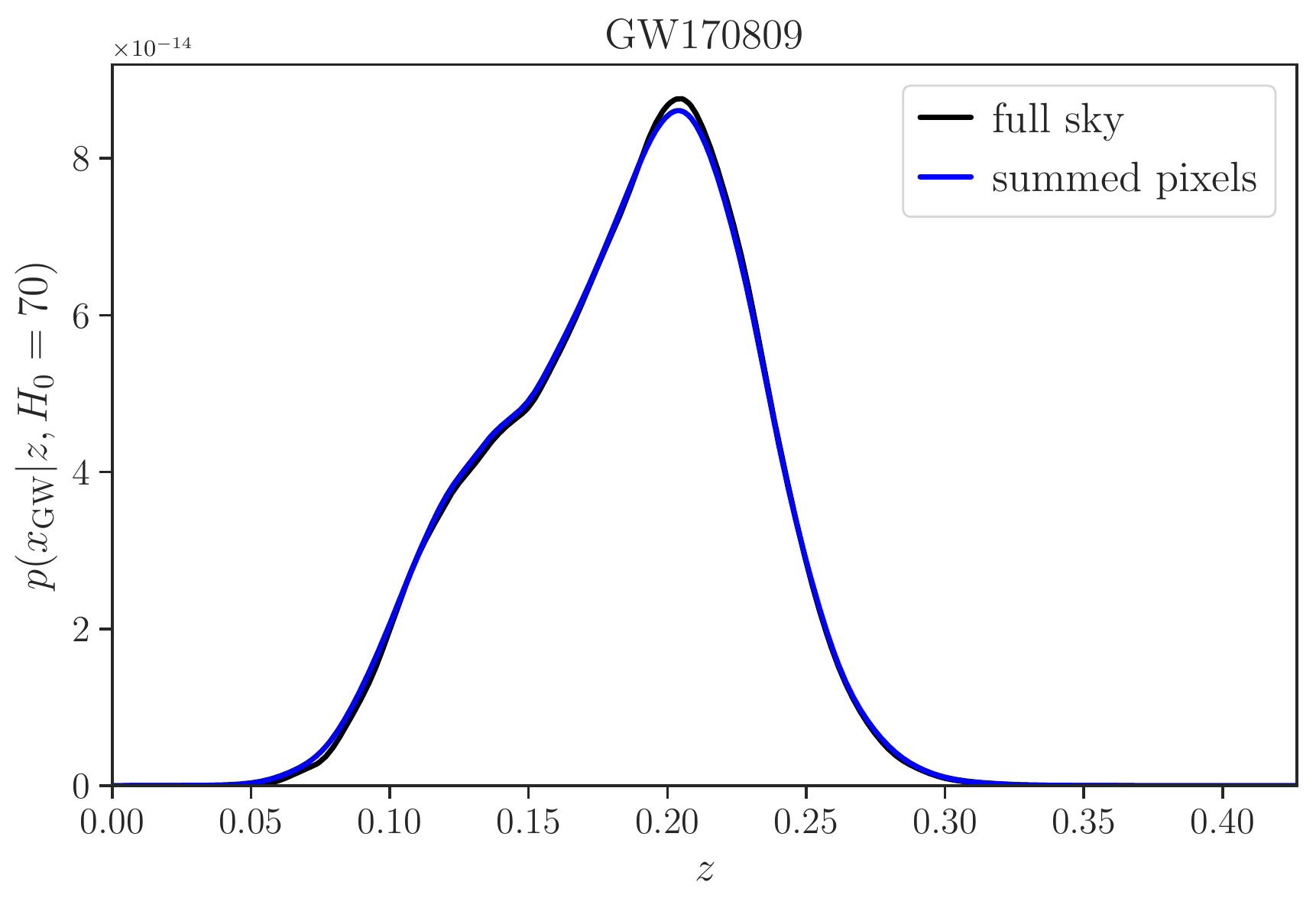}
\\
\includegraphics[width=0.3\linewidth]{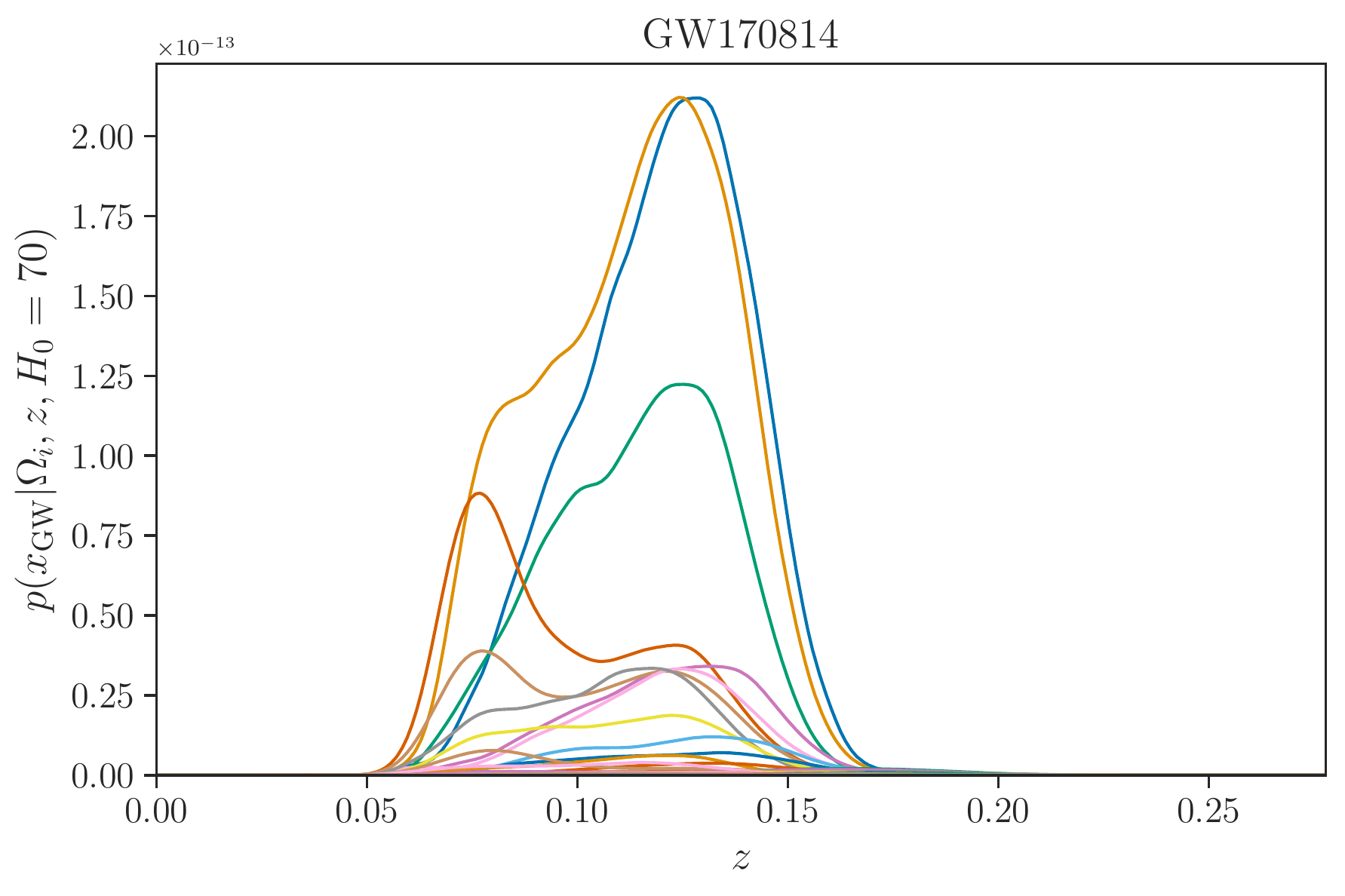}
\includegraphics[width=0.3\linewidth]{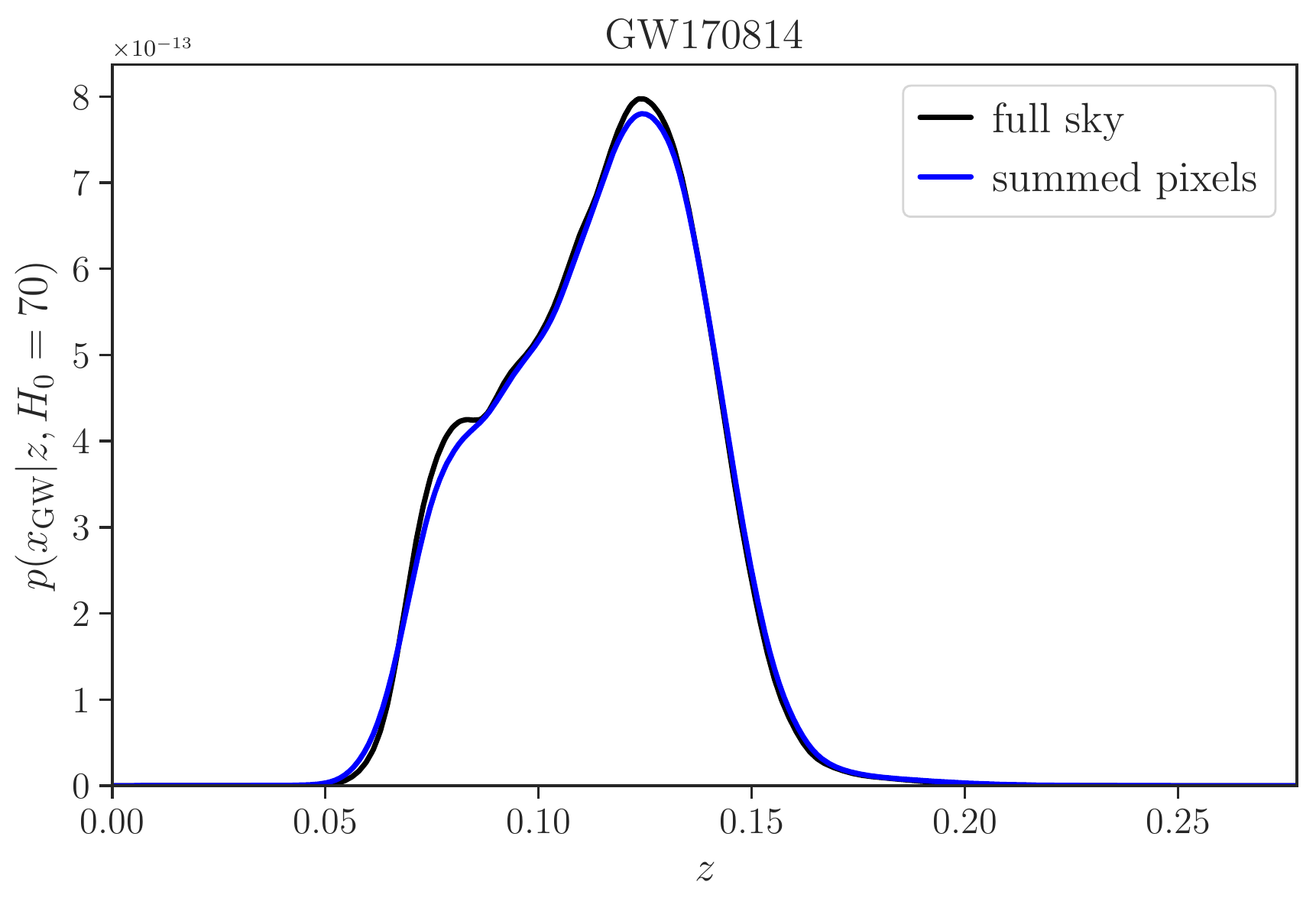}
\caption[Line-of-sight redshift estimates for GW150914, GW151226, GW170104, GW170608, GW170809 and GW170814.]{Line-of-sight redshift estimates for GW150914, GW151226, GW170104, GW170608, GW170809 and GW170814. \textit{Left-hand panels:} The line-of-sight redshift distribution of the event within each pixel, where the number of pixels and pixel nside for each event are specified in the $N_\text{pix}$ (low-res) and $n_\text{low}$ columns of Table \ref{tab: pixels}. \textit{Right-hand panels:} The full-sky redshift distribution for the event. The black curve shows the estimate from doing a \ac{KDE} on all the samples, while the blue curve shows the summed curves from the left-hand panel.}
\label{Fig:LOSredshift}
\end{figure*}

The choice of resolution used to represent the \ac{GW} data should depend on the area of the sky covered by the \ac{GW} within some probability threshold. Events with small sky areas will need smaller pixels to adequately represent the changing of their distance distribution across the sky, and vice versa. As such, for \gwcosmo the resolution is chosen by defining a threshold on the sky area of the event and determining the nside resolution that would be necessary to split that patch of the sky into (at least) a minimum number of pixels. Choosing a minimum pixel number of 30 to cover the 99.9\% sky area of each event produces the \ac{los} distance (redshift) estimates shown in Fig. \ref{Fig:LOSredshift}. The left-hand panels shows the breakdown of the redshift distribution along the line-of-sight of each pixel for the 6 \ac{BBH} events from GWTC-1 that were used in the main analysis result of \citet{O2H0paper} (GW150914, GW151226, GW170104, GW170608, GW170809 and GW170814).\footnote{Figure \ref{Fig:LOSredshift} is shown as a function of redshift, as opposed to luminosity distance, simply because in \gwcosmo the \ac{KDE} is done on redshift samples.} The redshift at which the distribution peaks varies from pixel to pixel, indicating that galaxies at different redshifts will be favoured for different lines of sight. The right-hand panel of Fig. \ref{Fig:LOSredshift} demonstrates that the overall distribution on redshift when summing the contribution from individual pixels is very close to the original distance distribution, estimated from all of the samples. As both are marginalised over $\Omega$ they should theoretically be identical, but in practice the pixelation process introduces minor variation. This variation is not large enough to cause noticeable impact to the inference of \ac{H0} (see section \ref{sec:emptycat}). Theoretically a larger number of pixels would more accurately capture the event's 3D localisation (see section \ref{sec:systematics resolution}), up to the limit where the smallness of the pixels would require samples to be reused for the \ac{los} estimates in the most probable pixels.

Table \ref{tab: pixels} summarises the \texttt{healpy} resolution ($\text{nside}_{\text{low}}$) and number of pixels ($N_{\text{pix}}$) required for each event in order to meet the criteria of at least 30 pixels to cover the event's 99.9\% sky area. Additionally, the number of posterior samples in the most probable pixel is recorded ($N_\text{samples}$) which shows that for every event the most probable pixels, which will be contributing most strongly to the result, have at least 1000 samples, and in some cases as high as tens of thousands.\footnote{While this was true for all of the GWTC-1 events it should not be assumed true in general, and number of pixels or the threshold for the sky area may need to be adjusted to meet this criteria.} The impact of changing the sky area under consideration, or the minimum number of pixels required to cover it (and hence the resolution with which the \ac{GW} data is treated) is examined in greater detail in Section \ref{sec:systematics resolution}.

\begin{table}
\centering
\renewcommand{\arraystretch}{1.5}
\setlength{\tabcolsep}{3pt}
\begin{tabular}{ccccccc}
\hline
\hline
Event & $\text{nside}_{\text{low}}$ & $N_{\text{pix}}$ & $N_\text{samples}$ & $N_{\text{sub-pix}}$/pixel & $N_{\text{sub-pix}}$ (total) \\ \hline
GW150914 & 8 & 35 & 7759 & 16 & 560  \\ 
GW151012 & 4 & 40 & 1941 & 64 & 2560  \\ 
GW151226 & 4 & 34 & 3737 & 64 & 2176  \\
GW170104 & 4 & 30 & 4804 & 64 & 1920  \\
GW170608 & 8 & 38 & 4010 & 16 & 608  \\
GW170729 & 4 & 32 & 22582 & 64 & 2048  \\
GW170809 & 8 & 34 & 30882 & 16 & 544 \\
GW170814 & 16 & 54 & 12750 & 4 & 216 \\
GW170818 & 16 & 72 & 6417 & 4 & 288 \\
GW170823 & 4 & 43 & 6130 & 64 & 2752 \\
\hline
\hline
\end{tabular}
\caption[A summary of the resolutions used for the pixelated analysis for the O1 and O2 BBHs.]{\label{tab: pixels} A summary of the resolutions used for the pixelated analysis for the O1 and O2 \acp{BBH}. The second and third column show the lowest nside which satisfies the criteria that at least 30 pixels must cover the 99.9\% sky area of the event, $\text{nside}_{\text{low}}$, and the number of pixels, $N_{\text{pix}}$, required to do so. The most probable pixel for each event contains $N_\text{samples}$ samples. Each of the pixels is broken into $N_{\text{sub-pix}}$ sub-pixels in order to reach a galaxy catalogue resolution of nside=32, and the total number of sub-pixels which cover the event's 99.9\% sky area is $N_{\text{sub-pix}}$ (total).}
\end{table}

\subsubsection{Sanity check: the empty catalogue case}\label{sec:emptycat}
\begin{figure*}
\includegraphics[width=0.32\linewidth]{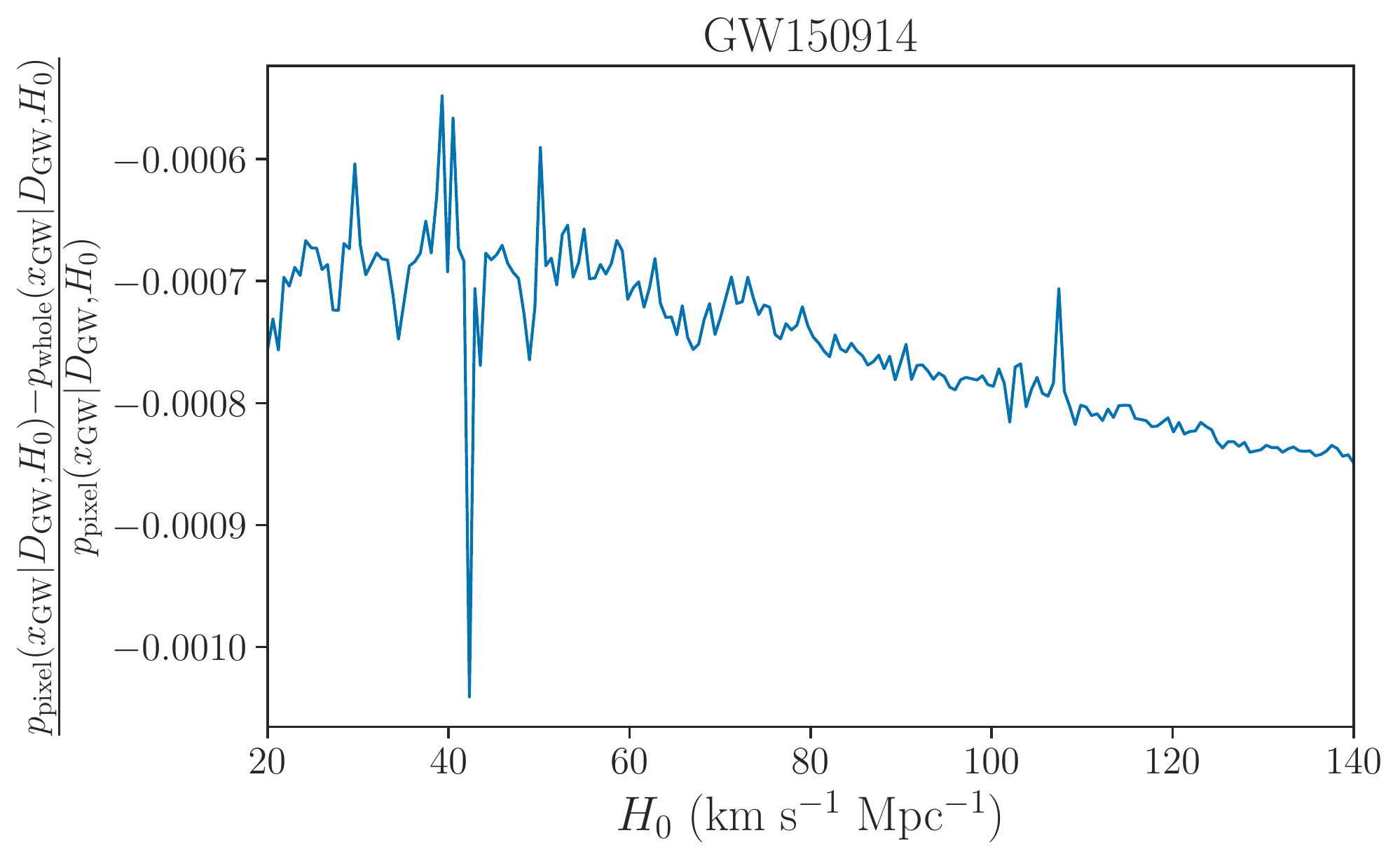}
\includegraphics[width=0.3\linewidth]{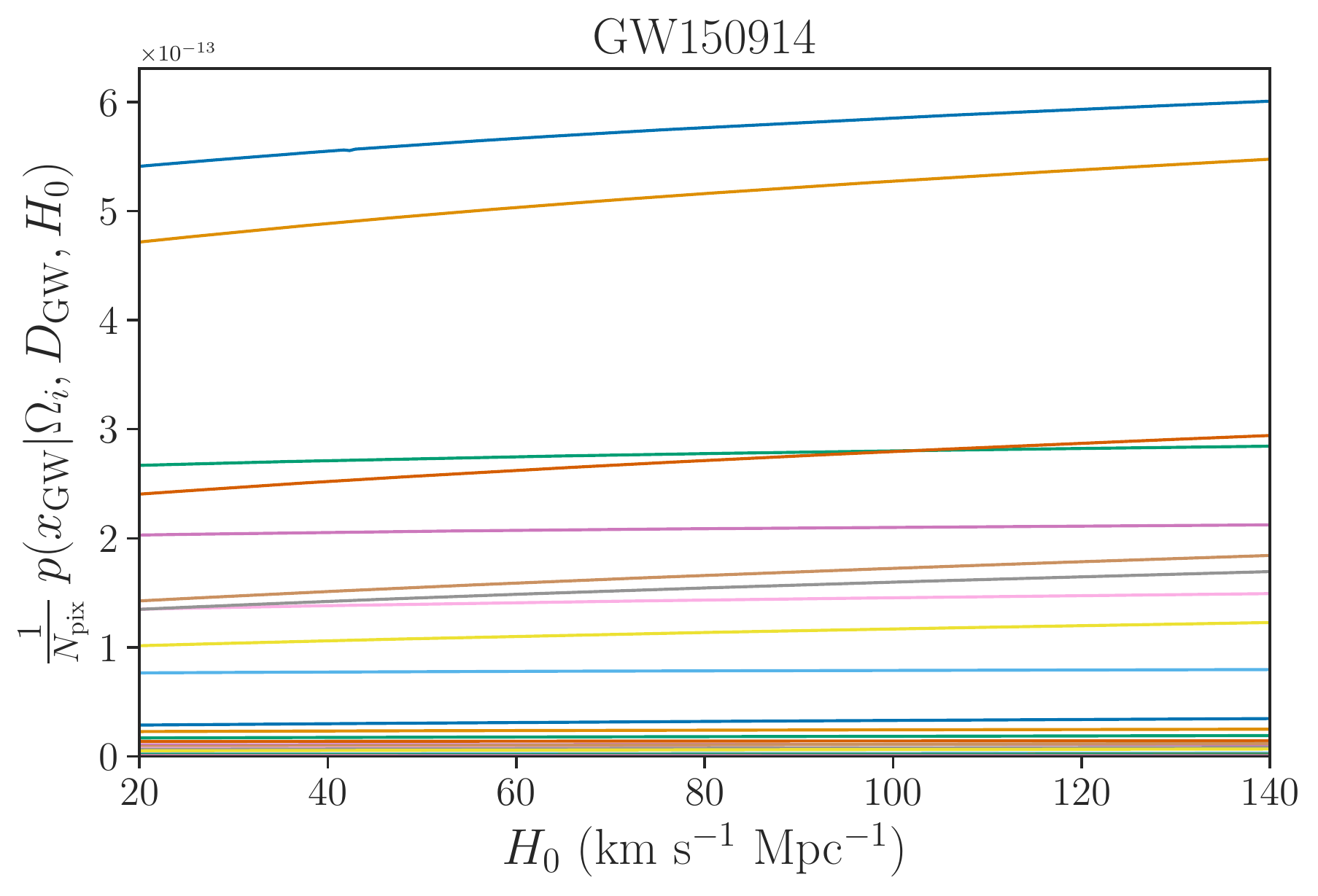}
\\
\includegraphics[width=0.32\linewidth]{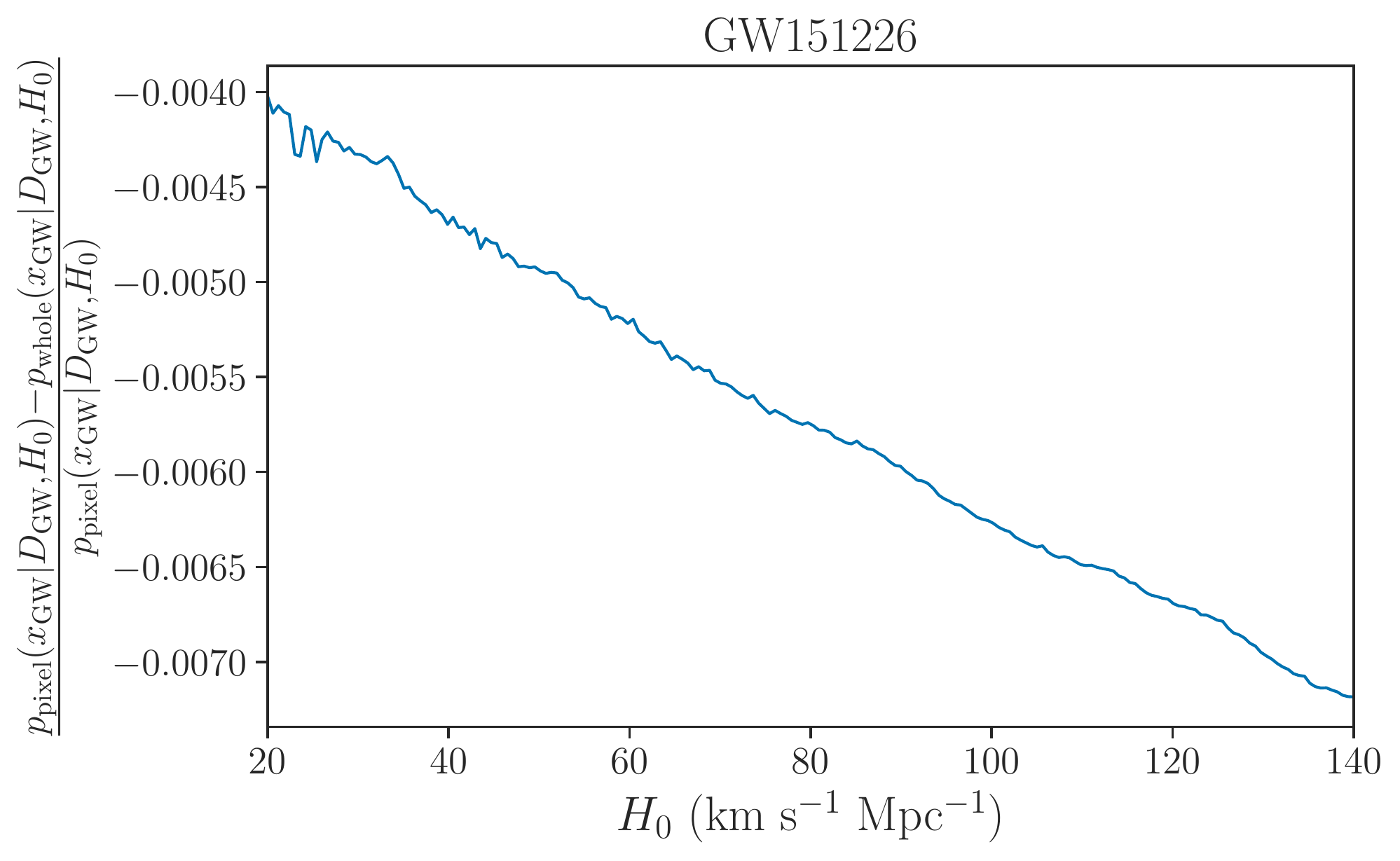}
\includegraphics[width=0.3\linewidth]{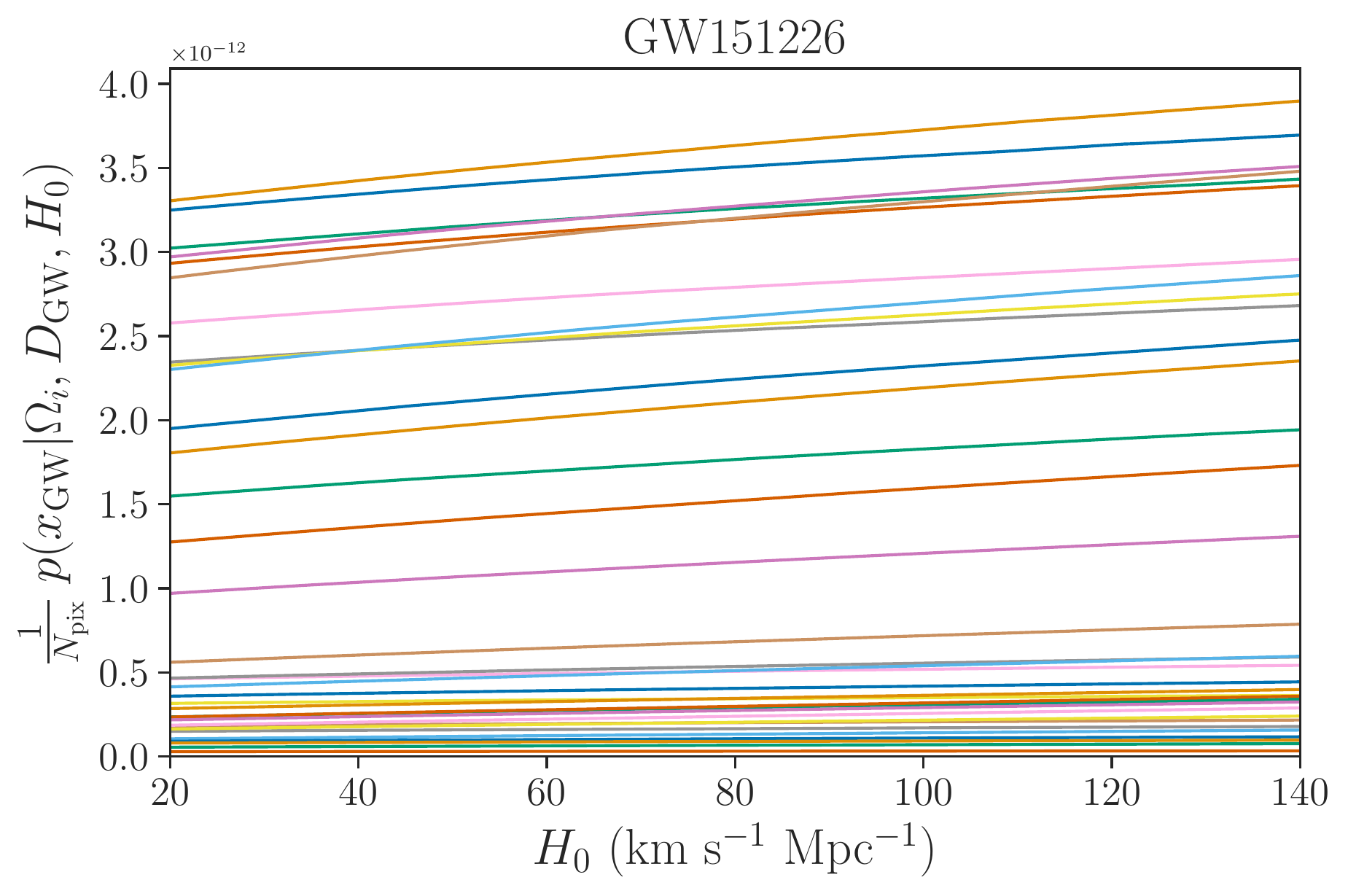}
\\
\includegraphics[width=0.32\linewidth]{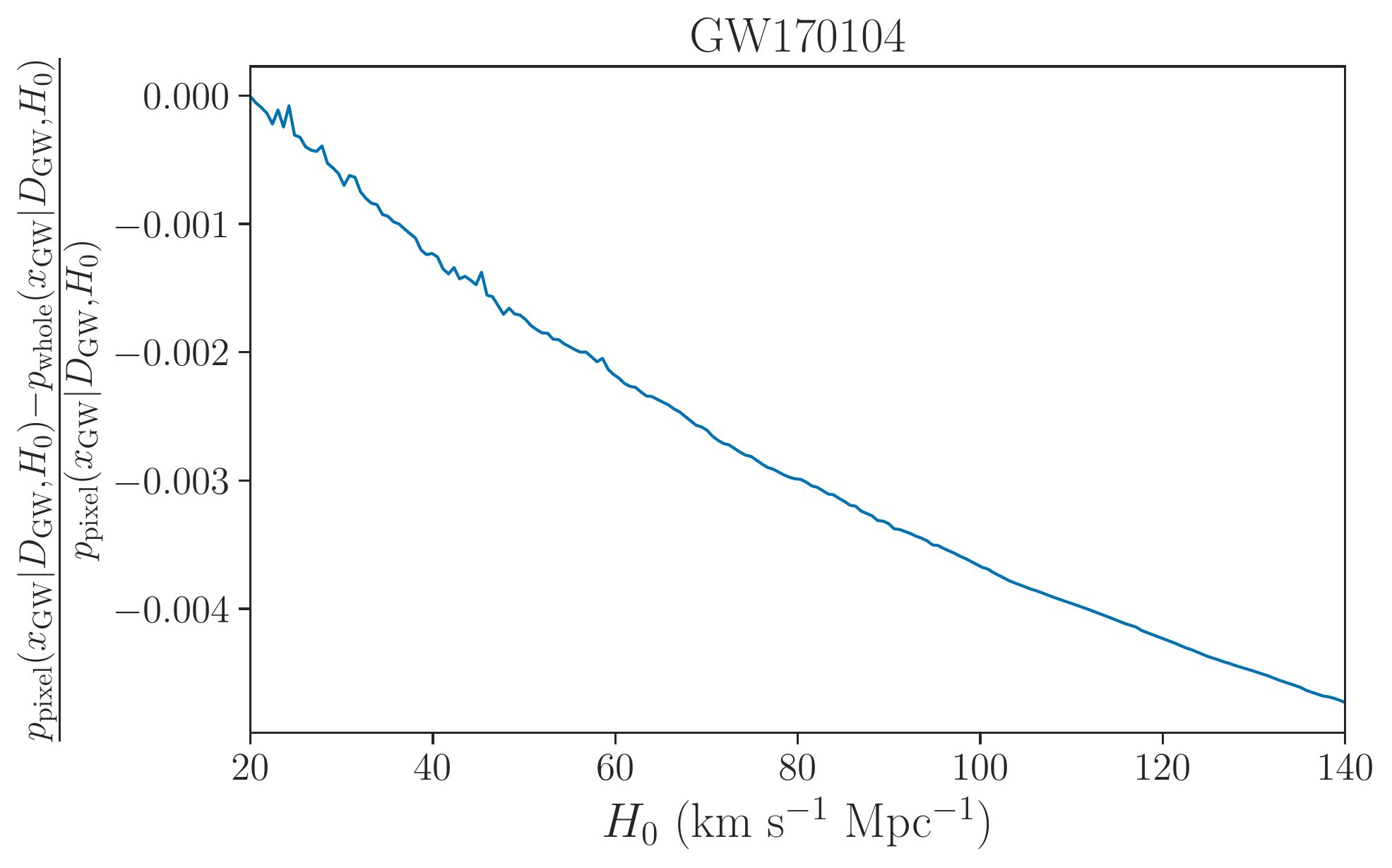}
\includegraphics[width=0.3\linewidth]{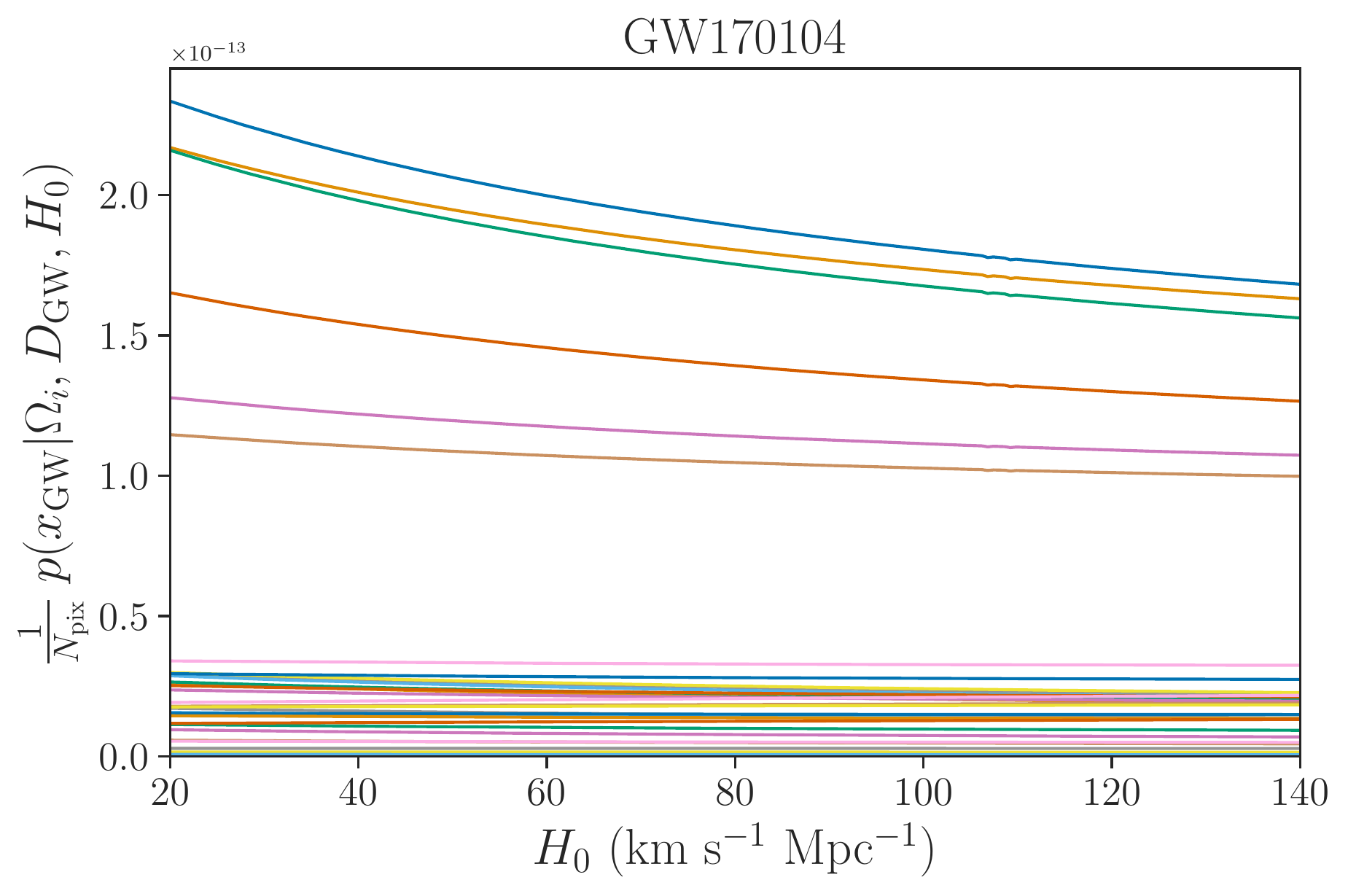}
\\
\includegraphics[width=0.32\linewidth]{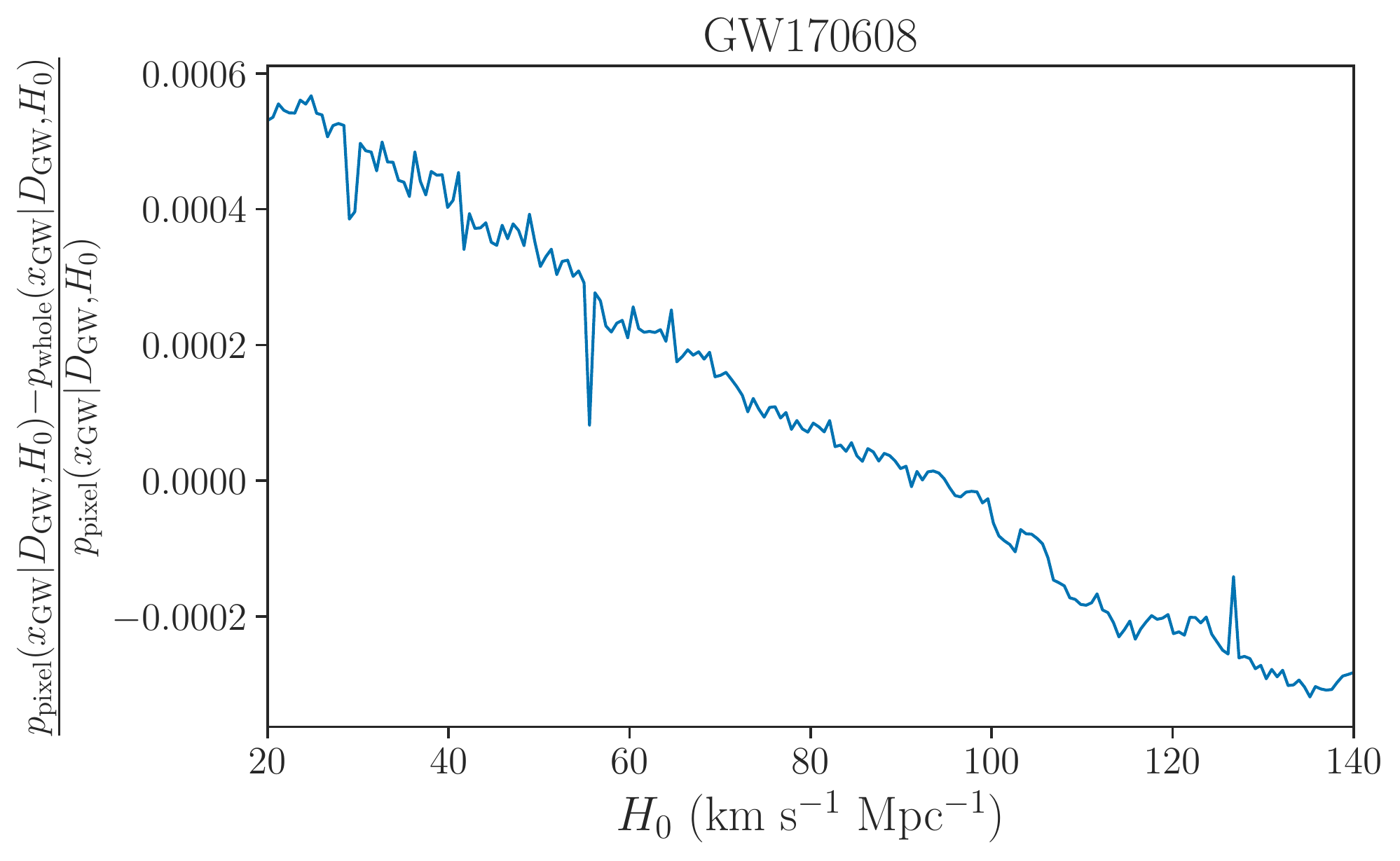}
\includegraphics[width=0.3\linewidth]{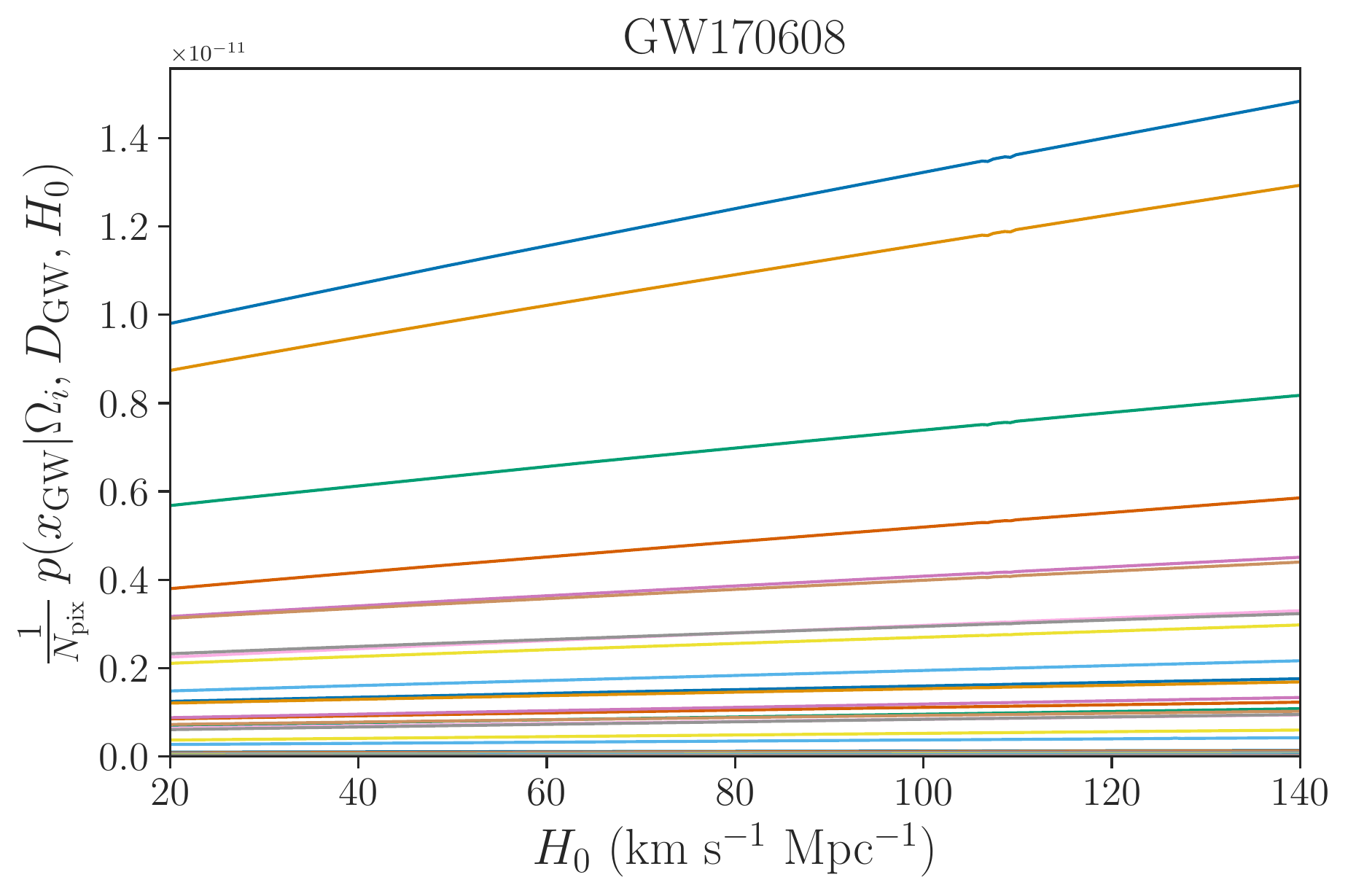}
\\
\includegraphics[width=0.32\linewidth]{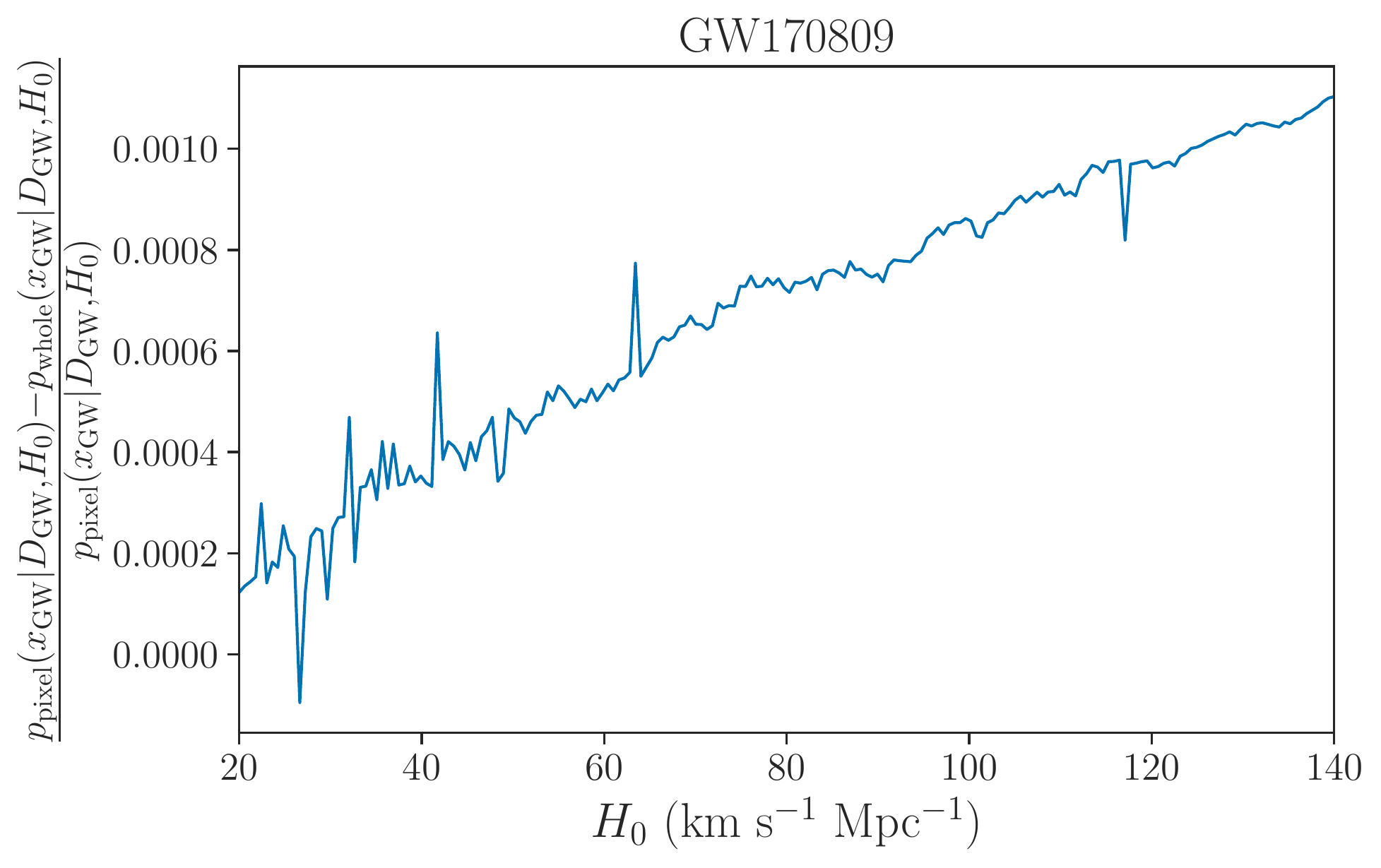}
\includegraphics[width=0.3\linewidth]{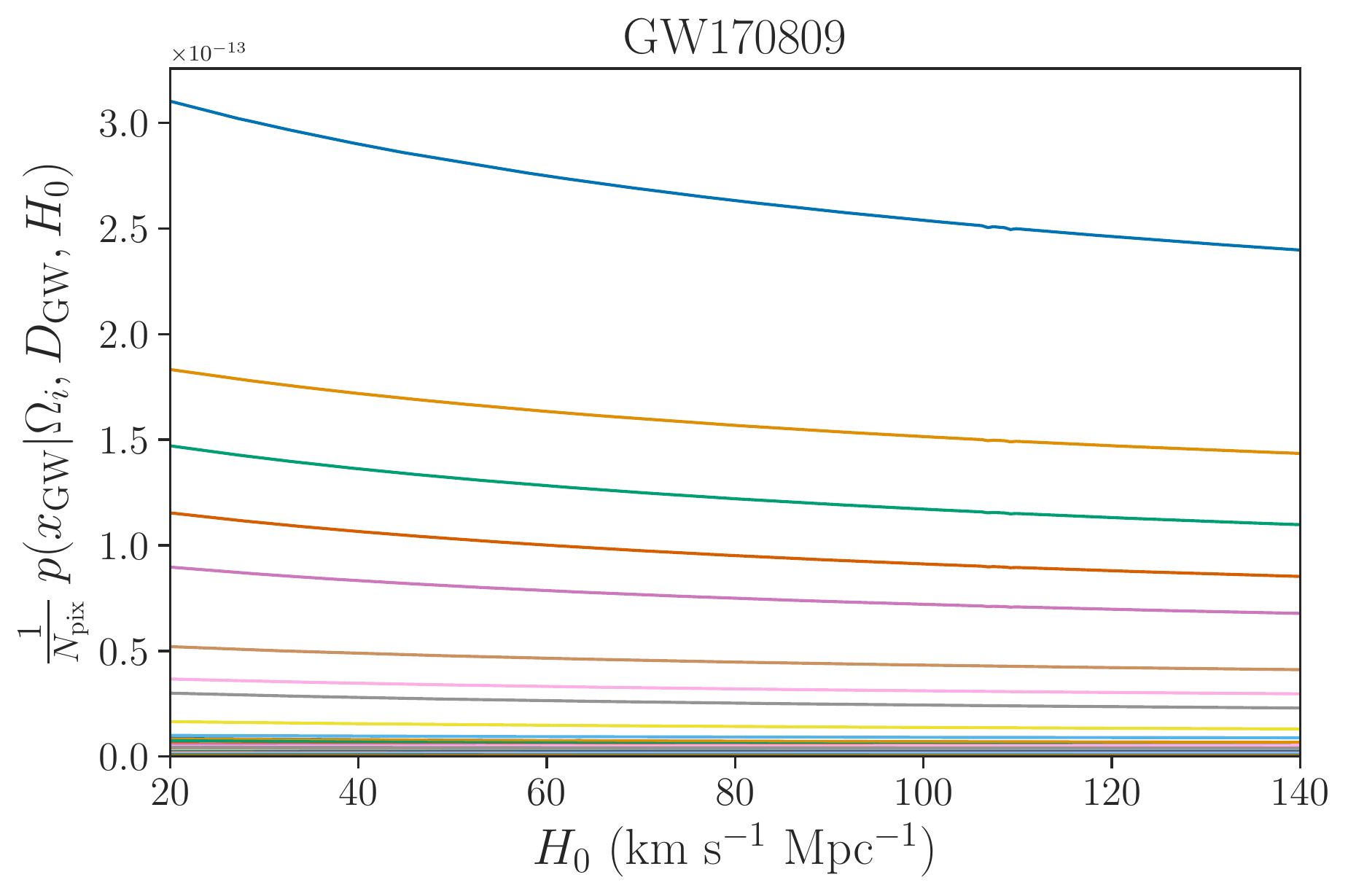}
\\
\includegraphics[width=0.32\linewidth]{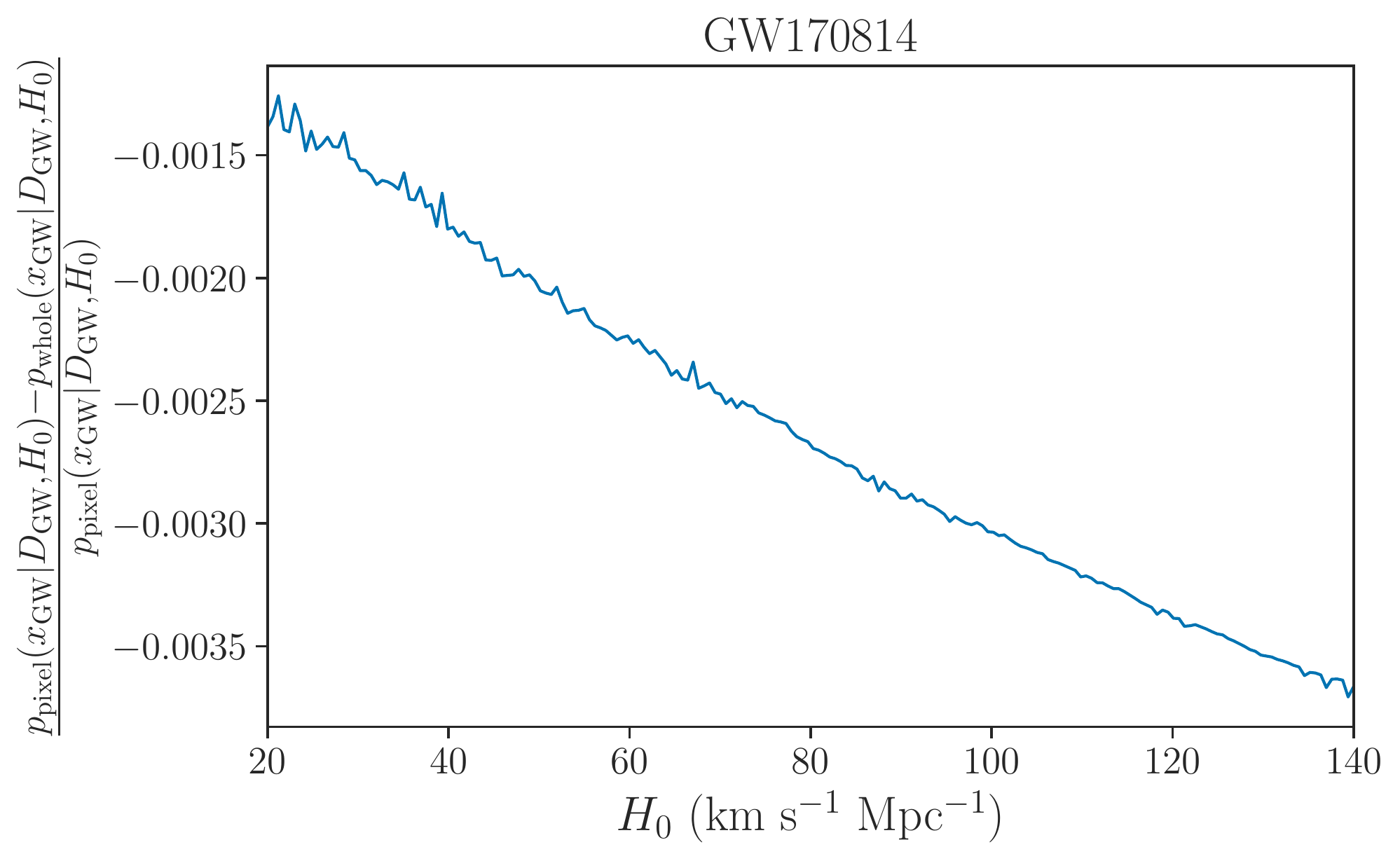}
\includegraphics[width=0.3\linewidth]{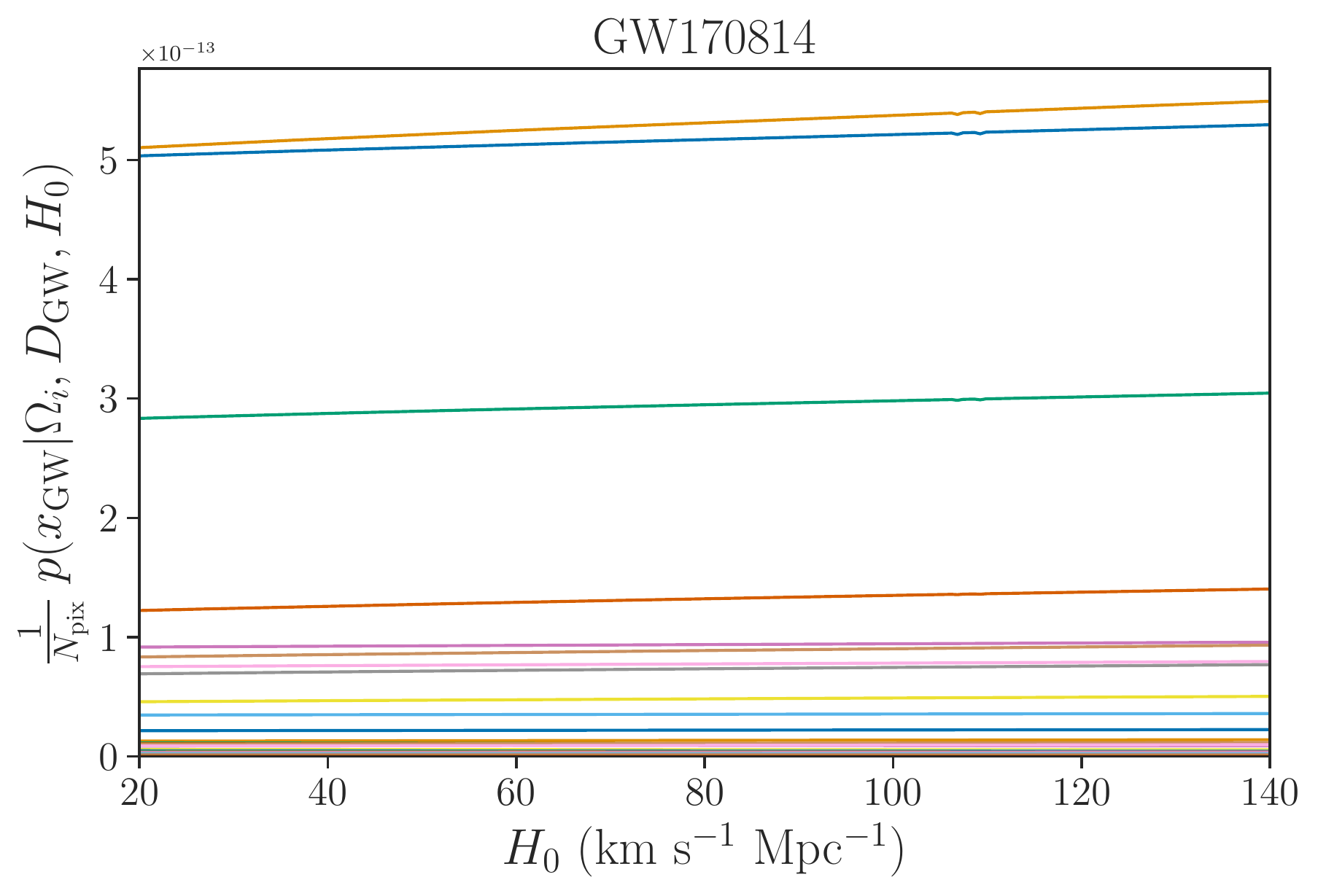}
\caption[Likelihoods on $H_0$ for the empty catalogue analysis with GW150914, GW151226, GW170104, GW170608, GW170809 and GW170814.]{Likelihoods on $H_0$ for the empty catalogue analysis with GW150914, GW151226, GW170104, GW170608, GW170809 and GW170814. \textit{Left-hand panels:} Fractional difference between the pixelated empty catalogue likelihood on $H_0$ and the non-pixelated (whole sky) likelihood, for each event. \textit{Right-hand panels:} A breakdown of the pixelated empty catalogue likelihood by pixel.}
\label{Fig:pixel_empty_likelihoods}
\end{figure*}

To further prove that this method of estimating the \ac{los} distance is robust, and to ensure that the \ac{H0}-dependence of the distribution has been correctly propagated, it is worth coming back to the empty catalogue analysis, in which no galaxy catalogue is used and the redshift prior is taken to be uniform in co-moving volume. In this case the results are relatively (but not completely) uninformative. Small amounts of information come from the mass and distance distributions of the \ac{GW} sources, leading to individual event likelihoods which are not completely flat in $H_0$. The pixelated empty catalogue analysis should return results equivalent an empty catalogue analysis which is done over the whole sky --- because no catalogue information enters this analysis, the order in which the marginalisation over $\Omega$ takes place is irrelevant. 

The analysis assumes a power-law source frame mass distribution for the \acp{BBH}, with the primary mass $p(m_1)\propto m_1^{-\alpha}$, where $\alpha=1.6$ and the secondary mass uniform between $M_\text{min}$ and $m_1$. Additional constraints are defined such that $M_\text{min}=5M_\odot$ and $M_\text{max}=100M_\odot$ and the network SNR threshold assumed for computing \ac{GW} selection effects is 12. Results for GW150914, GW151226, GW170104, GW170608, GW170809 and GW170814 can be seen in Fig. \ref{Fig:pixel_empty_likelihoods}. 
The left-hand panels show the fractional difference between the likelihoods computed using the pixelated method, and the likelihoods computed over the whole sky at once. For every event, the difference is less than 1\% across all values of \ac{H0}. The right-hand panels show the contribution to the pixelated empty catalogue likelihood from each pixel covering the 99.9\% sky area of the event. A close examination of the right hand panels shows that individual pixel contributions do not have the same \ac{H0} dependence, independent of the scale (GW150914 in particular shows a good example of this). This difference in the slope is due to the change in the GW \ac{los} distance distribution for different pixels. The fact that the different methods produce only fractional changes to the final likelihoods demonstrates that the \ac{GW} data is well-represented by this pixelated approach.

\subsection{Varying $m_\text{th}$ within an event's sky area}
\label{sec:varymth}

The second major improvement the pixelated method offers is the ability to treat the apparent magnitude threshold, \ac{mth}, of a galaxy catalogue as a function of sky direction.

\begin{figure*}
\includegraphics[width=0.49\linewidth]{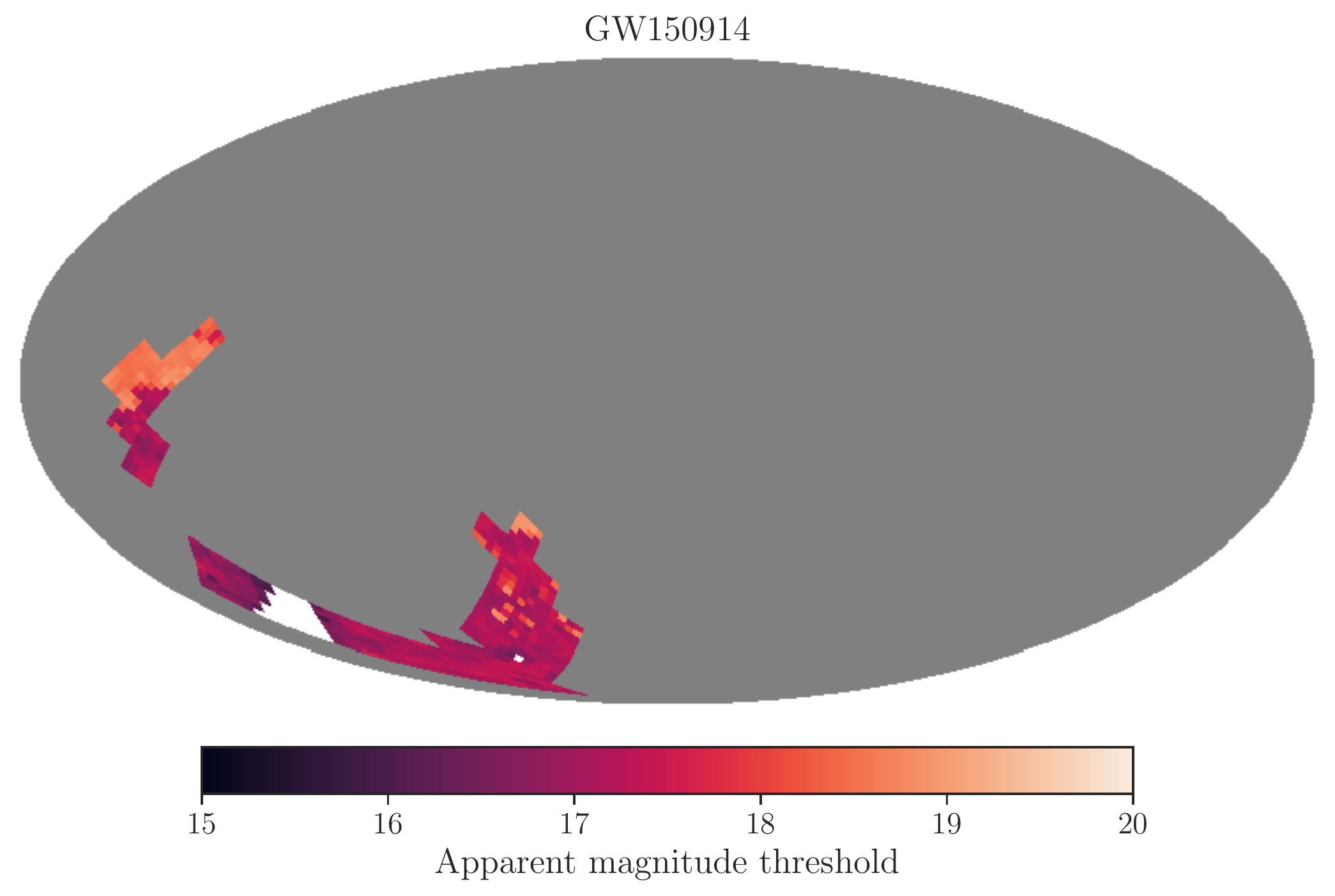}
\includegraphics[width=0.49\linewidth]{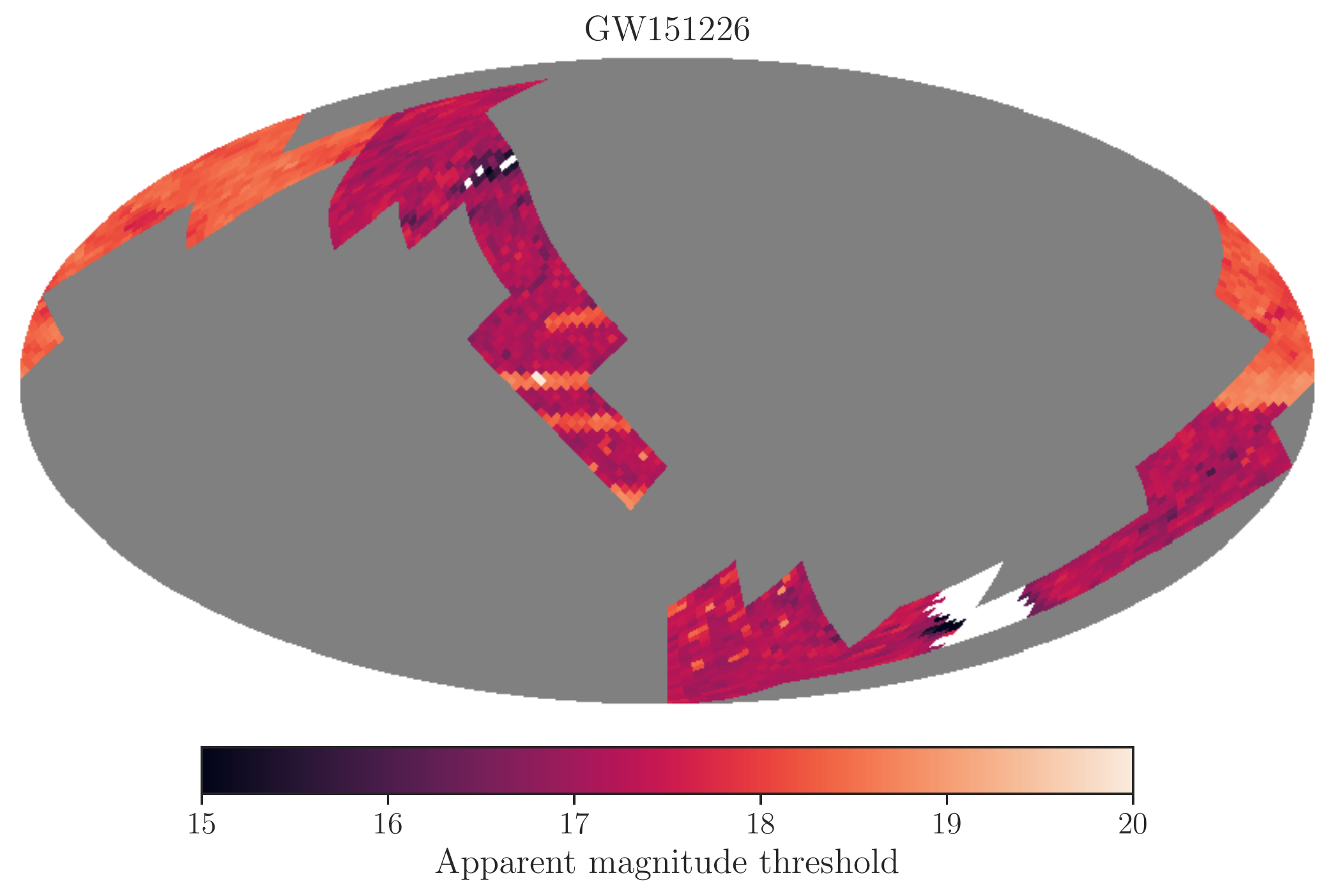}
\\
\includegraphics[width=0.49\linewidth]{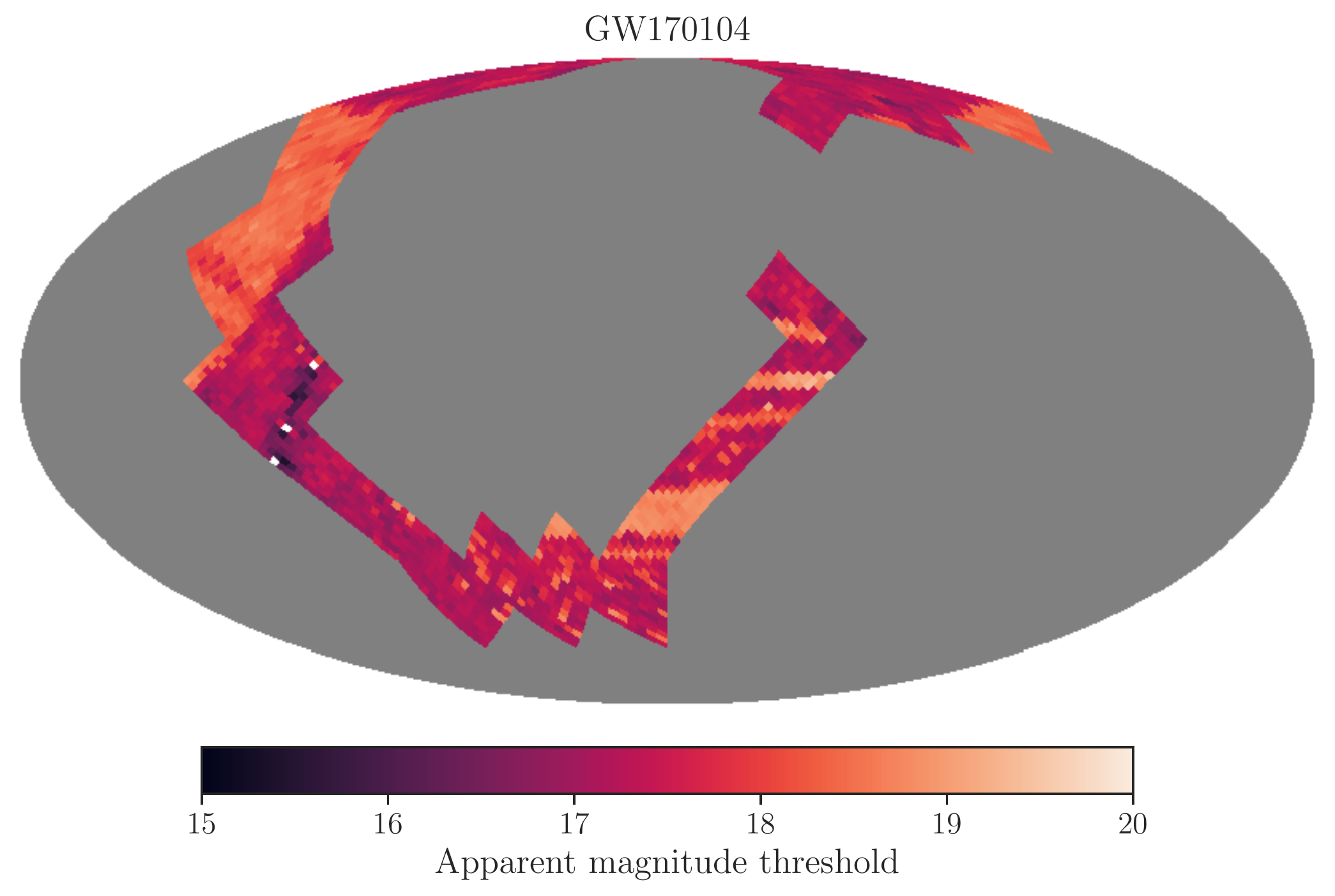}
\includegraphics[width=0.49\linewidth]{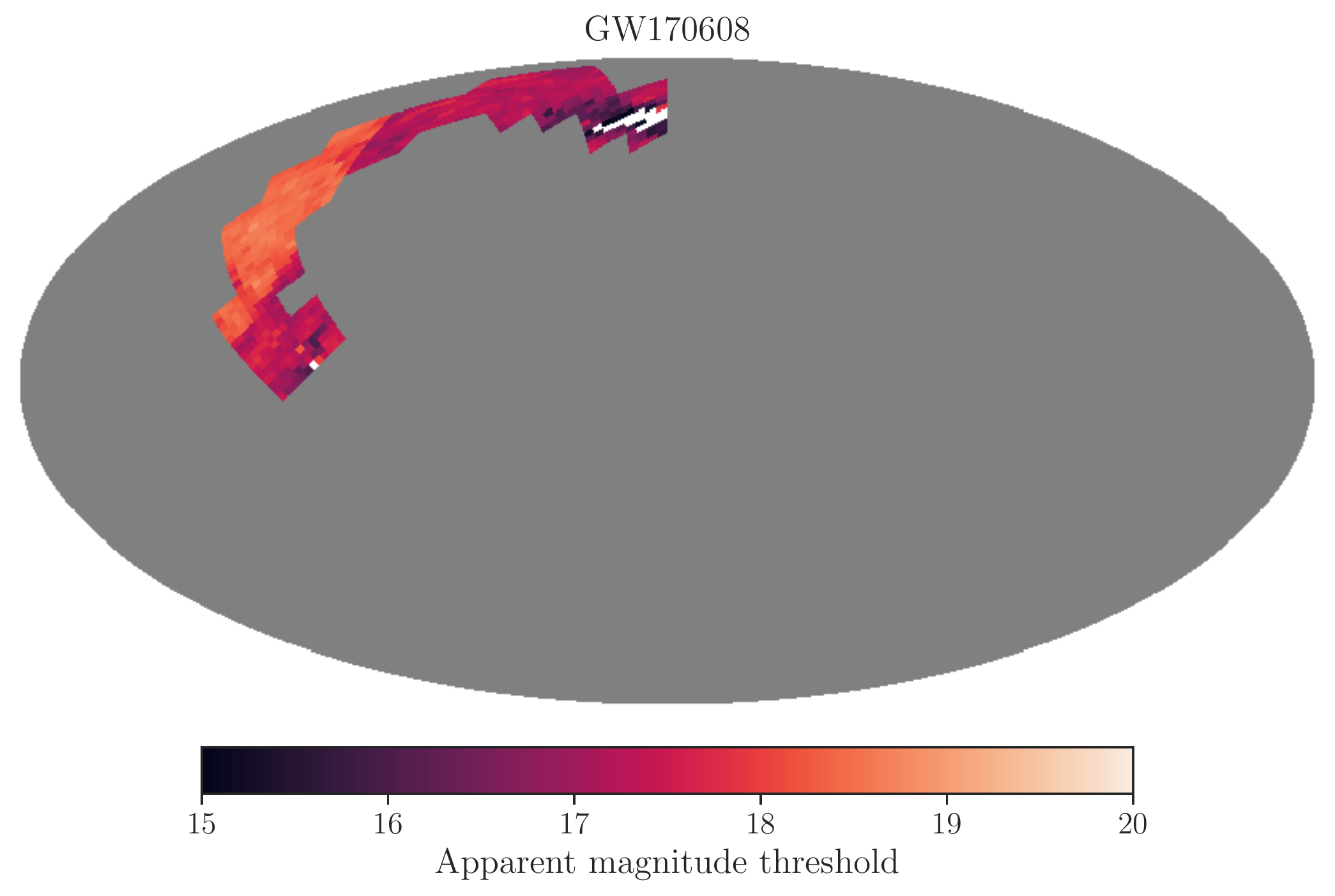}
\\
\includegraphics[width=0.49\linewidth]{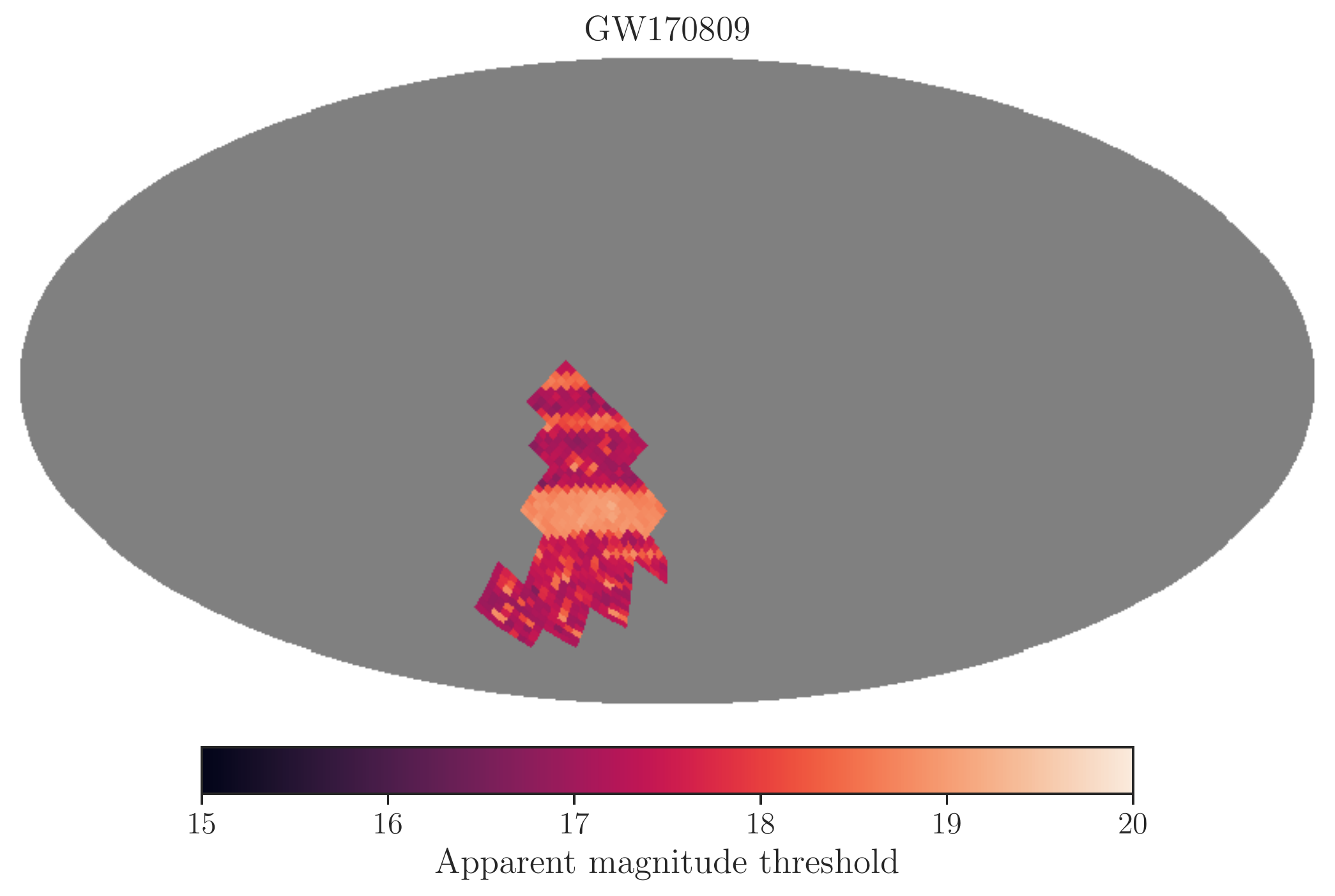}
\includegraphics[width=0.49\linewidth]{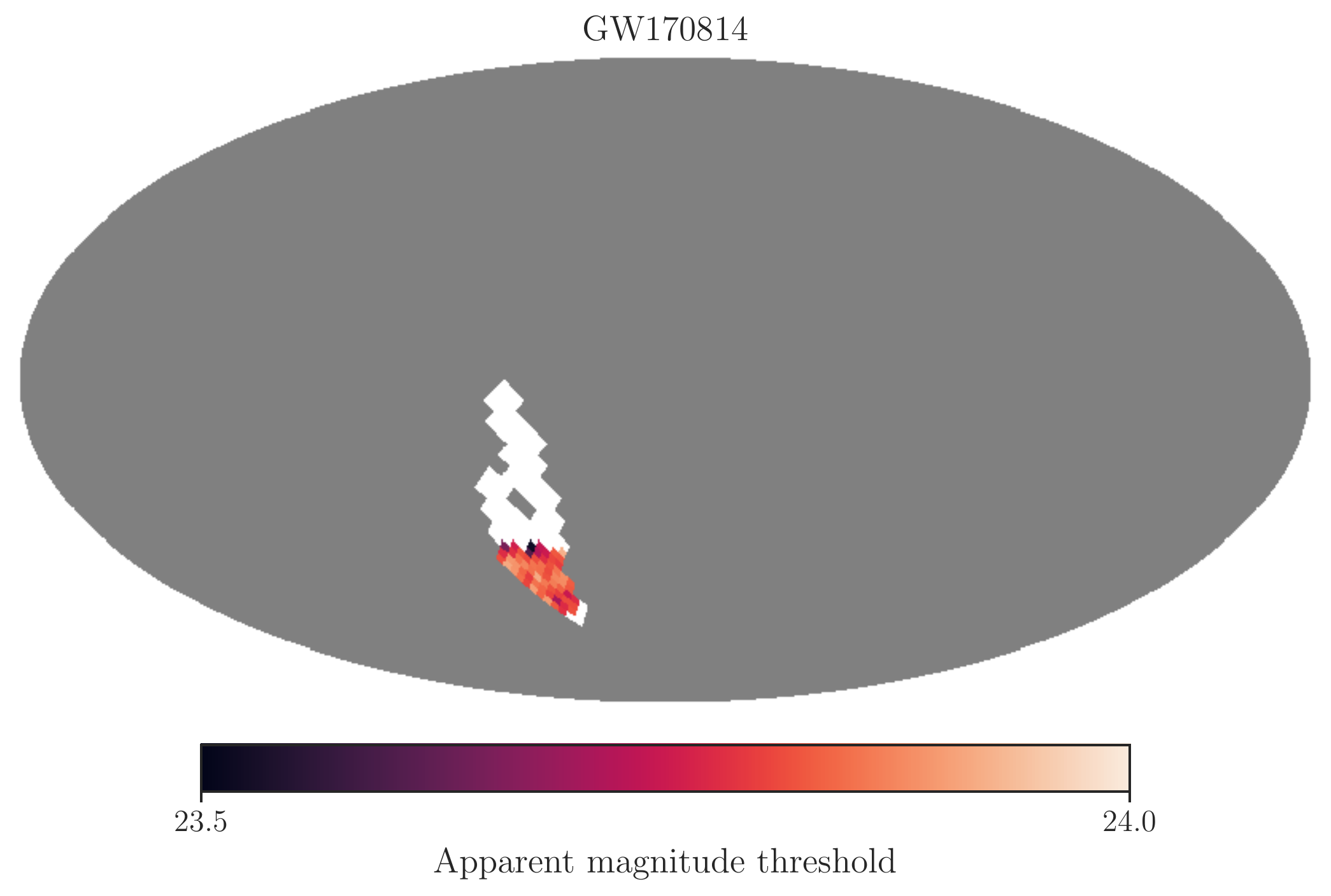}
\caption[Variation of the apparent magnitude threshold within the 99.9\% sky area of GW150914, GW151226, GW170104, GW170608 and GW170809 (using the GLADE catalogue), and GW170814 (using the DES-Y1 catalogue).]{Variation of the apparent magnitude threshold within the 99.9\% sky area of GW150914, GW151226, GW170104, GW170608 and GW170809 (using the \ac{GLADE} catalogue), and GW170814 (using the \ac{DES}-Y1 catalogue). The galaxy catalogue resolution (small pixels) is set with an nside of 32. The grey indicates a part of the sky which is not in the 99.9\% sky area of the event. The white pixels contain fewer than 10 galaxies either due to obscuration by the Milky Way band, or a lack of survey data, and are taken to be empty.}
\label{Fig:event_mthmap}
\end{figure*}

Figure \ref{Fig:event_mthmap} shows the variation in $B$-band apparent magnitude threshold across the 99.9\% sky area of GW150914, GW151226, GW170104, GW170608, GW170809 and GW170814. The resolution of the larger pixels, which determine which patch of the sky is considered for the analysis (in colour, where the grey is what is excluded), is determined by the resolution required to cover the 99.9\% sky area with at least 30 pixels. For GW150914, GW170608 and GW170809 this was an nside of 8. For GW151226 and GW170104, which were particularly poorly localised, this reduces to an nside of 4, while GW170814, which is much better-localised, adopts an nside of 16. An nside of 32 is chosen to represent the galaxy catalogue information, meaning that the pixels for each event needs to be split into a number of sub-pixels in order to reach the required resolution. If $\text{nside}_{\text{low}}$ is the low-resolution nside used to represent the \ac{GW} data, and $\text{nside}_{\text{high}}$ is the nside used to represent the galaxy catalogue, the number of sub-pixels that a single pixel must be divided into, $N_{\text{sub-pix}}$, is given by $4^k$, where $k = \log_2 (\text{nside}_{\text{high}}/\text{nside}_{\text{low}})$. This information is summarised for each event in the last columns of Table \ref{tab: pixels}.

\begin{figure*}
\includegraphics[width=0.65\linewidth]{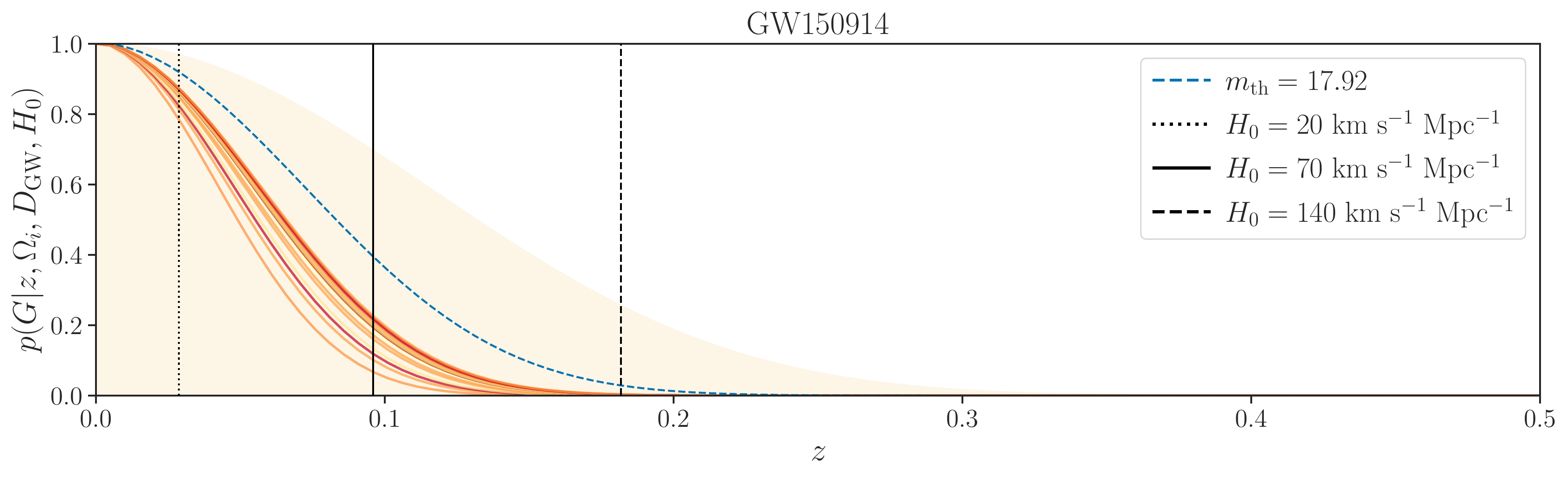}
\\
\includegraphics[width=0.65\linewidth]{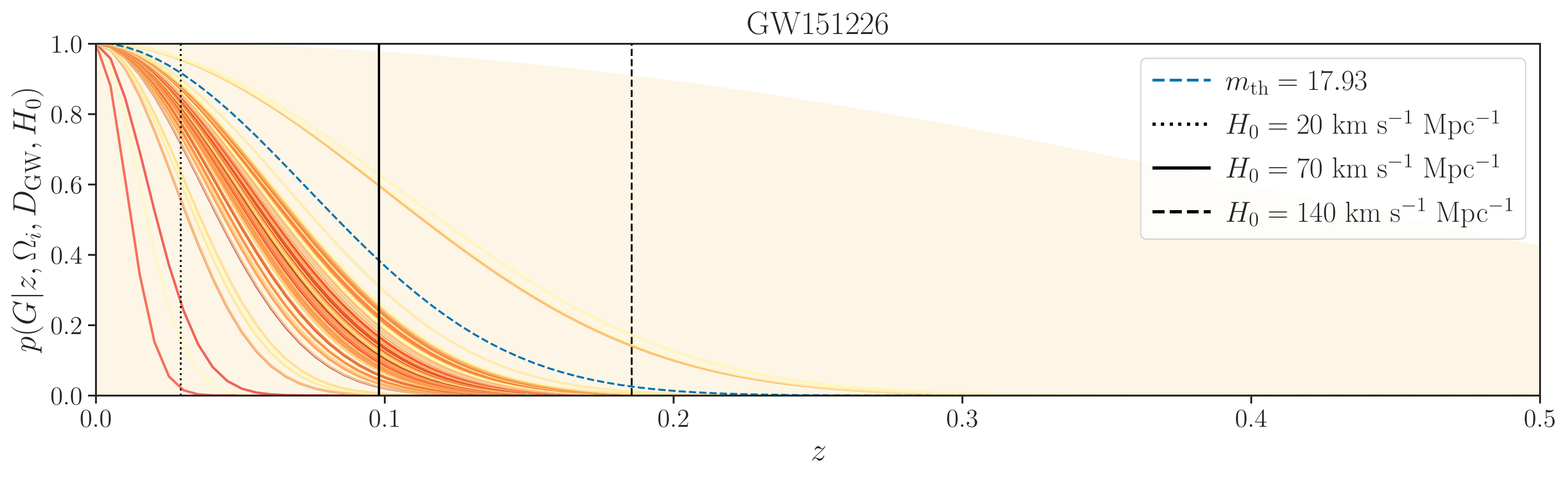}
\\
\includegraphics[width=0.65\linewidth]{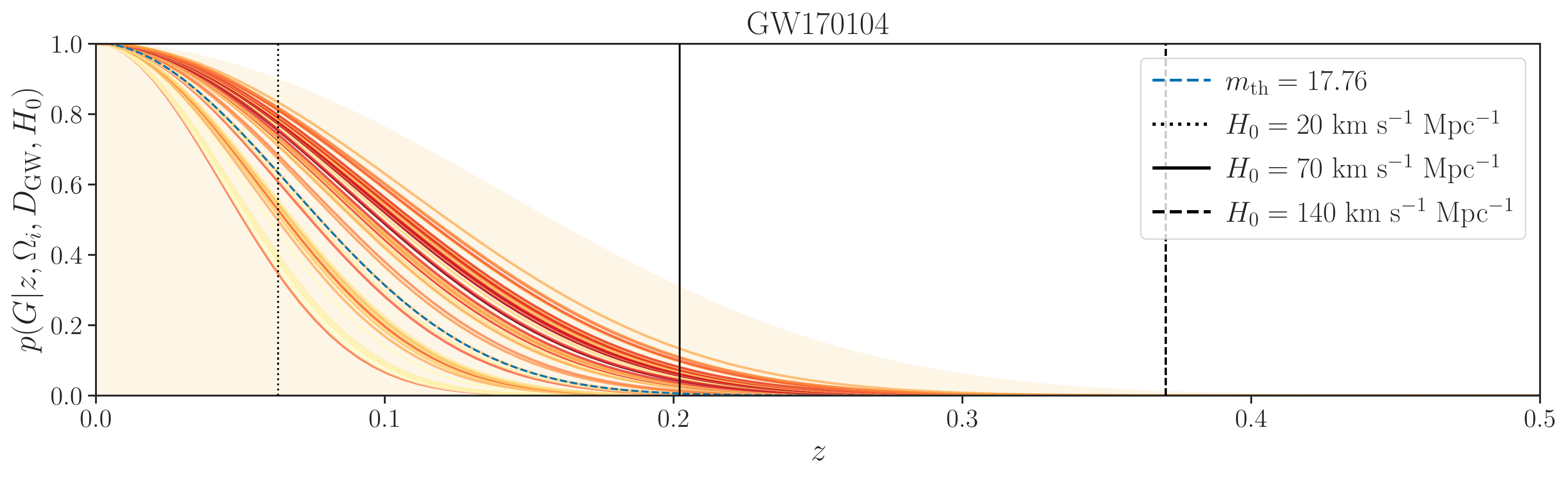}
\\
\includegraphics[width=0.65\linewidth]{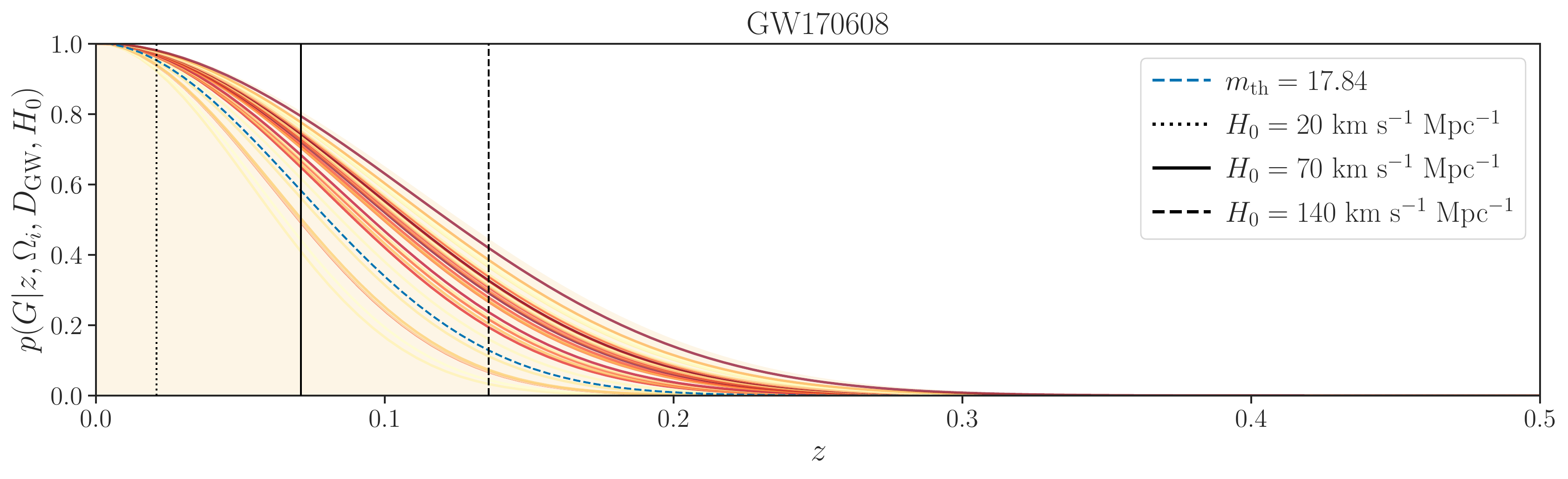}
\\
\includegraphics[width=0.65\linewidth]{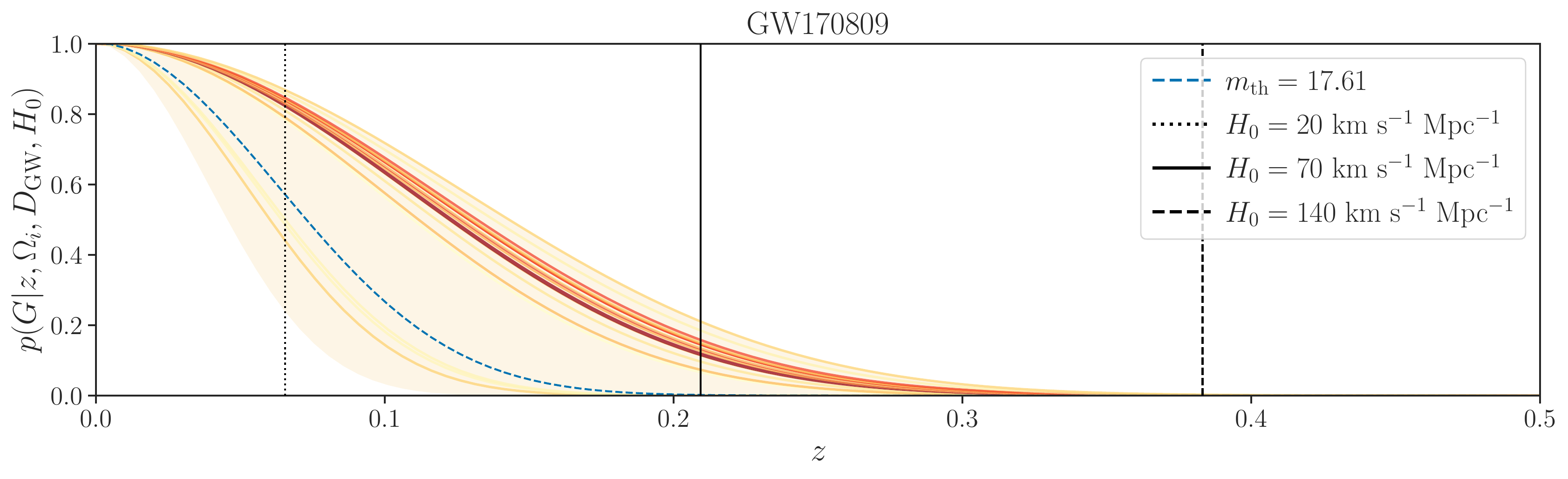}
\\
\includegraphics[width=0.65\linewidth]{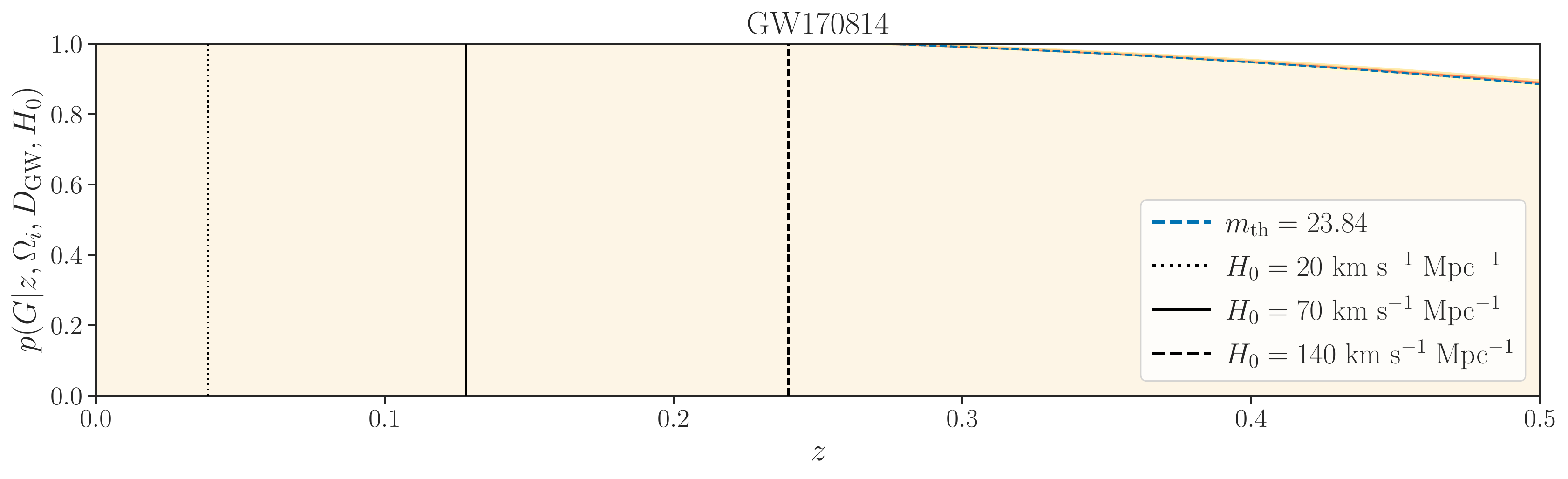}
\caption[Probability that the host galaxy is in the galaxy catalogue as a function of redshift, for the pixelated catalogue case (GW150914, GW151226, GW170104, GW170608, GW170809 and GW170814).]{Probability that the host galaxy is in the galaxy catalogue as a function of redshift, for the pixelated catalogue case. Solid orange curves show the probability that the host is inside the galaxy catalogue along the line-of-sight of the pixels which cover the 50\% sky-area of the event (darker orange corresponds to pixels which contain higher values of the \ac{GW} sky probability). The shaded yellow area covers the range between the minimum and maximum apparent magnitude threshold within the 99.9\% sky-area of the event. Vertical black lines show the median redshift of the event for different values of \ac{H0}. The blue dashed line gives the probability that the host is in the galaxy catalogue assuming the \ac{mth} value used in \citet{O2H0paper}.}
\label{Fig:pG_pixel}
\end{figure*}

The method of determining \ac{mth} within each sub-pixel is the same as \citet{O2H0paper} --- by taking the median $B$-band apparent magnitude of galaxies within the pixel. Again, looking at Fig. \ref{Fig:event_mthmap}, it is clear that this method captures a large variation in the apparent magnitude threshold within each event's sky-area.
Picking the median apparent magnitude is a conservative choice, and results which use this method will give less support to the in-catalogue part of the analysis than a method which calculates \ac{mth} more robustly. However, for consistency of comparisons with \citet{O2H0paper}, the same choice is made here. 

The variation of \ac{mth} over the sky translates directly into a variation in the probability that the host galaxy of the event is inside the catalogue, with lower thresholds corresponding to lower in-catalogue probabilities. Figure \ref{Fig:pG_pixel} shows how the probability that the host is in the catalogue varies within each event's sky area. The orange curves correspond to the pixels which contain 50\% of the \ac{GW} event's sky probability, with darker curves corresponding to the pixels with higher probabilities. The yellow shaded area shows the full range covered by the pixels which make up the 99.9\% sky area. This extends to a probability of 0 at $z=0$ if there are pixels within the event's 99.9\% sky area which are empty. In general, these plots show how the probability that the host is in the catalogue compares between the pixelated case and the \citet{O2H0paper} case which is shown by the dashed blue line. This curve always lies within the extremes of the pixelated case (within the yellow area).  However, the orange curves indicate where the bulk of the \ac{GW} probability is lying. For GW150914, for example, the bulk of the \ac{GW} probability corresponds to lower in-catalogue probability than previously, no doubt driven by the fact that part of the event is obscured by the Milky Way band, which will cause a low \ac{mth} in the adjacent pixels (pixels which are entirely empty do not show up on this plot). Conversely, GW170608's probability is clustered in an area of higher in-catalogue completeness than the average, boosting the in-catalogue support. For GW170814, the most probable pixels overlap with the \citet{O2H0paper} curve, as the \ac{mth} estimation for the DES-Y1 catalogue varies very little within the sky area of the event.

\section{Results}
\label{sec:pixelO2H0}

\begin{figure*}
\includegraphics[width=0.3\linewidth]{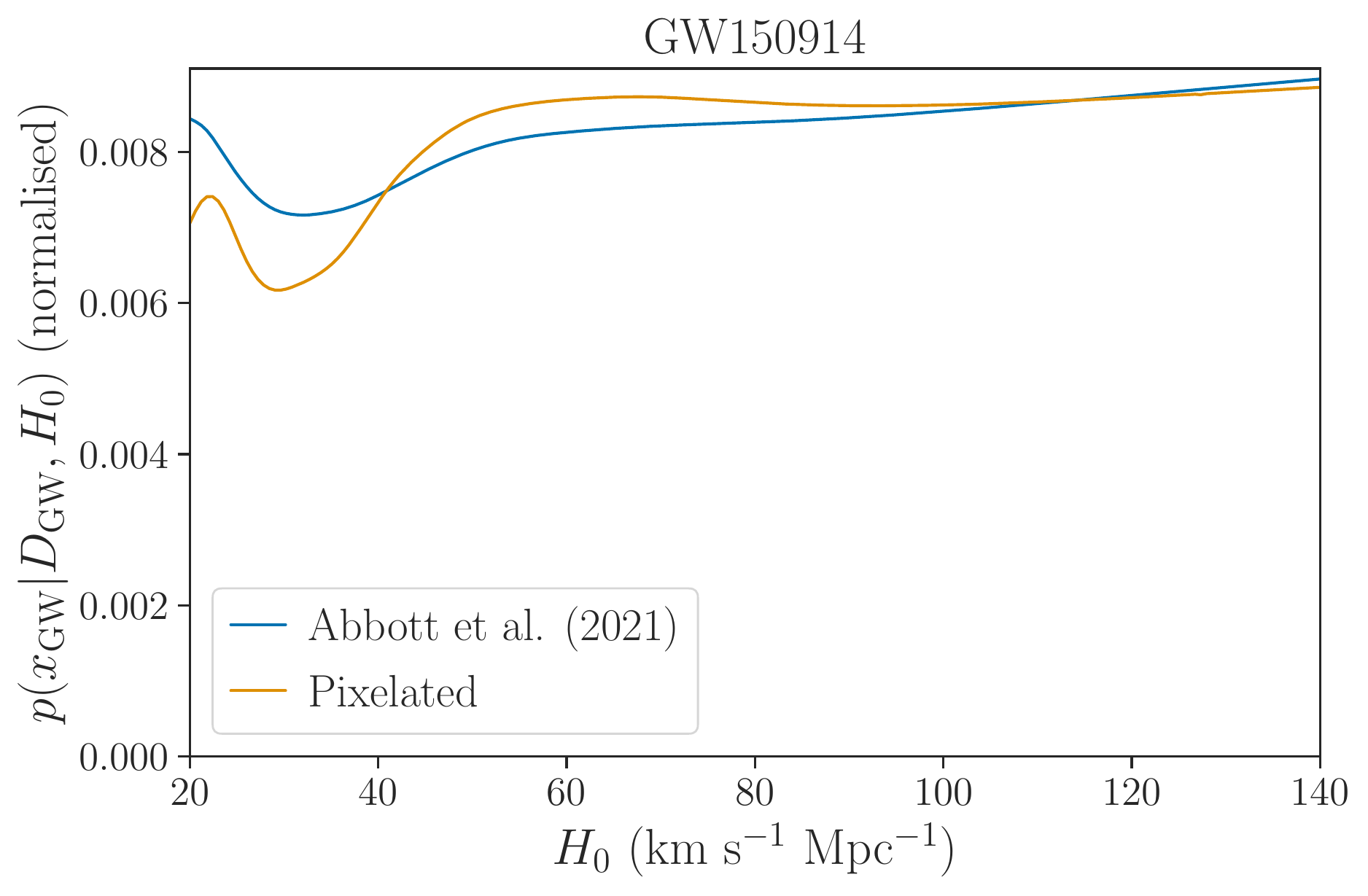}
\includegraphics[width=0.3\linewidth]{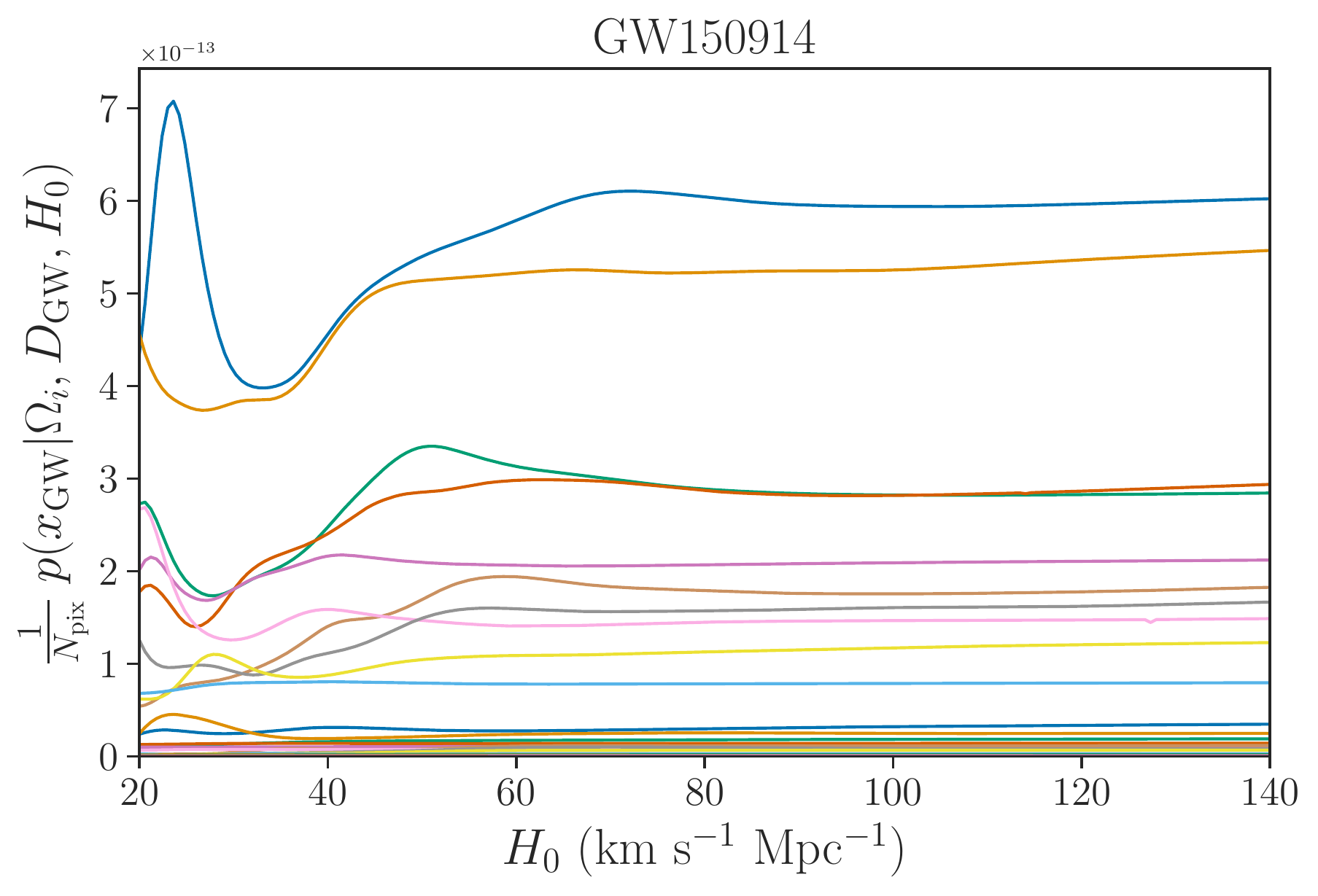}
\\
\includegraphics[width=0.3\linewidth]{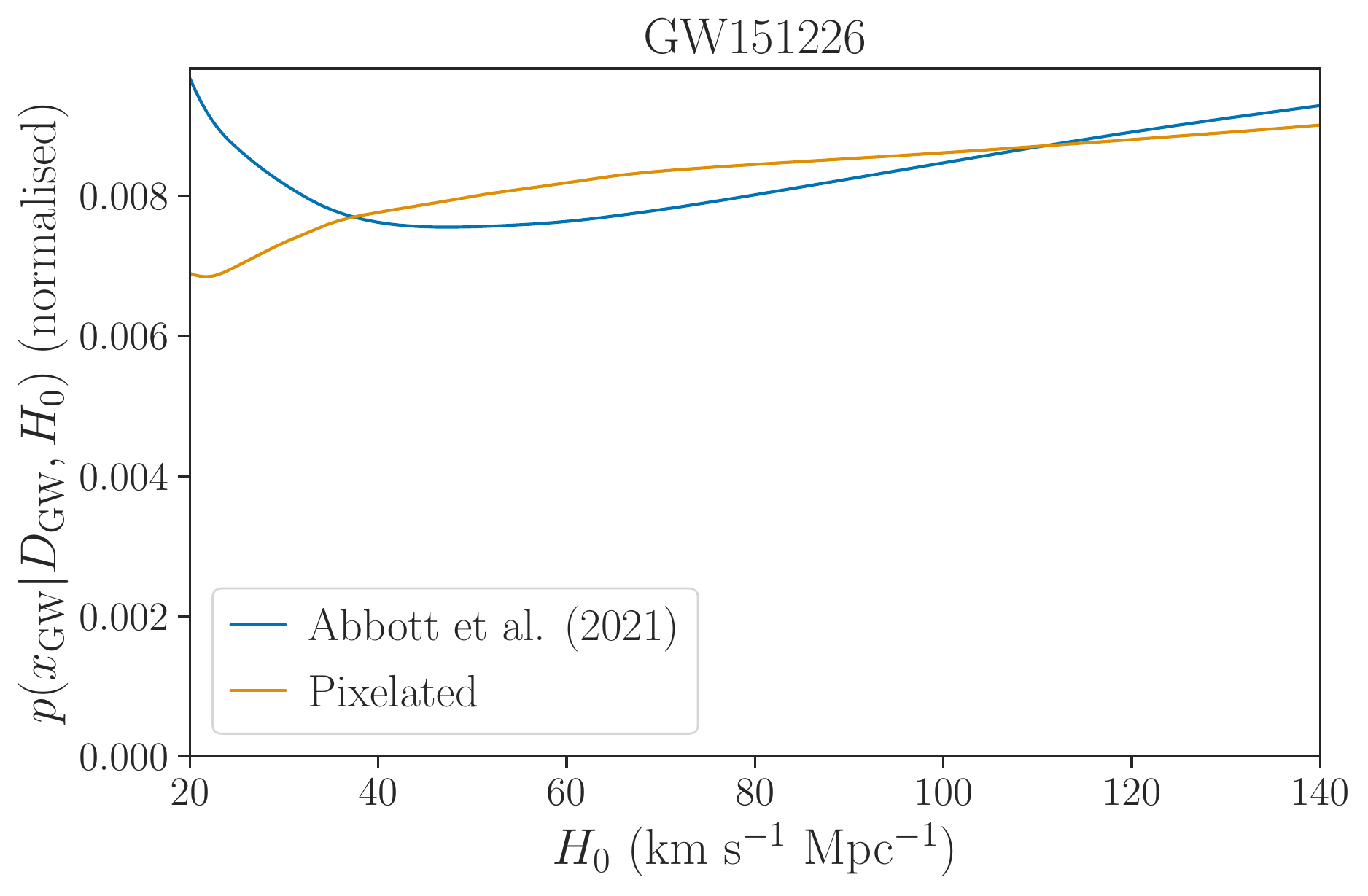}
\includegraphics[width=0.3\linewidth]{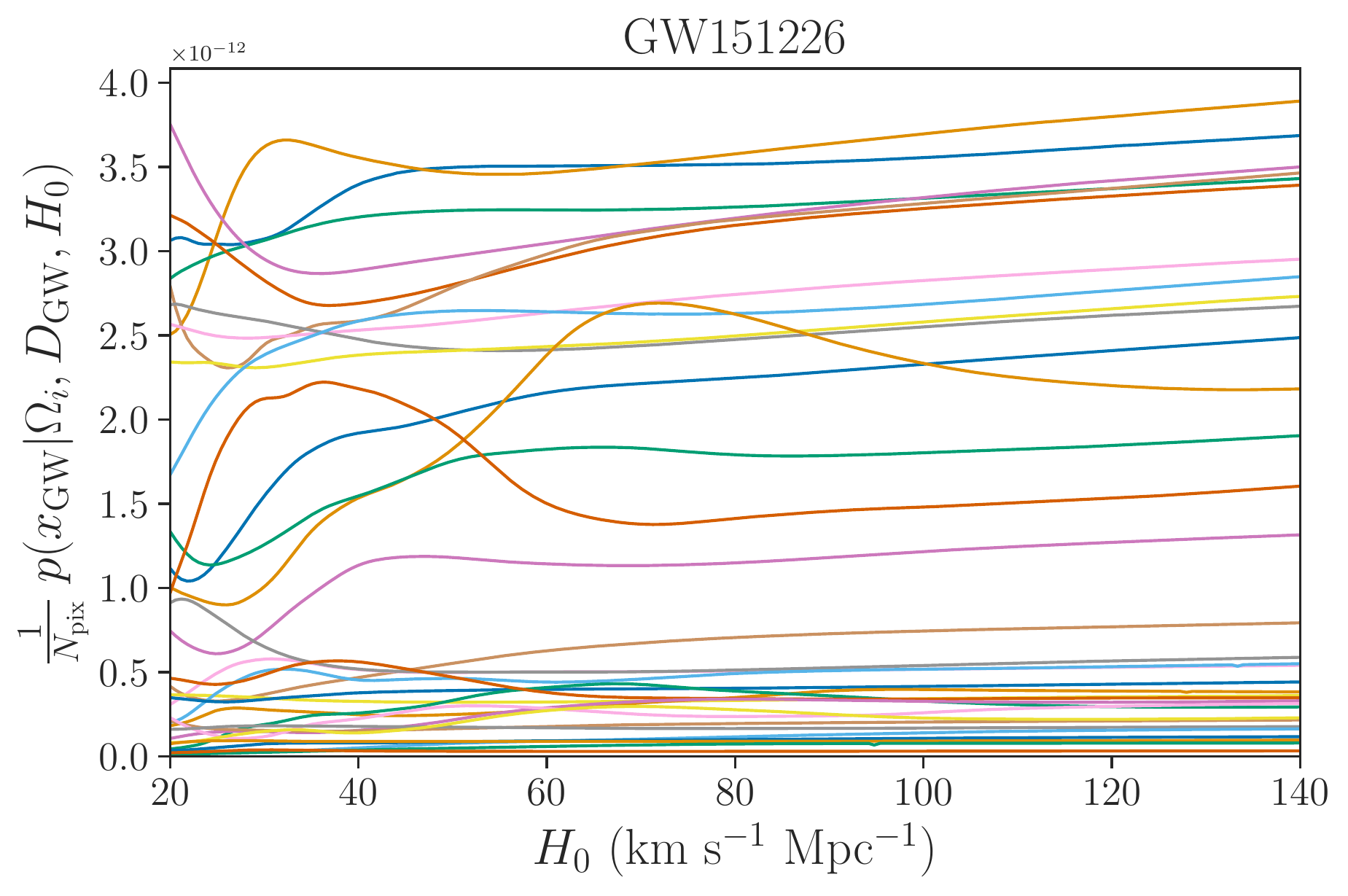}
\\
\includegraphics[width=0.3\linewidth]{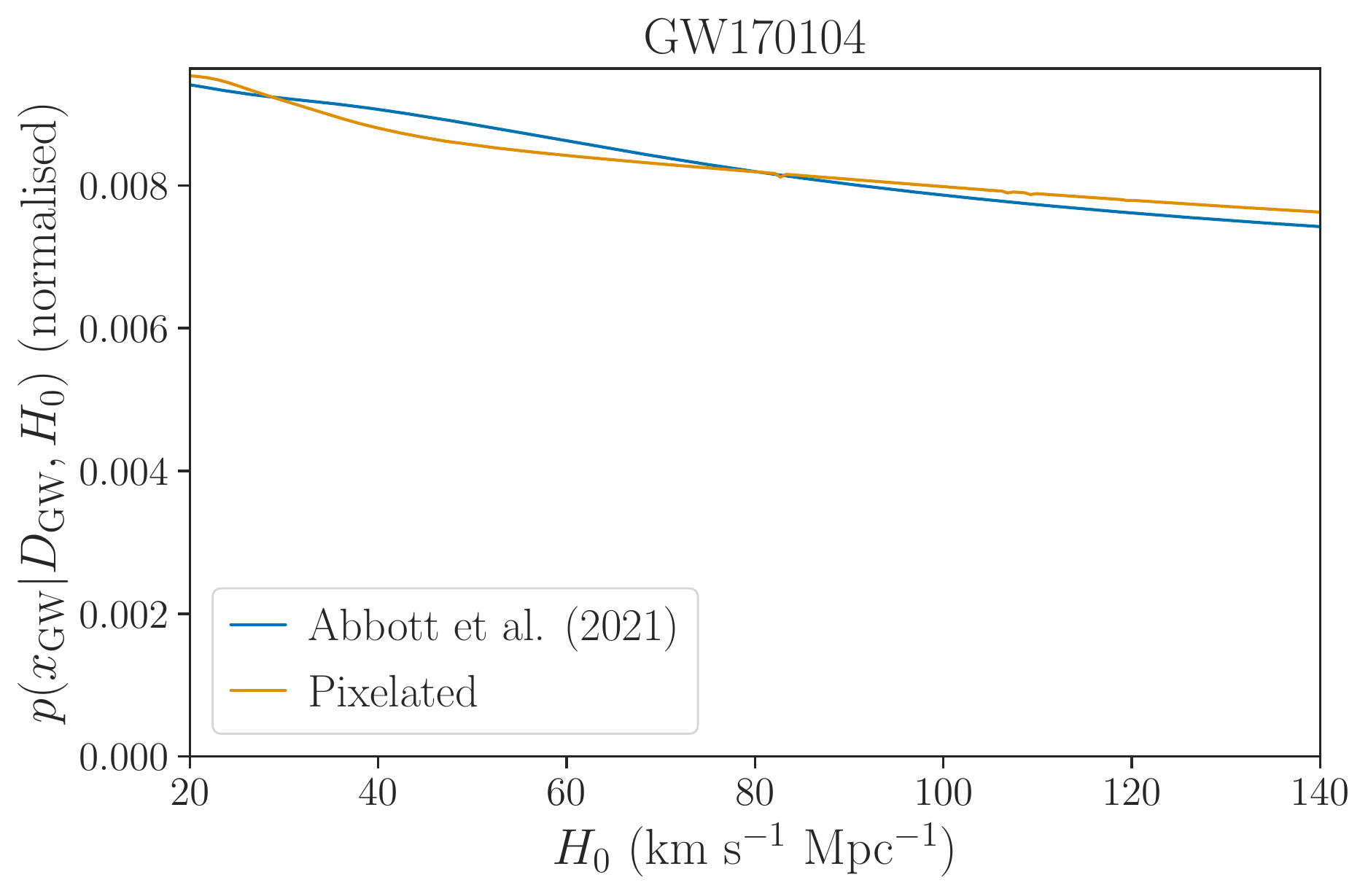}
\includegraphics[width=0.3\linewidth]{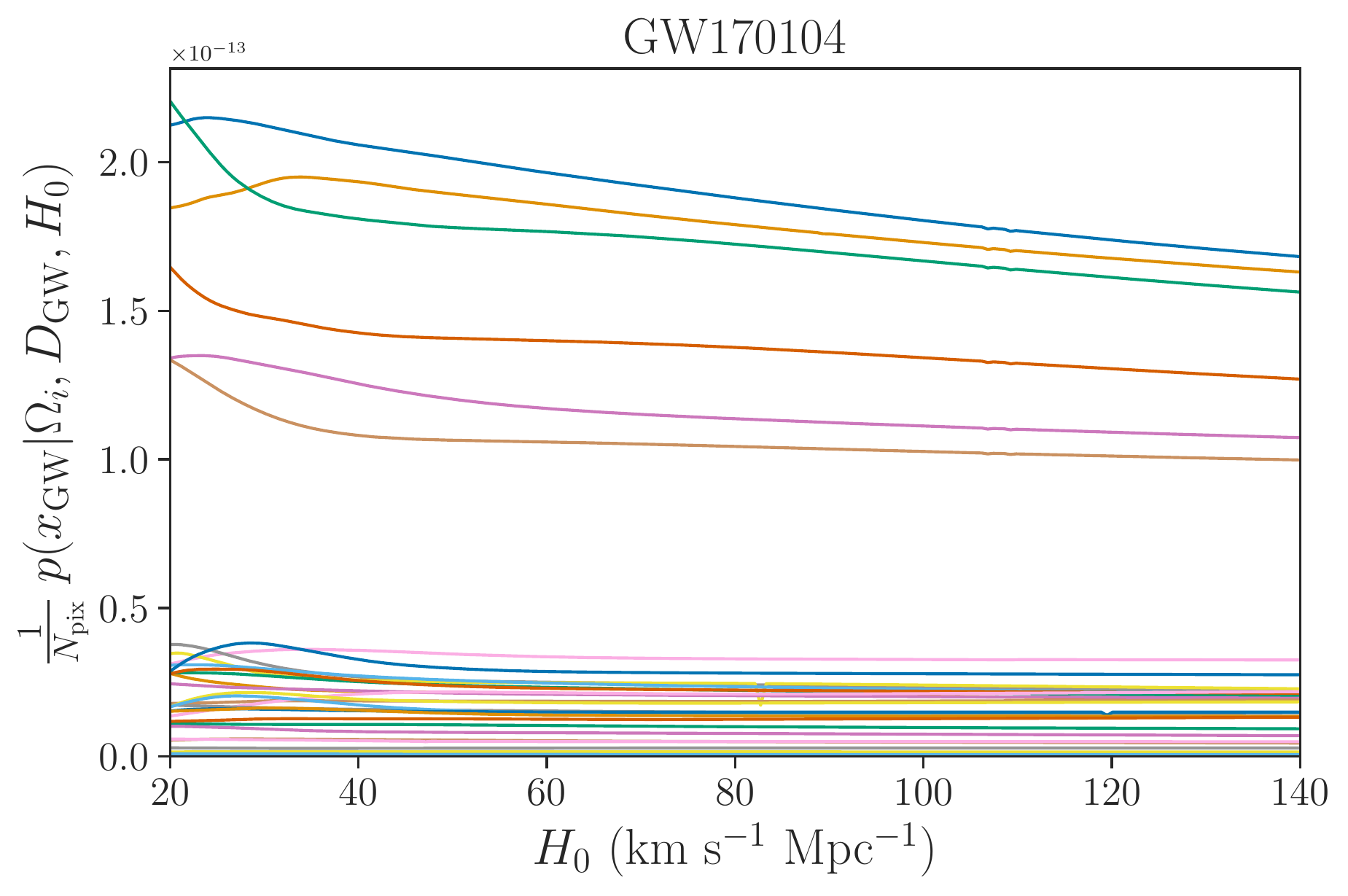}
\\
\includegraphics[width=0.3\linewidth]{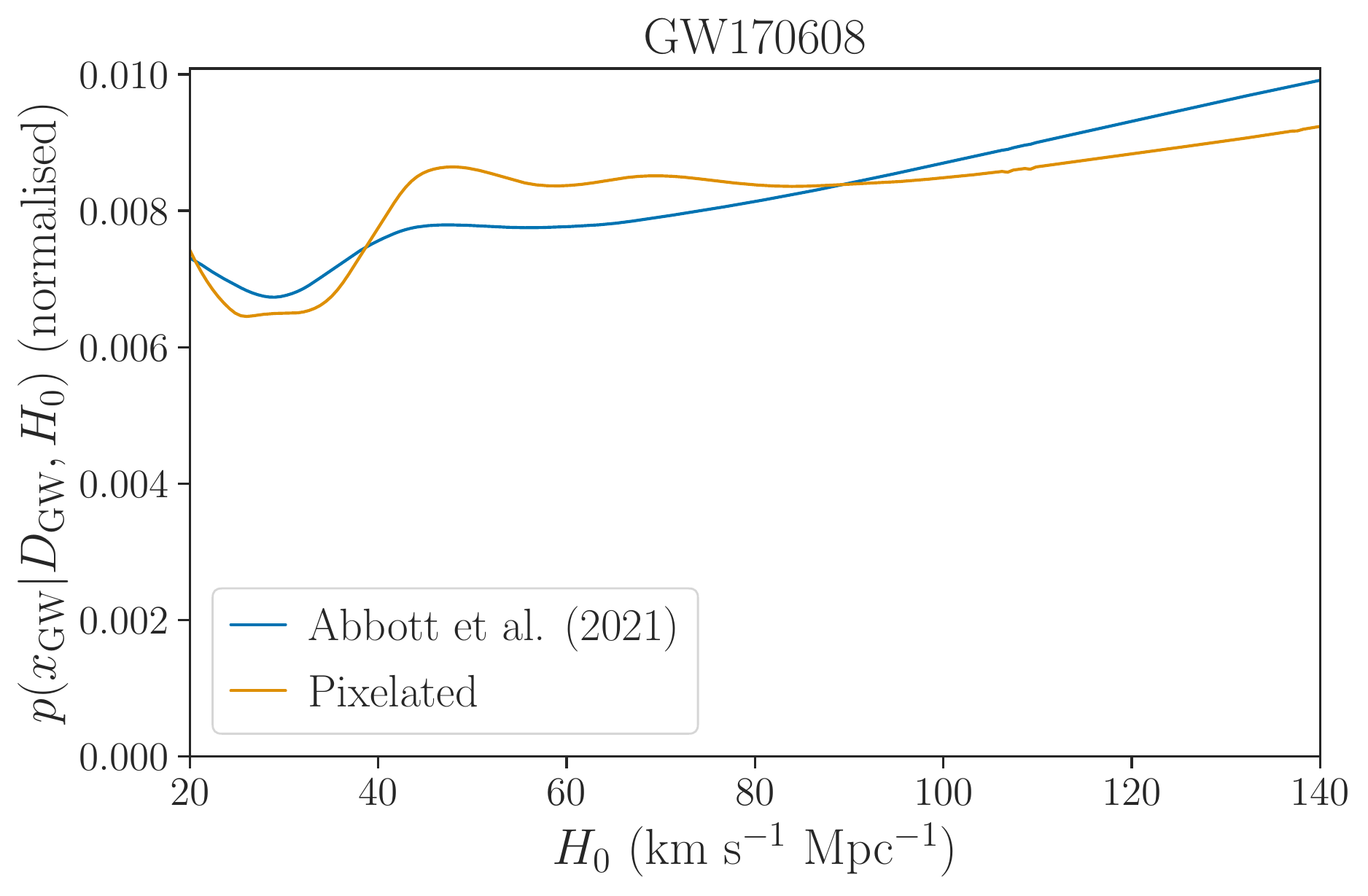}
\includegraphics[width=0.3\linewidth]{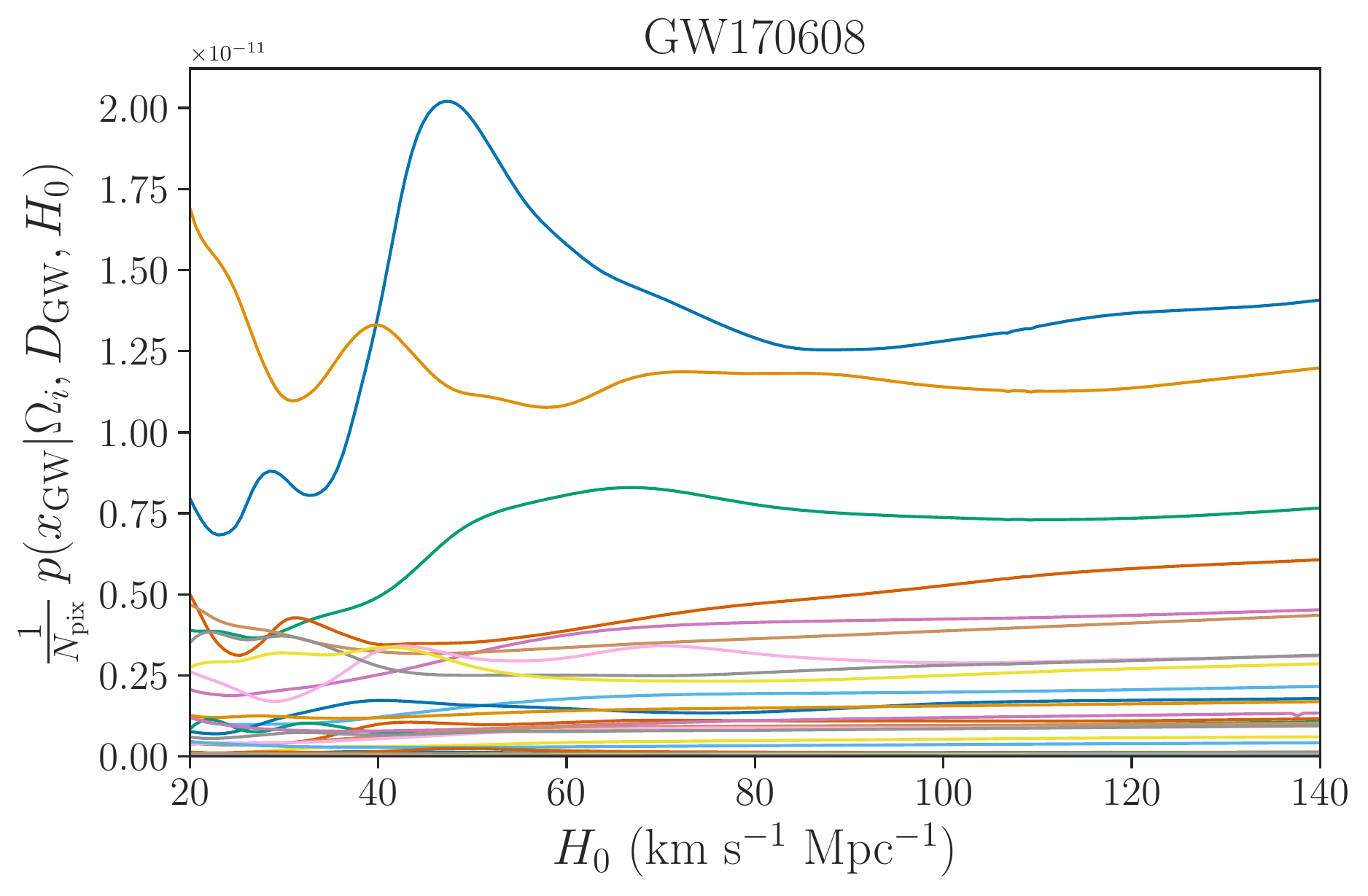}
\\
\includegraphics[width=0.3\linewidth]{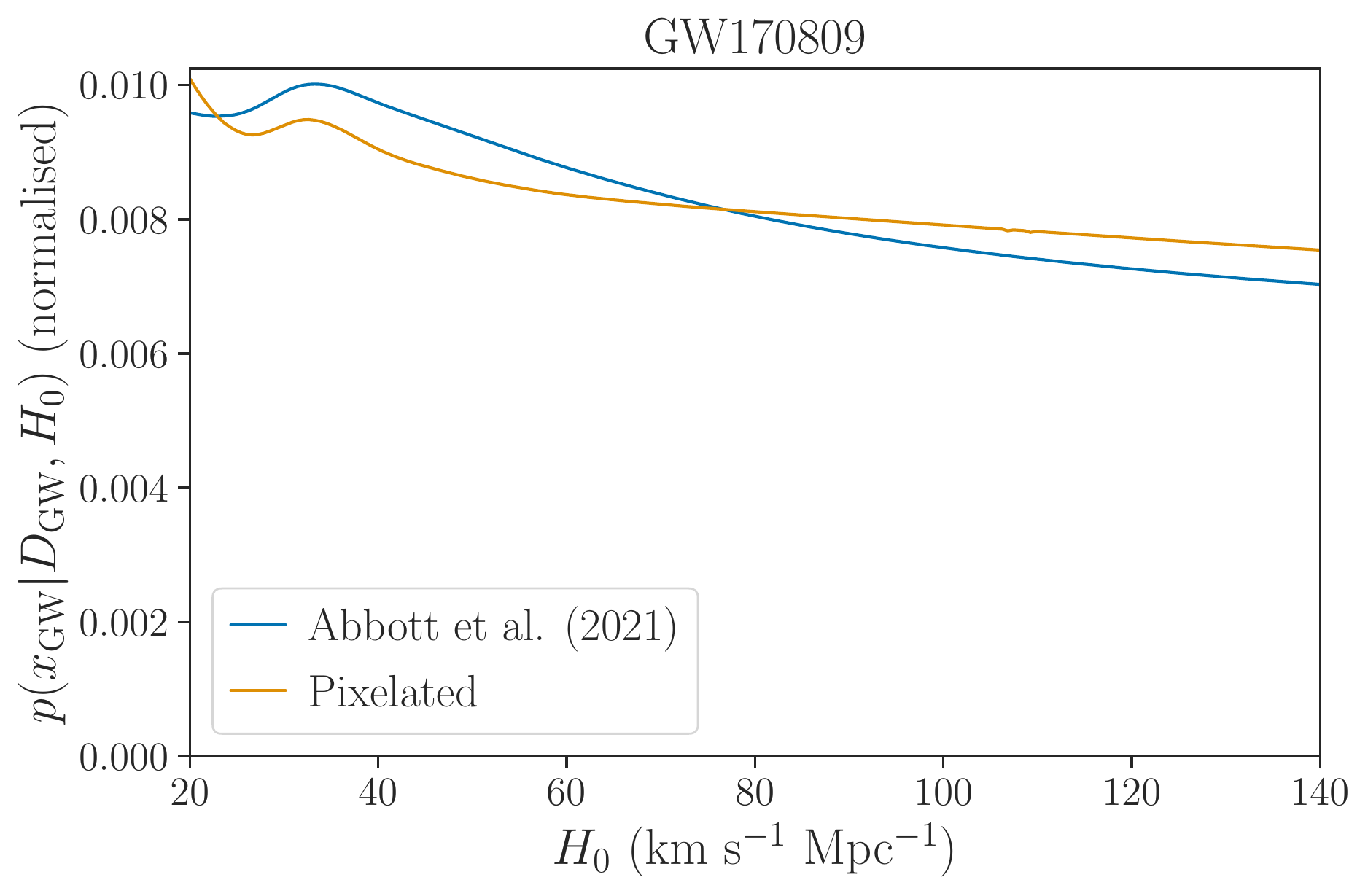}
\includegraphics[width=0.3\linewidth]{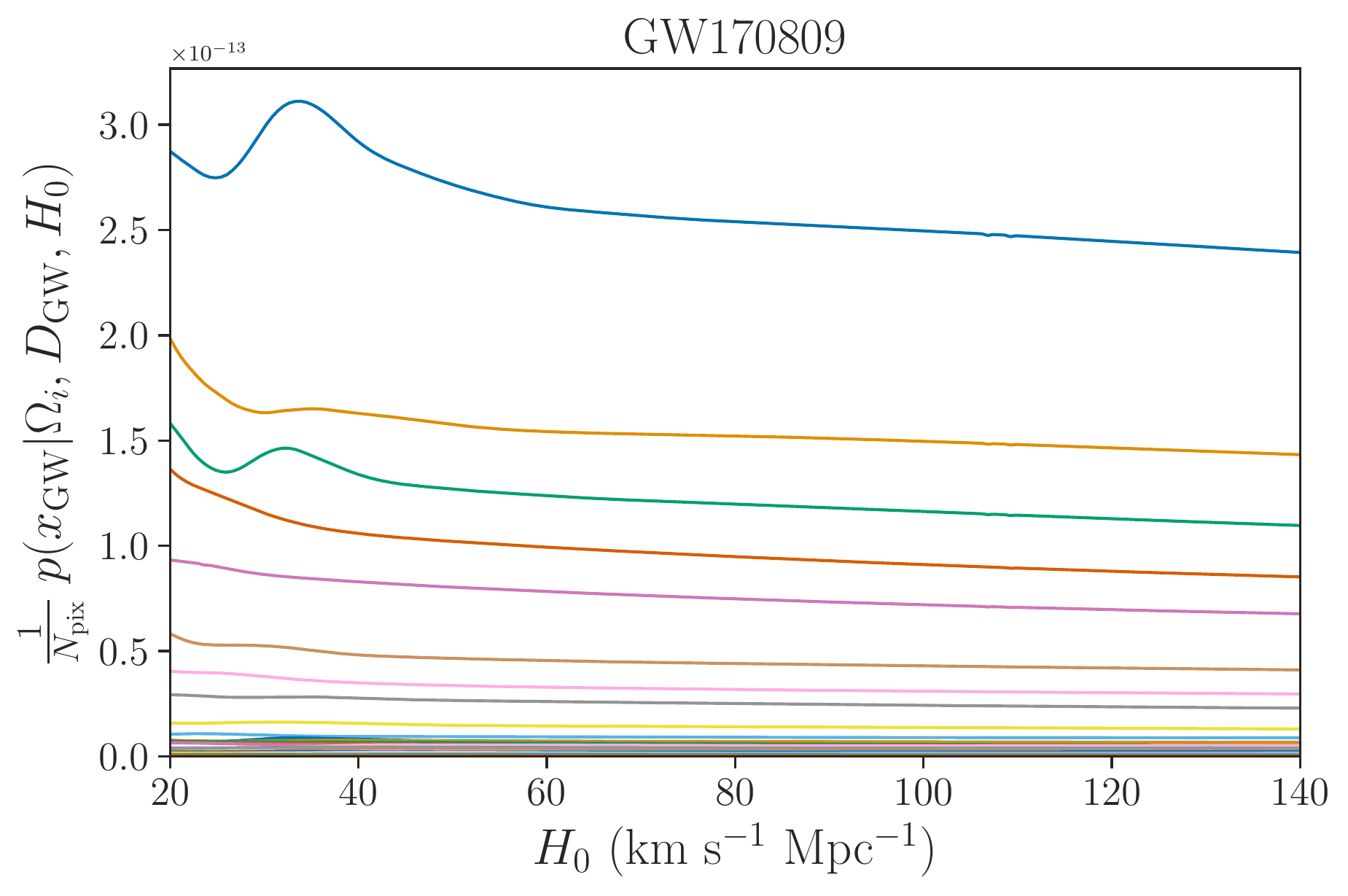}
\\
\includegraphics[width=0.3\linewidth]{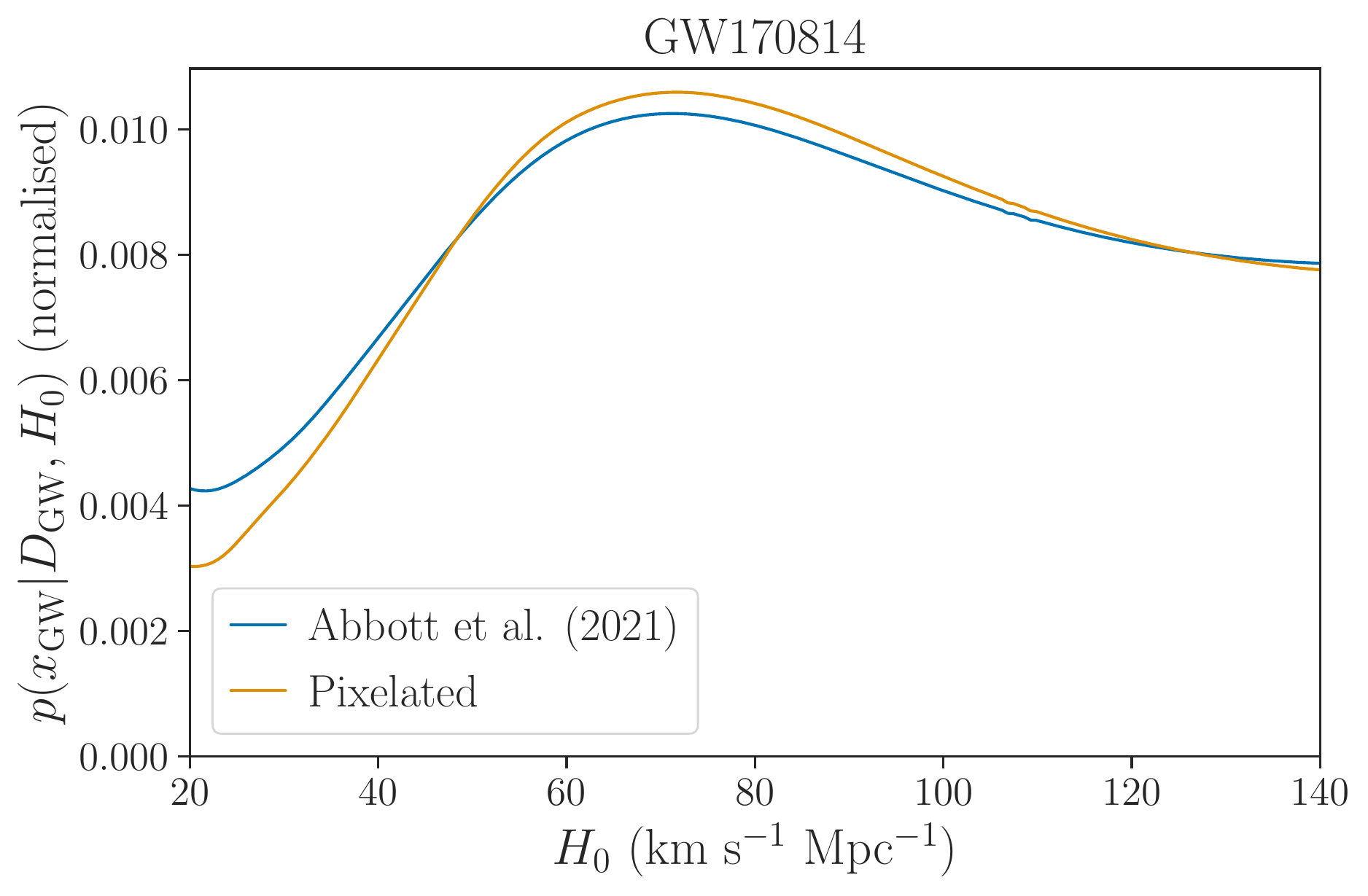}
\includegraphics[width=0.3\linewidth]{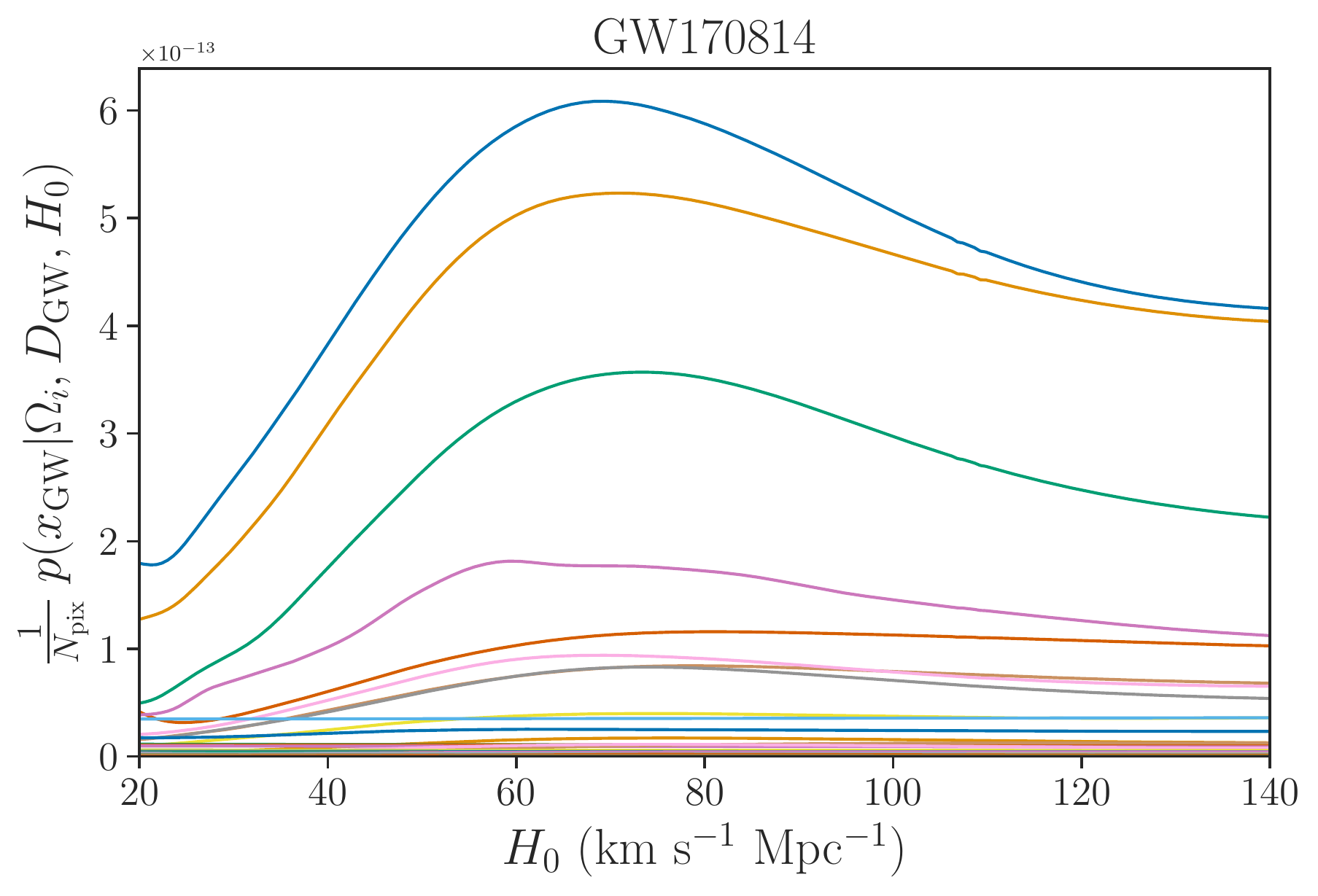}
\caption[Likelihoods on $H_0$ for the pixelated analysis with GW150914, GW151226, GW170104, GW170608, GW170809 and GW170814.]{Likelihoods on $H_0$ for the pixelated analysis with GW150914, GW151226, GW170104, GW170608, GW170809 and GW170814. \textit{Left-hand panels:} Comparison of the (normalised) likelihoods between the \citet{O2H0paper} result (blue) and the pixelated result (orange). \textit{Right-hand panels:} A breakdown of the pixelated likelihood by low resolution pixel. The pixelated likelihood in the left-hand panel is the sum of the curves in the right-hand panel (then normalised).}
\label{Fig:pixel_likelihoods}
\end{figure*}

Taking the method outlined in Section \ref{sec:pixelmethod} and applying it to the 6 GWTC-1 \acp{BBH} which pass a network SNR threshold of 12 (GW150914, GW151226, GW170104, GW170608, GW170809 and GW170814) produces the individual event likelihoods on \ac{H0} shown in Fig. \ref{Fig:pixel_likelihoods}.  The same mass distribution and detection threshold as outlined in section \ref{sec:emptycat} is used. The $B$-band luminosity function of galaxies (used to match those in the GLADE catalogue) is assumed to follow a Schechter function with slope $\alpha=-1.07$, a characteristic absolute magnitude of $M^*(H_0) = -19.7+5\log h$ and a maximum limit on the faintest galaxies of $-12.2+5\log h$, where $h \equiv H_0/100$, which follows \citet{Gehrels:2015uga}. The $g$-band luminosity function (used to match the DES-Y1 galaxy catalogue \citep{Abbott:2018jhe,Drlica-Wagner:2017tkk}) is modelled using a Schechter function with $\alpha=-0.89$, $M^*(H_0)=-19.39+5\log_{10}h$, and a limit on faint galaxies of $-16.1 + 5\log_{10}h$ based on \cite{Blanton_2003}. Both of these Schechter functions are chosen to match the assumptions made in \citet{O2H0paper}. GW170814 is analysed with the DES-Y1 catalogue,\footnote{The DES-Y1 catalogue is available at \url{https://des.ncsa.illinois.edu/releases/y1a1}.} while the remaining events are analysed with the GLADE 2.4 catalogue.

The most interesting event is GW170608, due to its high in-catalogue probability. Looking at its right-hand panel in Fig. \ref{Fig:pixel_likelihoods}, which shows the contribution to the final likelihood on \ac{H0} from each individual pixel, it is clear that for low values of \ac{H0} the in-catalogue contribution is dominating. The final likelihood shows more structure than the \citet{O2H0paper} result, including increased posterior support around $H_0 \sim 70$ \kmsMpc.

While the likelihoods on \ac{H0} from GW150914 and GW151226 do not show much structure, both have more support around central values of $H_0$, and have reduced support at high \ac{H0}, compared to the \citet{O2H0paper} case. For these events, the contributions from the individual pixels are more interesting than the combined likelihood as they show, especially at the low-\ac{H0} end (which corresponds to low redshifts and therefore increased catalogue support) structure which corresponds to real redshift and luminosity information from the GLADE catalogue galaxies. It is interesting that GW150914's most probable pixel peaks around $H_0 \sim 70$ {\kmsMpc}  which could indicate an over-density of galaxies at the relevant redshift, although it remains much more likely that the host galaxy for this event is not contained within the catalogue (based on the top panel of Fig. \ref{Fig:pG_pixel}).

The other event of interest is GW170814, which was analysed with the DES-Y1 catalogue, and for which the likelihood remains relatively unchanged with the pixelated method, though with a fraction more support around central values of $H_0$. Looking at the line-of-sight estimates for this event (Fig. \ref{Fig:LOSredshift}), it is clear that the bulk of the information is coming from a small number of pixels, which peak at approximately the same redshifts. The apparent magnitude threshold within the event's sky area remains relatively uniform (see Fig. \ref{Fig:event_mthmap}). From the likelihood breakdown of the event in Fig. \ref{Fig:pixel_likelihoods}, it seems likely that the most probable pixels overlap with the same over-density of galaxies in the catalogue, leading to a relatively unchanged result. 

\begin{figure*}
\includegraphics[width=0.7\linewidth]{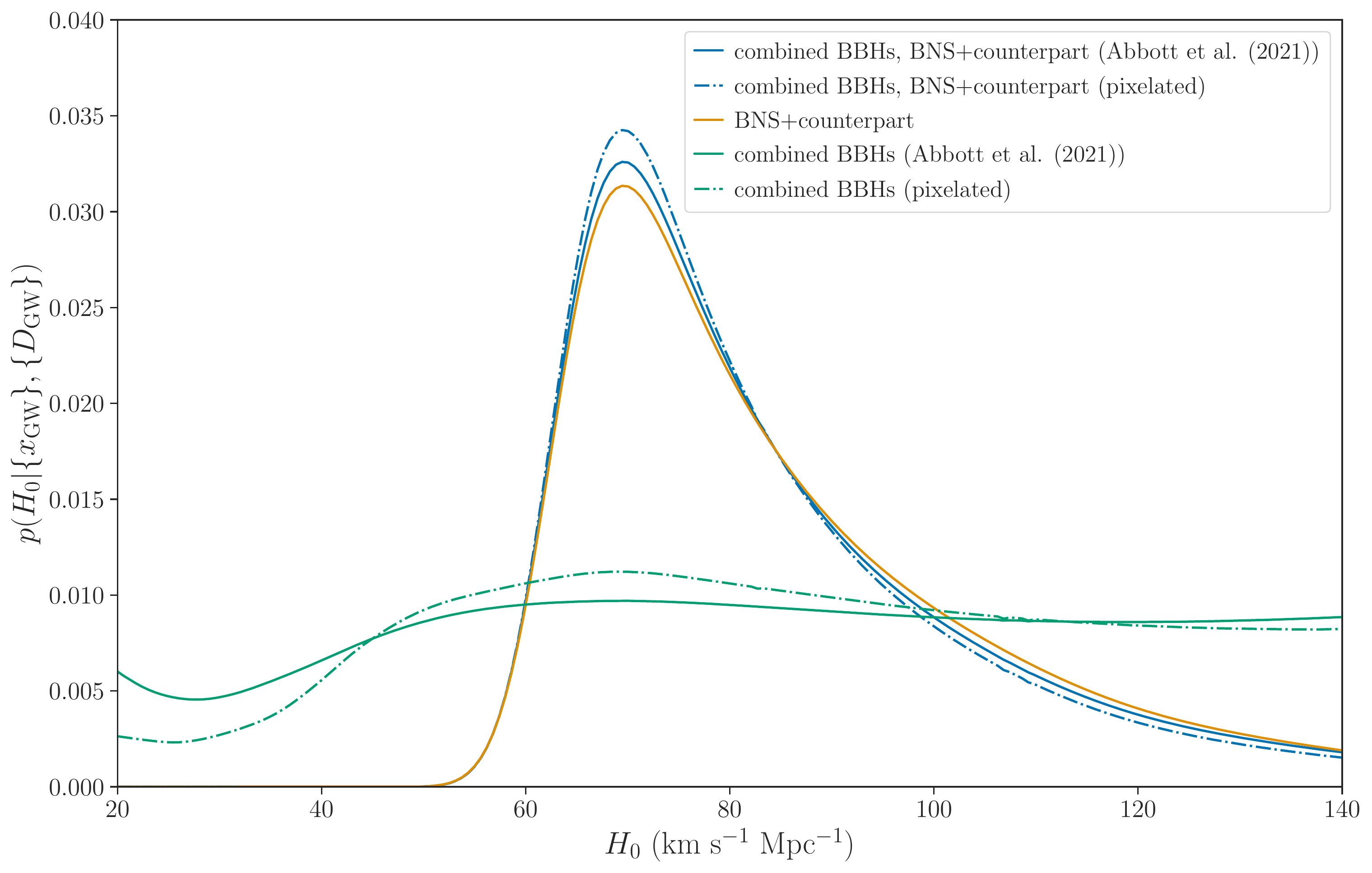}
\caption{Comparison of the posterior on $H_0$ between the pixelated and \citet{O2H0paper} analyses.  The green lines show the combined posterior from 6 \acp{BBH} (GW150914, GW151226, GW170104, GW170608 and GW170809 analysed with the GLADE 2.4 catalogue, and GW170814 analysed with the DES-Y1 catalogue). The orange line shows the contribution from GW170817 and its counterpart. Blue lines show the combined posterior on \ac{H0} from the 6 \acp{BBH} and the \ac{BNS} with counterpart. Solid lines correspond to the \citet{O2H0paper} results and dot-dashed lines show the pixelated results.}
\label{Fig:pixel_posterior}
\end{figure*}

Taking these 6 \acp{BBH} and combining them with the result from GW170817 and its counterpart \citep{GW170817:H0,GW170817:MMA} gives the posterior on $H_0$ shown in Fig. \ref{Fig:pixel_posterior}. The pixelated method gives a result of $H_0 = $ {\HubbleMeasCombinedCounterpartFLATLOGpixel} \kmsMpc, (maximum a posteriori and 68.3\% highest-density interval, with a flat-in-log prior on \ac{H0}). This is approximately a 5\% improvement over the \citet{O2H0paper} result. The improvement is driven by the additional information from GW170608, which has greater support around an $H_0$ of 70 \kmsMpc due to a more informative catalogue contribution, as well as by minor improvements from GW150914, GW151226 and GW170814, all of which have marginally increased support at middling values of \ac{H0} under this new method.

\subsection{The impact of resolution choices on the result}
\label{sec:systematics resolution}
The pixelated method introduces several choices to the analysis: the threshold for the sky area that will be analysed, the number of pixels that will be used to cover that sky area (and hence the size of those pixels) and the resolution of the galaxy catalogue. This section investigates the impact of varying those choices.

\begin{figure}
\centering
\includegraphics[width=0.95\linewidth]{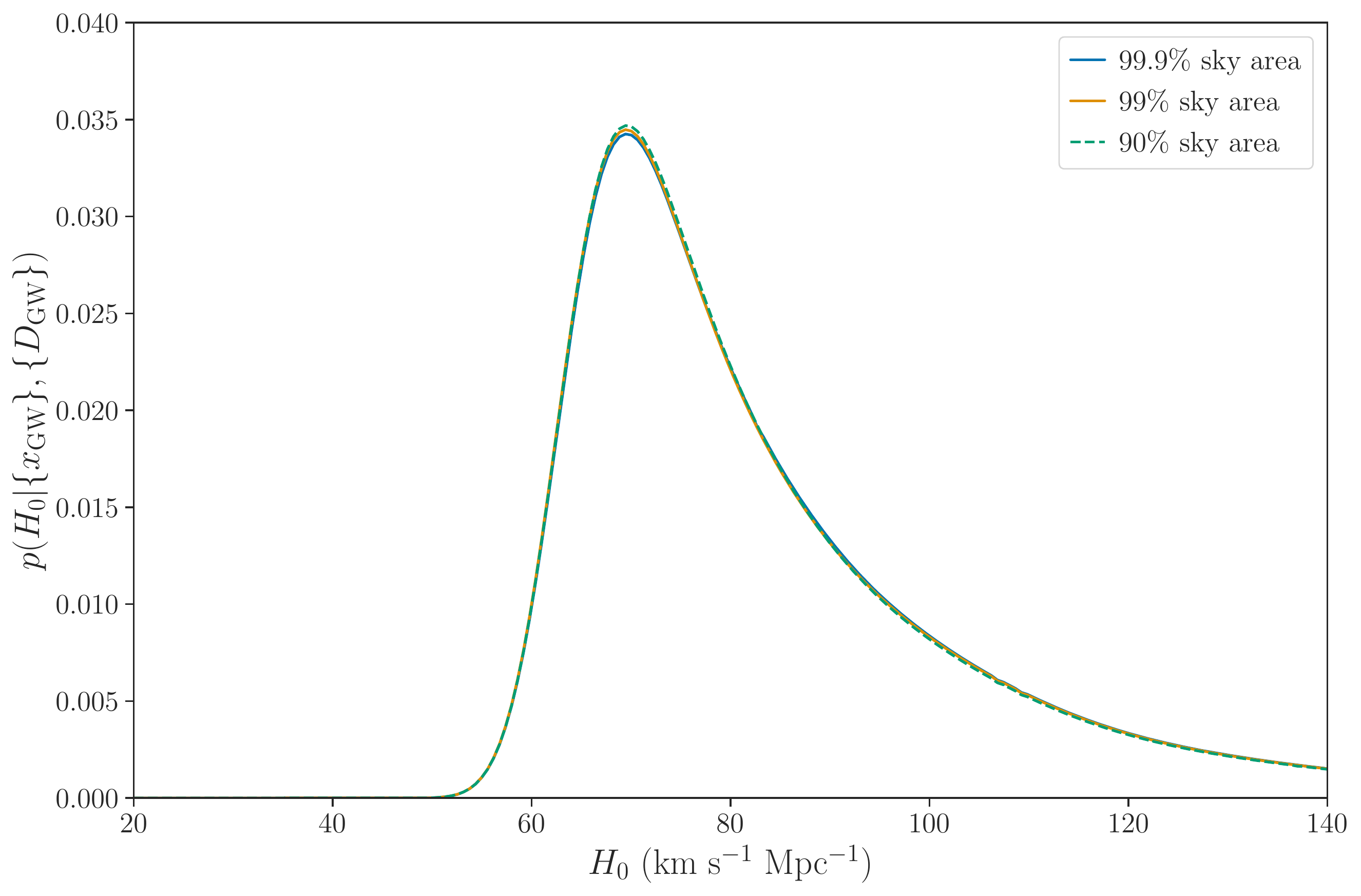}
\caption[Comparison of the posterior on $H_0$ when the threshold on the GW sky area used in the pixelated analysis is reduced]{Comparison of the posterior on $H_0$ when the threshold on the GW sky area used in the pixelated analysis is reduced.  The blue line shows the main result for 6 \acp{BBH} and the \ac{BNS}, where the 99.9\% sky area for each \ac{BBH} is analysed. The orange line corresponds to the 99\% sky area, and the green line to the 90\% sky area.}
\label{Fig:pixel_skyarea_comp}
\end{figure}

In the first instance, the threshold on the \ac{GW} sky area is varied. The results in Section \ref{sec:pixelO2H0} made use of the 99.9\% sky area. Figure \ref{Fig:pixel_skyarea_comp} demonstrates the change to the final posterior on \ac{H0} when the area is reduced to 99\% and 90\% for the \acp{BBH} under consideration. The change to the final posterior is marginal, with a slight increase in the height of the peak corresponding to the smaller sky areas. This is expected as the less informative edges of the \ac{GW} sky distribution are being discarded. In this case, more informative is not necessarily a good thing, as there is the possibility that discarding this additional information could eventually introduce some level of bias to the result. While not important here, that impact should be reassessed in the future, when results from large numbers of \ac{GW} events are being combined.

\begin{figure}
\centering
\includegraphics[width=0.95\linewidth]{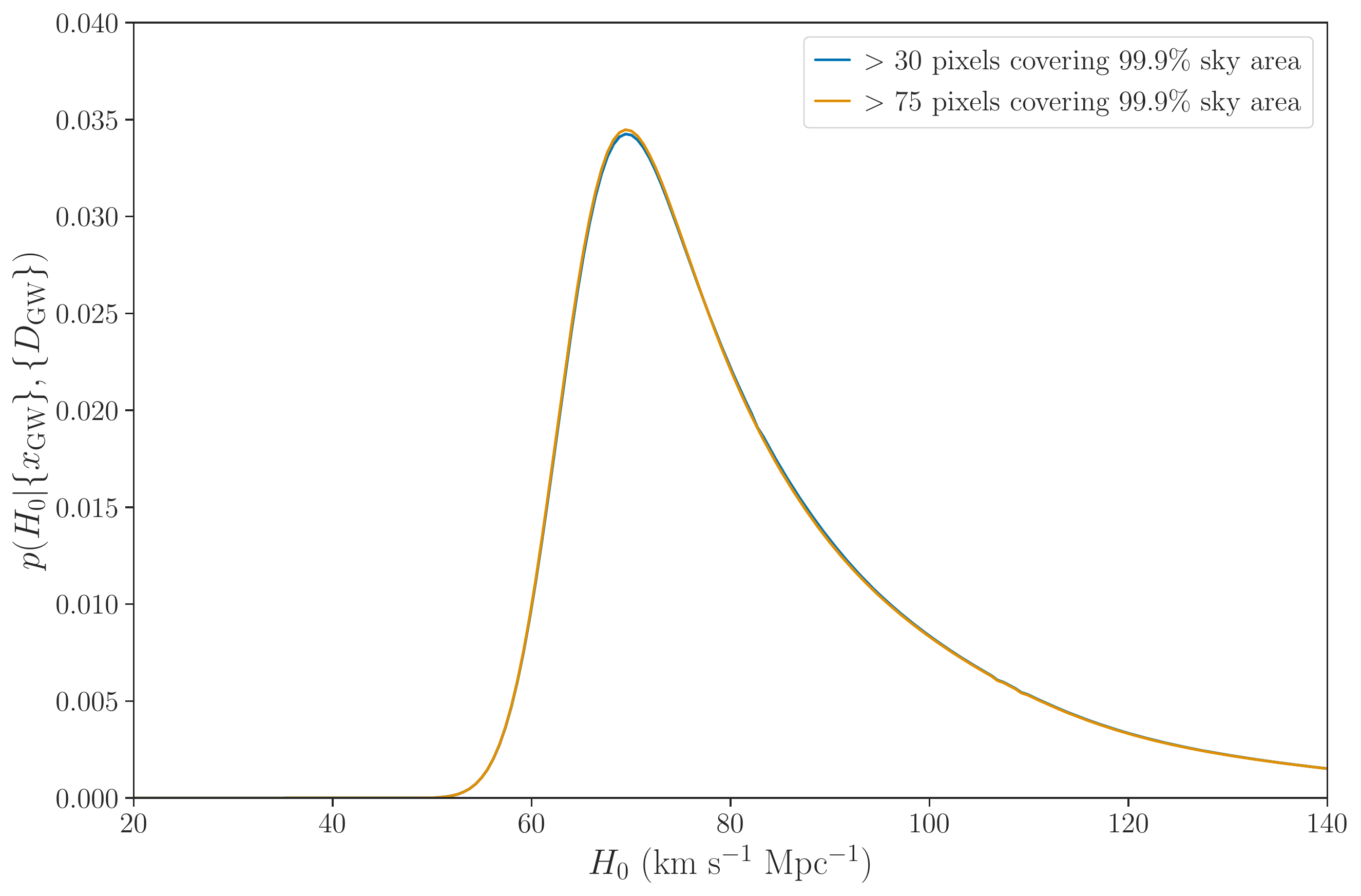}
\caption[Comparison of the posterior on $H_0$ when the minimum number of pixels covering the 99.9\% GW sky area is increased.]{Comparison of the posterior on $H_0$ when the minimum number of pixels covering the 99.9\% GW sky area is increased. The blue line shows the main result for 6 \acp{BBH} and the \ac{BNS}, where at least 30 pixels cover the 99.9\% sky area for each \ac{BBH} is analysed. The orange line corresponds to an analysis where the \ac{GW} data is analysed at one resolution step higher, where at least 75 pixels are used to cover the 99.9\% sky area.}
\label{Fig:pixel_minpix_comp}
\end{figure}

Next, the threshold is reverted to the 99.9\% sky area, but the number of pixels which covers it is increased. The nside which determines the resolution of the \ac{GW} data (column 2 of Table \ref{tab: pixels}) was doubled for each event, which leads to a factor of 4 increase in the number of pixels covering the 99.9\% sky area.\footnote{This is approximate, not exact, as variations in how the \ac{GW} sky probability is distributed between the higher resolution pixels means that not all will necessarily be required in order to reach the 99.9\% threshold.} The impact on the \ac{H0} posterior, shown in Fig. \ref{Fig:pixel_minpix_comp}, is a marginal increase in the height of the peak when the higher resolution is applied to the \ac{GW} data. The fact that the difference is so small should be taken as additional confirmation that a relatively low number of pixels can adequately represent the variation in the \ac{GW} \ac{los} distance distribution.\footnote{It is also worth noting that the results in Fig. \ref{Fig:pixel_skyarea_comp} and Fig. \ref{Fig:pixel_minpix_comp} are correlated, as reducing the sky area of the event under consideration results, for some events, in an increase in pixel resolution in order to keep the number of pixels covering the sky area roughly the same.}

\begin{figure}
\centering
\includegraphics[width=0.95\linewidth]{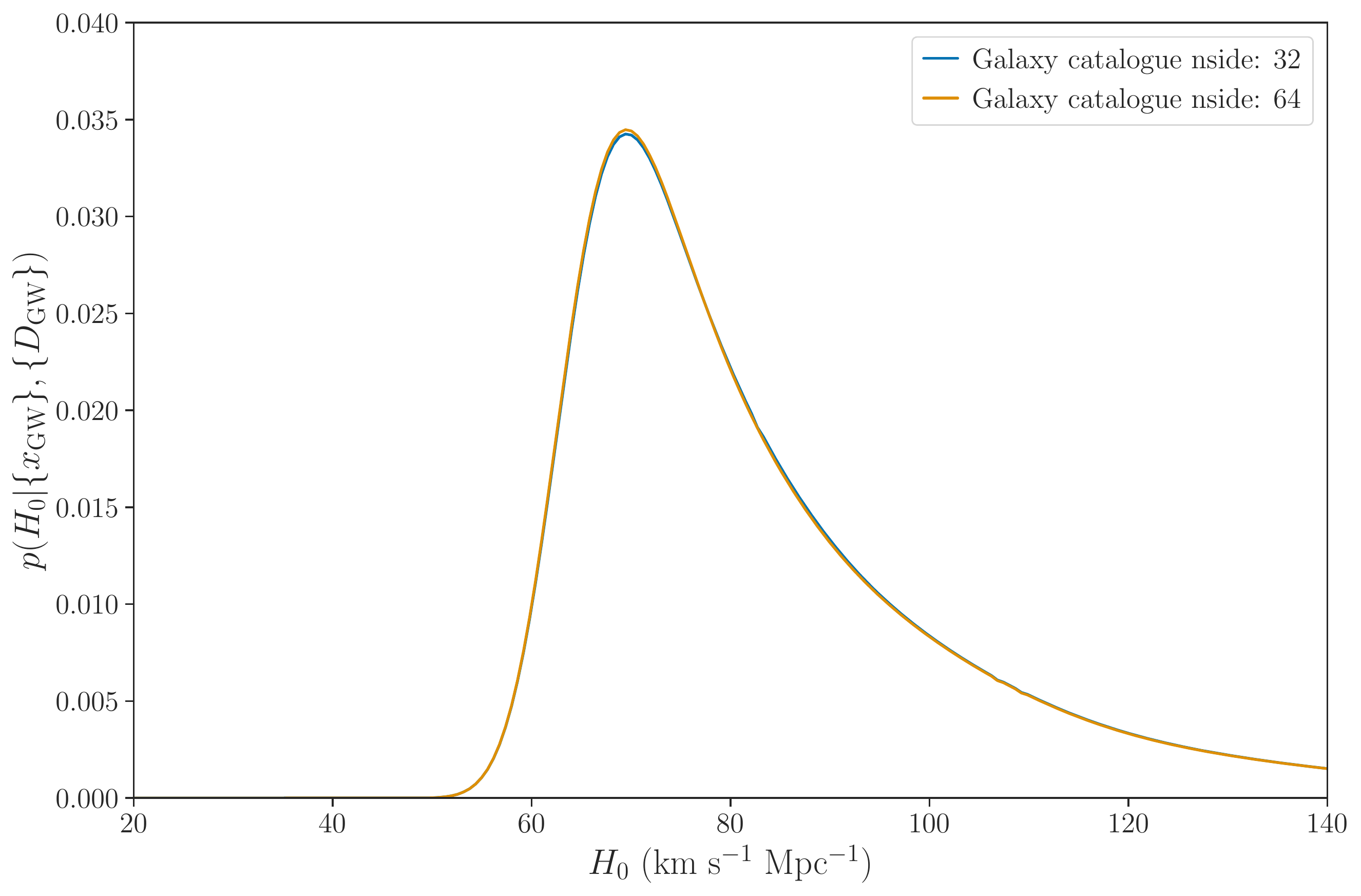}
\caption[Comparison of the posterior on $H_0$ when the galaxy catalogue resolution is increased.]{Comparison of the posterior on $H_0$ when the galaxy catalogue resolution is increased.  The blue line shows the main result for 6 \acp{BBH} and the \ac{BNS}, where an nside of 32 is assumed for both the GLADE and DES catalogue. The orange line shows the same, but where the nside of the both galaxy catalogues has been increased to 64.}
\label{Fig:pixel_nside_comp}
\end{figure}

Finally, the resolution of the galaxy catalogue is investigated, by increasing the nside of the sub-pixels from 32 to 64. This allows for a better representation of the hard edges of the GLADE catalogue where it is intersected by the Milky Way band, as well as better representation of the variation of \ac{mth} across different parts of the survey. The results are shown in Fig. \ref{Fig:pixel_nside_comp}, which shows a very small increase in the height of the peak for the higher resolution results. As the difference is again very small, it is safe to assume that the resolution from nside=32 is adequate to represent the variation in galaxy catalogue completeness over the sky.

\section{Conclusion}
\label{sec:pixelconclusion}
There are two major benefits of using the pixelated method presented in this paper for the measurement of \ac{H0} using standard sirens and galaxy catalogues. The first is that full use is now made of the \ac{GW} data by estimating a separate distance distribution for each pixel which makes up the event's sky area. The fact that these distributions will peak at different distances for different lines of sight -- and will therefore pick out galaxies at different redshifts depending on how they align -- increases the information available for the dark siren analysis. 
The second benefit is that the pixelated method allows for an accurate estimation of how the apparent magnitude threshold -- and hence the completeness -- of a galaxy catalogue changes across the sky. This is particularly important around the Milky Way band, where telescope visibility is limited, and at any boundaries between different observing surveys, between which telescope sensitivities may have changed.  Given how dominant the out-of-catalogue contribution is to the dark siren result for the majority of GWTC-1 events, and will continue to be for the the third observing run and beyond, this is an important milestone.

The results presented in Section \ref{sec:pixelO2H0} show a clear improvement over the results in \citet{O2H0paper} --- around that of 5\%. Not only does the pixelated method make use of the data in a way that is more technically correct, but this leads to a direct increase in the informativeness of the results it produces. These results are both more robust and have more to say. That said, it is worth remembering that both the results in this paper and in \citet{O2H0paper} are sensitive to the choice of \ac{GW} population model which, for a large number of dark sirens, can play a major role in the measurement if not suitably marginalised over. See, \eg, \citet{2021arXiv211103604T}, in which the impact of population assumptions on the \ac{GW} measurement of \ac{H0} using GWTC-3 events is explored in detail.

An investigation into how various choices for the resolution of the pixelated analysis impacts the result revealed that, for the most part, the results are insensitive to reasonable changes. Variations from increasing the resolution of pixels used to represent the \ac{GW} data were small, as were those from increasing the resolution of the galaxy catalogue. These changes may become more important as more events are analysed and the goal to reach 1\% measurement uncertainty on \ac{H0} comes into reach. In the future it may be worth improving the analysis outlined in this paper to use multi-resolution maps to represent the \ac{GW} data: 
well-localised highly probable areas could be treated with high resolution pixels in order to capture the fine detail, and the tail ends of the distribution -- the large areas with very little probability -- could be treated on a lower resolution. This would allow for the most information to be gained from the \ac{GW} data, without having to discard the low probability areas, which it may be necessary to include in order to avoid introducing bias to the result.

Of course, increasing the size of the \ac{GW} pixels means that, in order to reach the resolution set by the galaxy catalogue, each pixel needs to be split into more sub-pixels, which increases the computation time. Currently the computational cost of each pixel is approximately proportional to the number of sub-pixels which it is divided into (which means that the pixels from events with large sky areas have a longer run-time than those which are well-localised, even though these events are unlikely to contribute much information to the final result). 
 The solution to this, which will radically reduce computation time for large pixels, is to merge sub-pixels which have similar apparent magnitude thresholds.  Looking back at the top left panel of Fig. \ref{Fig:event_mthmap}: within the sky-area of GW150914, \ac{mth} varies between approximately 16 and 19 magnitudes - a huge variation in terms of completeness. However it is also clear that the majority of pixels have a threshold around 17.3 (reddish-pink pixels). Pixels in the top left have higher thresholds, around 18.8. Pixels on the edge of the Milky Way band dip below 17. Overall, the variation of \ac{mth} within GW150914's sky-area could be adequately represented by a handful of different thresholds. The costly out-of-catalogue contribution within each pixel, which has to be calculated for every value of \ac{mth}, would then be reduced from 16, in this case, down to only several.  And if the resolution of the catalogue was increased at some point in the future, the number of out-of-catalogue calculations required would not increase, allowing incredibly high resolution of features such as empty patches and boundaries between surveys, with no extra computational cost. Following a similar approach to \citet{Finke:2021aom}, and computing a completeness map (but in terms of \ac{mth}) that assigns pixels to groups of similar completeness would be one way to do this.

In summary, the pixelated analysis demonstrated in this paper shows a clear improvement on the analysis in \citet{O2H0paper}. It improves the final constraint on \ac{H0} through better, more effective use of the \ac{GW} and galaxy catalogue data available. The implementation demonstrated here has clear avenues for future development, both in terms of further improving the accuracy of the analysis, and improving its efficiency. This pixelated approach has potential to become the main method for any future \ac{H0} analysis with dark standard sirens and galaxy catalogues.

\section*{Acknowledgements}
The authors are grateful to members of the LIGO, Virgo and KAGRA collaborations and specifically those in the cosmology working group for their valuable input, and to Maciej Bilicki and Archisman Ghosh in particular for their helpful comments. The authors are grateful for computational resources provided by the LIGO Laboratory. RG was supported by the Science and Technology Facilities Council (award number 1947165) and by the European Research Council, starting grant SHADE 949572. JV and CM are supported by the Science and Technology Research Council (grant No. ST/V005634/1).


\section*{Data Availability}
GWTC-1 data is available at \url{https://www.gw-openscience.org/GWTC-1}. The GLADE 2.4 catalogue is available at \url{http://glade.elte.hu}, and the DES-Y1 catalogue is available at \url{https://des.ncsa.illinois.edu/releases/y1a1}. \gwcosmo is available at \url{https://git.ligo.org/lscsoft/gwcosmo}.
 



\bibliographystyle{mnras}
\bibliography{masterbib} 

\begin{thebibliography}{}
\makeatletter
\relax
\def\mn@urlcharsother{\let\do\@makeother \do\$\do\&\do\#\do\^\do\_\do\%\do\~}
\def\mn@doi{\begingroup\mn@urlcharsother \@ifnextchar [ {\mn@doi@}
  {\mn@doi@[]}}
\def\mn@doi@[#1]#2{\def\@tempa{#1}\ifx\@tempa\@empty \href
  {http://dx.doi.org/#2} {doi:#2}\else \href {http://dx.doi.org/#2} {#1}\fi
  \endgroup}
\def\mn@eprint#1#2{\mn@eprint@#1:#2::\@nil}
\def\mn@eprint@arXiv#1{\href {http://arxiv.org/abs/#1} {{\tt arXiv:#1}}}
\def\mn@eprint@dblp#1{\href {http://dblp.uni-trier.de/rec/bibtex/#1.xml}
  {dblp:#1}}
\def\mn@eprint@#1:#2:#3:#4\@nil{\def\@tempa {#1}\def\@tempb {#2}\def\@tempc
  {#3}\ifx \@tempc \@empty \let \@tempc \@tempb \let \@tempb \@tempa \fi \ifx
  \@tempb \@empty \def\@tempb {arXiv}\fi \@ifundefined
  {mn@eprint@\@tempb}{\@tempb:\@tempc}{\expandafter \expandafter \csname
  mn@eprint@\@tempb\endcsname \expandafter{\@tempc}}}

\bibitem[\protect\citeauthoryear{{Abbott} et~al.,}{{Abbott}
  et~al.}{2017a}]{GW170817:discovery}
{Abbott} B.~P.,  et~al., 2017a, \mn@doi [Phys. Rev. Lett.]
  {10.1103/PhysRevLett.119.161101}, 119, 161101

\bibitem[\protect\citeauthoryear{{Abbott} et~al.,}{{Abbott}
  et~al.}{2017b}]{GW170817:H0}
{Abbott} B.~P.,  et~al., 2017b, \mn@doi [Nature] {10.1038/nature24471}, \href
  {http://adsabs.harvard.edu/abs/2017Natur.551...85A} {551, 85}

\bibitem[\protect\citeauthoryear{{Abbott} et~al.,}{{Abbott}
  et~al.}{2017c}]{GW170817:MMA}
{Abbott} B.~P.,  et~al., 2017c, \mn@doi [Astrophys. J.]
  {10.3847/2041-8213/aa91c9}, \href
  {https://ui.adsabs.harvard.edu/#abs/2017ApJ...848L..12A} {848, L12}

\bibitem[\protect\citeauthoryear{Abbott et~al.}{Abbott
  et~al.}{2018}]{Abbott:2018jhe}
Abbott T. M.~C.,  et~al., 2018, \mn@doi [Astrophys. J. Suppl.]
  {10.3847/1538-4365/aae9f0}, 239, 18

\bibitem[\protect\citeauthoryear{Abbott et~al.}{Abbott
  et~al.}{2019}]{O2:catalog}
Abbott B.~P.,  et~al., 2019, \mn@doi [Phys. Rev. X]
  {10.1103/PhysRevX.9.031040}, 9, 031040

\bibitem[\protect\citeauthoryear{Abbott et~al.}{Abbott
  et~al.}{2020}]{GW1908142020}
Abbott R.,  et~al., 2020, \mn@doi [Astrophys. J. Lett.]
  {10.3847/2041-8213/ab960f}, 896, L44

\bibitem[\protect\citeauthoryear{Abbott et~al.}{Abbott
  et~al.}{2021a}]{PhysRevX.11.021053}
Abbott R.,  et~al., 2021a, \mn@doi [Phys. Rev. X] {10.1103/PhysRevX.11.021053},
  11, 021053

\bibitem[\protect\citeauthoryear{{Abbott} et~al.}{{Abbott}
  et~al.}{2021b}]{O2H0paper}
{Abbott} B.~P.,  et~al., 2021b, \mn@doi [Astrophys. J.]
  {10.3847/1538-4357/abdcb7}, \href
  {https://ui.adsabs.harvard.edu/abs/2021ApJ...909..218A} {909, 218}

\bibitem[\protect\citeauthoryear{Abbott et~al.}{Abbott
  et~al.}{2021c}]{NSBH2021}
Abbott R.,  et~al., 2021c, \mn@doi [Astrophys. J. Lett.]
  {10.3847/2041-8213/ac082e}, 915, L5

\bibitem[\protect\citeauthoryear{Blanton et~al.,}{Blanton
  et~al.}{2003}]{Blanton_2003}
Blanton M.~R.,  et~al., 2003, \mn@doi [Astrophys. J.] {10.1086/375776}, 592,
  819

\bibitem[\protect\citeauthoryear{{Chen}, {Essick}, {Vitale}, {Holz}  \&
  {Katsavounidis}}{{Chen} et~al.}{2017}]{2017ApJ...835...31C}
{Chen} H.-Y.,  {Essick} R.,  {Vitale} S.,  {Holz} D.~E.,   {Katsavounidis} E.,
  2017, \mn@doi [apj] {10.3847/1538-4357/835/1/31}, \href
  {https://ui.adsabs.harvard.edu/abs/2017ApJ...835...31C} {835, 31}

\bibitem[\protect\citeauthoryear{{Chen}, {Fishbach}  \& {Holz}}{{Chen}
  et~al.}{2018}]{Chen:2017rfc}
{Chen} H.-Y.,  {Fishbach} M.,   {Holz} D.~E.,  2018, \mn@doi [Nature]
  {10.1038/s41586-018-0606-0}, \href
  {https://ui.adsabs.harvard.edu/abs/2018Natur.562..545C} {562, 545}

\bibitem[\protect\citeauthoryear{{Del Pozzo}}{{Del
  Pozzo}}{2012}]{DelPozzo:2012}
{Del Pozzo} W.,  2012, \mn@doi [Phys. Rev. D] {10.1103/PhysRevD.86.043011},
  \href {http://adsabs.harvard.edu/abs/2012PhRvD..86d3011D} {86, 043011}

\bibitem[\protect\citeauthoryear{Drlica-Wagner et~al.}{Drlica-Wagner
  et~al.}{2018}]{Drlica-Wagner:2017tkk}
Drlica-Wagner A.,  et~al., 2018, \mn@doi [Astrophys. J. Suppl.]
  {10.3847/1538-4365/aab4f5}, 235, 33

\bibitem[\protect\citeauthoryear{Dálya et~al.,}{Dálya
  et~al.}{2018}]{Dalya:2018cnd}
Dálya G.,  et~al., 2018, \mn@doi [Mon. Not. R. Astron. Soc.]
  {10.1093/mnras/sty1703}, 479, 2374

\bibitem[\protect\citeauthoryear{Farr, Fishbach, Ye  \& Holz}{Farr
  et~al.}{2019}]{Farr:2019twy}
Farr W.~M.,  Fishbach M.,  Ye J.,   Holz D.,  2019, \mn@doi [Astrophys. J.
  Lett.] {10.3847/2041-8213/ab4284}, 883, L42

\bibitem[\protect\citeauthoryear{Finke, Foffa, Iacovelli, Maggiore  \&
  Mancarella}{Finke et~al.}{2021}]{Finke:2021aom}
Finke A.,  Foffa S.,  Iacovelli F.,  Maggiore M.,   Mancarella M.,  2021,
  \mn@doi [JCAP] {10.1088/1475-7516/2021/08/026}, 08, 026

\bibitem[\protect\citeauthoryear{Fishbach, Holz  \& Farr}{Fishbach
  et~al.}{2018}]{Fishbach:2018edt}
Fishbach M.,  Holz D.~E.,   Farr W.~M.,  2018, \mn@doi [Astrophys. J.]
  {10.3847/2041-8213/aad800}, 863, L41

\bibitem[\protect\citeauthoryear{Fishbach, Gray, Magaña~Hernandez, Qi, Sur
  et~al.}{Fishbach et~al.}{2019}]{Fishbach:2018gjp}
Fishbach M.,  Gray R.,  Magaña~Hernandez I.,  Qi H.,  Sur A.,   et~al., 2019,
  \mn@doi [Astrophys. J.] {10.3847/2041-8213/aaf96e}, 871, L13

\bibitem[\protect\citeauthoryear{Gehrels, Cannizzo, Kanner, Kasliwal, Nissanke
  \& Singer}{Gehrels et~al.}{2016}]{Gehrels:2015uga}
Gehrels N.,  Cannizzo J.~K.,  Kanner J.,  Kasliwal M.~M.,  Nissanke S.,
  Singer L.~P.,  2016, \mn@doi [Astrophys. J.] {10.3847/0004-637X/820/2/136},
  820, 136

\bibitem[\protect\citeauthoryear{{G{\'o}rski}, {Hivon}, {Banday}, {Wandelt},
  {Hansen}, {Reinecke}  \& {Bartelmann}}{{G{\'o}rski}
  et~al.}{2005}]{2005ApJ...622..759G}
{G{\'o}rski} K.~M.,  {Hivon} E.,  {Banday} A.~J.,  {Wandelt} B.~D.,  {Hansen}
  F.~K.,  {Reinecke} M.,   {Bartelmann} M.,  2005, \mn@doi [Astrophys. J.]
  {10.1086/427976}, \href {http://adsabs.harvard.edu/abs/2005ApJ...622..759G}
  {622, 759}

\bibitem[\protect\citeauthoryear{Gray et~al.}{Gray et~al.}{2020}]{Gray:2019ksv}
Gray R.,  et~al., 2020, \mn@doi [Phys. Rev. D] {10.1103/PhysRevD.101.122001},
  101, 122001

\bibitem[\protect\citeauthoryear{MacLeod \& Hogan}{MacLeod \&
  Hogan}{2008}]{MacLeod:2007jd}
MacLeod C.~L.,  Hogan C.~J.,  2008, \mn@doi [Phys. Rev.]
  {10.1103/PhysRevD.77.043512}, D77, 043512

\bibitem[\protect\citeauthoryear{Mastrogiovanni et~al.,}{Mastrogiovanni
  et~al.}{2021}]{PhysRevD.104.062009}
Mastrogiovanni S.,  et~al., 2021, \mn@doi [Phys. Rev. D]
  {10.1103/PhysRevD.104.062009}, 104, 062009

\bibitem[\protect\citeauthoryear{{Mukherjee}, {Wandelt}, {Nissanke}  \&
  {Silvestri}}{{Mukherjee} et~al.}{2021}]{2021PhRvD.103d3520M}
{Mukherjee} S.,  {Wandelt} B.~D.,  {Nissanke} S.~M.,   {Silvestri} A.,  2021,
  \mn@doi [prd] {10.1103/PhysRevD.103.043520}, \href
  {https://ui.adsabs.harvard.edu/abs/2021PhRvD.103d3520M} {103, 043520}

\bibitem[\protect\citeauthoryear{Palmese et~al.,}{Palmese
  et~al.}{2020}]{GW190814:DES}
Palmese A.,  et~al., 2020, \mn@doi [Astrophys. J. Lett.]
  {10.3847/2041-8213/abaeff}, 900, L33

\bibitem[\protect\citeauthoryear{{Planck Collaboration} et~al.,}{{Planck
  Collaboration} et~al.}{2020}]{Aghanim:2018eyx}
{Planck Collaboration} et~al., 2020, \mn@doi [Astron. Astrophys.]
  {10.1051/0004-6361/201833910}, \href
  {https://ui.adsabs.harvard.edu/abs/2020A&A...641A...6P} {641, A6}

\bibitem[\protect\citeauthoryear{Riess, Casertano, Yuan, Macri  \&
  Scolnic}{Riess et~al.}{2019}]{Riess:2019cxk}
Riess A.~G.,  Casertano S.,  Yuan W.,  Macri L.~M.,   Scolnic D.,  2019,
  \mn@doi [Astrophys. J.] {10.3847/1538-4357/ab1422}, 876, 85

\bibitem[\protect\citeauthoryear{{Schechter}}{{Schechter}}{1976}]{Schechter:1976}
{Schechter} P.,  1976, \mn@doi [Astrophys. J.] {10.1086/154079}, \href
  {https://ui.adsabs.harvard.edu/abs/1976ApJ...203..297S} {203, 297}

\bibitem[\protect\citeauthoryear{{Schutz}}{{Schutz}}{1986}]{Schutz:1986}
{Schutz} B.~F.,  1986, \mn@doi [Nature] {10.1038/323310a0}, \href
  {http://adsabs.harvard.edu/abs/1986Natur.323..310S} {323, 310}

\bibitem[\protect\citeauthoryear{Scott}{Scott}{1992}]{Scott}
Scott D.~W.,  1992, Multivariate Density Estimation: Theory, Practice and
  Visualization.
John Wiley \& Sons, Inc.

\bibitem[\protect\citeauthoryear{Soares-Santos et~al.}{Soares-Santos
  et~al.}{2019}]{Soares-Santos:2019irc}
Soares-Santos M.,  et~al., 2019, \mn@doi [Astrophys. J.]
  {10.3847/2041-8213/ab14f1}, 876, L7

\bibitem[\protect\citeauthoryear{{The LIGO Scientific Collaboration}
  et~al.,}{{The LIGO Scientific Collaboration}
  et~al.}{2021}]{2021arXiv211103604T}
{The LIGO Scientific Collaboration} et~al., 2021, arXiv e-prints, \href
  {https://ui.adsabs.harvard.edu/abs/2021arXiv211103604T} {p. arXiv:2111.03604}

\bibitem[\protect\citeauthoryear{Vasylyev \& Filippenko}{Vasylyev \&
  Filippenko}{2020}]{Vasylyev2020}
Vasylyev S.~S.,  Filippenko A.~V.,  2020, \mn@doi [Astrophys. J. Lett.]
  {10.3847/1538-4357/abb5f9}, 902, 149

\bibitem[\protect\citeauthoryear{Virtanen et~al.,}{Virtanen
  et~al.}{2020}]{2020SciPy-NMeth}
Virtanen P.,  et~al., 2020, \mn@doi [Nature Methods]
  {10.1038/s41592-019-0686-2}, \href {https://rdcu.be/b08Wh} {17, 261}

\bibitem[\protect\citeauthoryear{Zonca, Singer, Lenz, Reinecke, Rosset, Hivon
  \& Gorski}{Zonca et~al.}{2019}]{Zonca2019}
Zonca A.,  Singer L.,  Lenz D.,  Reinecke M.,  Rosset C.,  Hivon E.,   Gorski
  K.,  2019, \mn@doi [Journal of Open Source Software] {10.21105/joss.01298},
  4, 1298

\makeatother
\end{thebibliography}





\bsp	
\label{lastpage}
\end{document}